\documentclass[twocolumn]{aa} 
\usepackage{graphicx}
\usepackage{amsmath}
\usepackage{txfonts}
\usepackage{natbib}
\usepackage{textcomp}
\usepackage{subfigure}
\newcommand{\tablenotea}[1]{\parbox{8.8cm}{ \indent
\footnotesize{\textsc{Note.--}~#1}}}
\begin{document}
\title{Nitrogen isotopic ratios in Barnard 1: a consistent study of the N$_2$H$^+$, NH$_3$, CN, HCN and HNC isotopologues .}
\titlerunning{N$_2$H$^+$, NH$_3$, CN, HCN and HNC isotopologues in B1}
\authorrunning{F. Daniel et al.}
\author{F. Daniel\inst{1}, M. Gerin\inst{2}, E. Roueff\inst{3}, J. Cernicharo\inst{1}, N. Marcelino \inst{4}, 
F. Lique\inst{5}, D.C.  Lis\inst{6}, D. Teyssier\inst{7}, N. Biver\inst{8}, D. Bockel\'ee--Morvan\inst{8}}
\institute{Departamento de Astrof\'isica, Centro de
Astrobiolog\'ia, CSIC-INTA, Ctra. de Torrej\'on a Ajalvir km 4,
28850 Madrid, Spain; \email{danielf@cab.inta-csic.es} 
\and
LERMA, UMR 8112 du CNRS, Observatoire de Paris, Ecole Normale Sup\'erieure, France
\and 
Observatoire de Paris, LUTH UMR CNRS 8102, 5 Place Janssen, 92195 Meudon, France
\and
NRAO, 520 Edgemont Road, Charlottesville, VA 22902, USA
\and
LOMC-UMR 6294, CNRS-Universit\'e du Havre, 25 rue Philippe Lebon, BP 540 76058 Le Havre France
\and
California Institute of Technology, Cahill Center for Astronomy and Astrophysics 301-17, Pasadena, CA 91125, USA
\and
European Space Astronomy Centre, ESA, PO Box 78, 28691, Villanueva de la Ca\~nada, Madrid, Spain
\and
LESIA, Observatoire de Paris, CNRS, UPMC, Universit\'e Paris-Diderot, 5 place Jules Janssen, 92195, Meudon, France
}

\date{Received; accepted}

\abstract
{The $^{15}$N isotopologue abundance ratio measured today in different bodies of the solar system
is thought to be connected to $^{15}$N--fractionation effects that would have occured in the protosolar nebula.}
{The present study aims at putting constraints on the degree of $^{15}$N--fractionation that occurs during the prestellar phase,
through observations of D, $^{13}$C and $^{15}$N--substituted isotopologues towards B1b. Both molecules from
the nitrogen hydride family, i.e. N$_2$H$^+$ and NH$_3$, and from the nitrile family, i.e. HCN, HNC and CN, are considered in the analysis.}
{As a first step, we model the continuum emission in order to derive the physical structure of the cloud, i.e. gas temperature
and H$_2$ density. These parameters are subsequently used as an input in a non--local radiative transfer model to infer the 
radial abundances profiles of the various molecules.}
{Our modeling shows that all the molecules are affected by depletion onto dust grains, in the region that encompasses 
the B1--bS and B1--bN cores. While high levels of deuterium fractionation are derived, we conclude that no fractionation
occurs in the case of the nitrogen chemistry. Independently of the chemical family, the molecular 
abundances are consistent with
$^{14}$N/$^{15}$N$\sim$300, a value representative of the elemental atomic abundances 
of the parental gas.}
{The inefficiency of the $^{15}$N--fractionation effects in the B1b region can be linked to the relatively high gas temperature 
$\sim$17K which is representative of the innermost part of the cloud. Since this region shows signs of depletion
onto dust grains, we can not exclude the possibility that the molecules were previously enriched 
in $^{15}$N, earlier in the B1b history, and that such an enrichment could have 
been incorporated into the ice mantles. It is thus necessary to repeat this kind of study in colder sources 
to test such a possibility.}


\maketitle
%

\section{Introduction}

Characterizing the nitrogen chemistry is of great interest in studies of star 
forming regions. Moreover, the possible existence of $^{15}$N--fractionation effects could 
have strong implications for our understanding of the current solar system. Indeed,  
large variations have been found in the $^{14}$N/$^{15}$N ratio among different solar system bodies  
\citep[see e.g.][]{jehin2009}. 
While this elemental ratio 
is $\sim$270 on Earth, it is found to be of the order of 450 in the solar wind and Jupiter atmosphere.
The latter value is thought to be representative of the protosolar nebula \citep[see e.g.][]{jehin2009}. Finally, this ratio is 
in the range 100--150 in cometary ices, as deduced from the 
analysis of HCN or CN observations. In the coma of comets, most of the CN radicals  
come from the photodissociation of HCN molecules and, as a consequence, the 
$^{14}$N/$^{15}$N ratios derived from both molecules are necessarily identicals.
To date, remote estimates of the nitrogen isotopic elemental ratio in comets only rely on these two 
molecules.
To the contrary, the analysis of the samples of the \textit{Stardust} mission did not report such high
differences. Indeed, the $^{14}$N/$^{15}$N ratio was only found to differ by $\sim$20\% with respect
to the Earth value \citep{stadermann2008}. 
An explanation for such a discrepancy could come from a possible enrichment of the HCN
molecule in $^{15}$N atoms, due to zero--point chemistry. This would lead to a lower value 
for the HC$^{14}$N/HC$^{15}$N ratio compared to the elemental $^{14}$N/$^{15}$N ratio. Since the nitriles are 
not the main career of nitrogen atoms, such a process could reconcile the values obtained from 
the HCN or CN observations with the in--situ measurements.
  
Still today, the nitrogen fractionation chemistry is poorly constrained. Pioneering astrochemical models were 
computed by \citet{terzieva2000} who predicted that the $^{15}$N--enrichment should remain low
enough to prevent detection. However, it was 
subsequently shown by \citet{charnley2002} and \citet{rodgers2003,rodgers2004,rodgers2008} that, in the regions where CO is highly depleted, the level of fractionation could
be high for some species. These models predicted that N$_2$H$^+$ should show the highest degree of fractionation
and more generally, that the nitrogen hydride family should be highly fractioned. 
More recently, \citet{wirstrom2012} extended the previous studies and included the dependence on the ortho and para 
symmetries of H$_2$ in the chemical network. They showed that, contrary to what was predicted earlier, the nitriles
should show the highest degree of fractionation. This latest result is particularly interesting since it would explain the low 
values obtained for the HC$^{14}$N/HC$^{15}$N ratio in comets. Additionally, since the nitrogen hydrides, which are 
the main carriers of nitrogen atoms,
are not highly fractioned, this would enable to reconcile the in--situ and remote observations.

Observationally, evidence in favor or against possible fractionation effects 
are still rare. In part, this comes from the fact 
that few studies have been dedicated to this problem. Principally, previous works concerned
NH$_3$ \citep{lis2010}, NH$_2$D \citep{gerin2009}, 
HCN \citep{dahmen1995,ikeda2002,hily-blant2013}, HCN and HNC \citep{tennekes2006}, 
HNC and CN \citep{adande2012}, or N$_2$H$^+$ \citep{bizzocchi2010,bizzocchi2013}.
Moreover, most of the studies performed to date
focused on a single molecular species (or a combination of two species from the same chemical family),
which prevents from distinguishing source--to--source variations of the elemental
$^{14}$N/$^{15}$N abundance ratio from fractionation effects related to the species under study. Another uncertainty 
concerns the methodology used. Indeed, many studies rely on the so--called double isotope method in order
to retrieve the column density of the main isotopologue, which leads, to some extent, to an uncertainty linked to the 
value assumed for the fundamental $^{12}$C/$^{13}$C ratio. Additionally, this method should only work as long as the 
$^{13}$C fractionation effects remain small for the species considered. 

Here, we focus on the interpretation of rotational lines observed towards B1b 
for the $^{13}$C, $^{15}$N and D--substituted isotopologues of 
N$_2$H$^+$, NH$_3$, CN, HCN and HNC. 
This source is located in the Perseus molecular cloud, which has been the subject of many studies. 
The whole molecular cloud has been mapped, leading to a cartography of its 
associated extinction \citep{cernicharo1985,carpenter2000,ridge2006}. 
Additionally, large--scale maps have been obtained of the CO isotopologues \citep{bachiller1986,ungerechts1987,hatchell2005,ridge2006}
and of the continuum emission, at various wavelengths \citep{hatchell2005,enoch2006,jorgensen2006}.
From these studies, it appears that this region contains a large number of active star--forming sites, with a few hundreds of protostellar objects identified. Moreover, it was found that the stars tend to form within clusters.
Apart from these extended maps, a large number of dense cores were mapped in other molecular tracers, 
like NH$_3$, C$_2$S \citep{rosolowsky2008,foster2009}, or N$_2$H$^+$ \citep{kirk2007}.

The source B1b has been the subject of many observational studies, because of its molecular richness. 
Indeed, many molecules were observed for the first time in this object, like HCNO \citep{marcelino2009} 
or CH$_3$O \citep{cernicharo2012}. Additionally, B1b shows a high degree of deuterium fractionation and 
has been associated with first detections of multiply deuterated molecules, 
like ND$_3$ \citep{lis2002} or D$_2$CS \citep{marcelino2005}.
The B1b region was mapped in many molecular tracers, e.g. CS, NH$_3$, $^{13}$CO \citep{bachiller1990},
N$_2$H$^+$ \citep{huang2013}, H$^{13}$CO$^+$ \citep{hirano1999}, or CH$_3$OH \citep{hiramatsu2010,oberg2010}.
This source consists of two cores, B1--bS and B1--bN, which were first identified by \citet{hirano1999} and initially
classified as class 0 objects. However, since none of these cores seems to be the driving source of molecular 
outflows \citep{walawender2005,hiramatsu2010} and because of the properties of their spectral energy distribution \citep{pezzuto2012}, 
they may correspond to less evolved objects and were proposed as candidates for the first hydrostatic core stage.

In the present study, we report observations of various isotopologues of N$_2$H$^+$, NH$_3$, CN, HCN and HNC
toward B1b (Sect. \ref{section:observations}). In Sect. \ref{modelling}, the analysis of these
observations is described. The physical model of the source is based on the analysis of previously published continuum data 
(Sect. \ref{modelling:continuum}), which serves as input for the non--local radiative transfer analysis of the molecular data 
(Sect. \ref{modelling:molecules}). The current results are then discussed in Sect. \ref{discussion} and the concluding 
remarks are given in Sect. \ref{conclusion}.

\section{Observations}\label{section:observations}

The list of molecular transitions observed as well as the parameters that describe the beam of the 
telescope at the corresponding line frequencies are summarized in Table \ref{table-observations}.
Note that this dataset contains previously published data of  
the NH$_3$ and NH$_2$D isotopologues, which are respectively taken from \citet{lis2010} and \citet{gerin2009}.
The central position of the observations is $\alpha$ = 3$^h$33$^m$20.90$^s$, $\delta$ = 31$\degr$07$\arcmin$34.0$\arcsec$ (J2000).

The observations described in this paper have been performed with
the IRAM--30m telescope, except for the p-NH$_3$ and p-$^{15}$NH$_3$ 
data which were obtained with the Green Bank telescope and
described by \citet{lis2010}.
The IRAM--30m data have been obtained in various observing sessions
in November 2001, August 2004, December 2007, using 
the heterodyne receivers A100/B100, C150/D150, A230/B230, C270/D270 
for spectral line transitions in the 80 -- 115~GHz, 130 -- 170~GHz, 
210 -- 260~GHz  and 250 -- 280~GHz frequency ranges, respectively.
The tuning of the receivers was optimized to allow simultaneous observations 
at two different frequencies simultaneously. These receivers were operated
in lower sideband, with a rejection of the upper sideband
larger than 15dB. The data obtained at frequencies lower than 80~GHz required 
a specific calibration, as the receivers were operated in a double sideband 
mode at these low frequencies. The sideband ratio
was calibrated using planets (Mars, Uranus, Neptune) as gain calibrators. 
The spectra were analysed with the
Versatile SPectrometer Array, achieving a spectral resolution ranging from 20kHz for the lower 
frequency lines (0.07 km s$^{-1}$) to 80 kHz (0.09 km s$^{-1}$) for the higher frequency transitions 
like N$_2$H$^+$(3--2). For the weakest lines,
the spectra have been smoothed to a spectral resolution of $\sim$0.12 km s$^{-1}$
to increase the S/N ratio. We used two observing modes. Frequency switching with
a frequency throw of $\sim 7.2$~MHz at 100 and 150 GHz and of
  $\sim 14.4$~MHz at 230 and 270 GHz, during the 2001 and 2004 observation campaigns 
(i.e. for the N$_2$H$^+$ and NH$_2$D maps as well as for the deuterated species). To limit the contamination
of the line profiles by baseline ripples and ensure flat baselines, 
we also used the wobbler switching mode during the 2007 campaign, 
using a throw of 4$\arcmin$, since the line and continuum maps
showed that most of the emission of the B1b core is concentrated in a region
of less than 2$\arcmin$.

The pointing was checked on
nearby quasars, while we used Uranus for checking the focus of the telescope. 
The spectra have been calibrated using the applicable forward and 
main beam efficiencies of 0.95/0.75, 0.93/0.64, 0.91/0.58 , 0.88/0.5 for
the A100/B100, C150/D150, A230/B230, C270/D270 mixers respectively. 
As the radiative transfer model includes the convolution with the telescope beam, we present all spectra in the antenna temperature
scale T$_A^*$. 

The spectra have been processed with the CLASS (Continuum and Line Analysis Single-dish Software) 
software (www.iram.fr/IRAMFR/GILDAS). We removed linear baselines  from the wobbler--switched 
spectra, while higher order polynomials were necessary for the frequency switched data. 

\begin{table*}
\caption{Observed line parameters}
\label{table-observations} \centering
\begin{tabular}{lcrccclc}
\hline \hline 
\multicolumn{1}{c}{}           &  \multicolumn{1}{c}{}           &  \multicolumn{1}{c}{$\nu_{\rm 0}$} &  \multicolumn{1}{c}{HPBW} & \multicolumn{1}{c}{$\eta$} & $\delta v$  & $\overline{T}_{ex}$ & $\tau$ \\
\multicolumn{1}{c}{molecule} & \multicolumn{1}{c}{transition} & \multicolumn{1}{c}{(GHz)}         &  \multicolumn{1}{c}{($\arcsec$)} & \multicolumn{1}{c}{} & \multicolumn{1}{c}{(km s$^{-1}$)}  & (K) &  \\
\hline
\hline  \\ 

p--NH$_3$ & (1,1) & 23.694506 &  33  &   0.80 & 0.08 & $8.11^{+0.34}_{-0.61}$ & 6.4  \\ [0.08cm]
                  & (2,2) & 23.722634 &  33  &   0.80 & 0.08 & $8.88^{+0.01}_{- 0.01}$ & 0.3 \\ [0.08cm]
p--$^{15}$NH$_3$ & (1,1) & 22.624929 &  33  &   0.80 & 0.08 & 7.53 & 0.02 \\ [0.08cm]
o--NH$_2$D & $1_{1,1}s - 1_{0,1}a$ & 85.926278 &    29 &  0.80 & 0.14 & $6.50^{+ 0.28}_{-0.45}$ & 6.4 \\ [0.08cm]
o--$^{15}$NH$_2$D & $1_{1,1}s - 1_{0,1}a$ & 86.420128 &    29 &  0.80 & 0.13 & 5.04 & 0.02 \\ [0.08cm]
\hline
N$_2$H$^+$ & 1-0 & 93.173772  &    27 &  0.79 & 0.13 & $6.80^{+0.92}_{-0.40}$ & 8.3 \\[0.08cm]
& 3-2 & 279.511858  &    9 &  0.57 & 0.08 & $5.82^{+1.38}_{-0.91}$ & 5.7 \\[0.08cm]

N$_2$D$^+$ & 1-0 & 77.109622 &    32 &  0.83 & 0.08 & $8.41^{+0.45}_{-0.45}$ & 1.7 \\[0.08cm]
& 3-2 & 231.321857 &    10 &  0.60 & 0.05 & $6.42^{+0.32}_{-0.39}$ & 2.3 \\[0.08cm]

$^{15}$NNH$^+$ & 1-0 & 90.263833 &    28 &  0.80 & 0.13 & $5.78^{+0.00}_{-0.00}$ & 0.03 \\[0.08cm]

N$^{15}$NH$^+$ & 1-0 & 91.205741&    28 &  0.80 & 0.13 & $5.73^{+0.00}_{-0.00}$ & 0.03 \\ [0.08cm]
\hline
HCN & 1-0 & 88.631847 &    29 &  0.79 & 0.13 & $5.77^{+2.35}_{-1.65}$ & 55.6 \\ [0.08cm]
 & 2-1 & 177.261222 &    14 &  0.69 & 0.07 &  $4.90^{+1.33}_{-1.35}$ & 89.7 \\ [0.08cm]
 & 3-2 & 265.886499 &     9 &  0.57 & 0.05 &  $4.42^{+0.57}_{-0.92}$ & 36.9 \\ [0.08cm]

H$^{13}$CN & 1-0 & 86.340163 &    29 &  0.79 & 0.14 & $3.28^{+0.13}_{-0.10}$ & 4.2 \\ [0.08cm]
& 2-1 & 172.677956 &    14 &  0.69 & 0.07 &  $3.11^{+0.11}_{-0.13}$ & 3.2 \\ [0.08cm]
& 3-2 & 259.011866 &      9 &  0.60 & 0.18 &  $3.45^{+0.10}_{-0.10}$ & 0.4 \\ [0.08cm]

HC$^{15}$N & 1-0 & 86.054966 &    29 &  0.79 & 0.14 & $3.18^{+0.00}_{-0.00}$ & 0.9  \\ [0.08cm]
 & 2-1 & 172.107957 &    14 &  0.69 & 0.07 & $3.07^{+0.00}_{-0.00}$ & 0.6 \\ [0.08cm]
 & 3-2 & 258.156996 &      9 &  0.60 & 0.05 & $3.47^{+0.00}_{-0.00}$ & 0.1 \\ [0.08cm]
 
DCN & 1-0 & 72.414933 &    34 &  0.84 & 0.08 & $4.53^{+0.44}_{-0.24}$ & 3.3 \\ [0.08cm]
 & 2-1 & 144.828111 &   17 &  0.69 & 0.08 &  $3.71^{+0.38}_{-0.38}$ & 4.9 \\ [0.08cm]
 & 3-2 & 217.238612 &    11 &  0.64 & 0.05 & $3.79^{+0.23}_{-0.29}$ & 1.4 \\ [0.08cm]

D$^{13}$CN & 3-2 & 213.519936 &    12 &  0.66 & 0.11 & $4.04^{+0.09}_{-0.05}$ & 0.8 \\[0.08cm]
\hline 
HNC & 1-0 & 90.663568 &    28 &  0.79 & 0.13 & $8.29^{+0.21}_{-0.24}$ & 8.6 \\[0.08cm]
& 3-2 & 271.981142 &      9 &  0.60 & 0.09 & $6.58^{+0.04}_{-0.06}$ & 11.7 \\[0.08cm]

HN$^{13}$C & 1-0 & 87.090850 &    29 &  0.79 & 0.13 & $5.84^{+0.07}_{-0.07}$ & 0.8  \\[0.08cm]
 & 2-1 & 174.179408 &    14 &  0.69 & 0.13 & $4.51^{+0.02}_{-0.03}$ & 1.2 \\[0.08cm]
 & 3-2 & 261.263310 &     9 &  0.60 & 0.05 & $4.37^{+0.02}_{-0.02}$ & 0.3 \\[0.08cm]

H$^{15}$NC & 1-0 & 88.865717 &    28 &  0.79 & 0.13 & $6.05^{+0.00}_{-0.00}$ & 0.2 \\[0.08cm]
 & 2-1 & 177.729094 &    14 &  0.69 & 0.13 & $4.42^{+0.00}_{-0.00}$ & 0.3 \\[0.08cm]
 & 3-2 & 266.587800 &      9 &  0.60 & 0.18 & $4.62^{+0.00}_{-0.00}$ & 0.06 \\[0.08cm]

DNC & 2-1 & 152.609737 &   16 &  0.69 & 0.08 & 6.98 & 7.0 \\[0.08cm]
& 3-2 & 228.910489 &    11 &  0.66 & 0.05 &  5.06 & 5.0 \\[0.08cm]

DN$^{13}$C & 2-1 & 146.734002 &    17 &  0.69 & 0.16 & 3.96 & 0.4 \\ [0.08cm]
 & 3-2 & 220.097238 &   11 &  0.66 & 0.05 & 3.92 & 0.1 \\ [0.08cm]

\hline

CN & 1-0 & 113.490982 &    21 &  0.79 & 0.20 & $4.62^{+ 1.57}_{-0.50}$ &  34.5 \\ [0.08cm]
     & 2-1 & 226.874781 &    11 &  0.66 & 0.11 & $3.86^{+1.29}_{-0.67}$ & 47.6 \\ [0.08cm]

$^{13}$CN & 1-0 &   108.780203 &    21 &  0.79 & 0.21 & $3.42^{+0.03}_{-0.01}$ & 1.6 \\ [0.08cm]
 & 2-1 & 217.467150 &    11 &  0.64 & 0.11 & $3.43^{+0.03}_{-0.02}$ & 0.8 \\ [0.08cm]

C$^{15}$N & 1-0 & 110.024590 &   21 &  0.79 & 0.11 & $3.39^{+0.01}_{-0.00}$ & 0.3 \\ [0.08cm]

\hline

\hline
\end{tabular}
\tablenotea{Summary of the nitrogen containing species observed at the IRAM telescope (except for the
NH$_3$ isotopologues that were observed at the Green Bank Telescope). 
The NH$_3$ and NH$_2$D isotopologue observations are respectively taken from \citet{lis2010} and \cite{gerin2009}.
All the observations were obtained at a single position except for N$_2$H$^+$ J=1-0 and NH$_2$D, for which we obtained maps.
For each line, we indicate the central frequency of the line ($\nu_0$) as well as the quantum numbers of the corresponding transition. 
The HPBW and efficiency ($\eta$) of the the telescope at this frequency and the spectral resolution ($\delta \nu$) of the observations are also indicated. The two last columns indicate the averaged excitation temperature and the opacity of the rotational lines (summed over the hyperfine components), as derived from the results of the radiative transfer modeling. The upper/lower subscripts for $\overline{T}_{ex}$ indicate the highest/lowest excitation temperatures found when considering all the hyperfine components. These quantities are zero when all the hyperfine components are described by the same excitation temperature. In the case of the DNC isotopologues, we neglected the hyperfine structure and the subscripts are thus absent.}
\end{table*}

\section{Continuum and molecular line modelling}\label{modelling}

The distance of the source is taken to be 235 pc \citep{hirota2008}. At this distance, an angular size of 
10$\arcsec$ corresponds to a physical size of 0.0114 pc.

\subsection{Continuum}\label{modelling:continuum}

\begin{figure}
\begin{center}
\includegraphics[angle=0,scale=0.4]{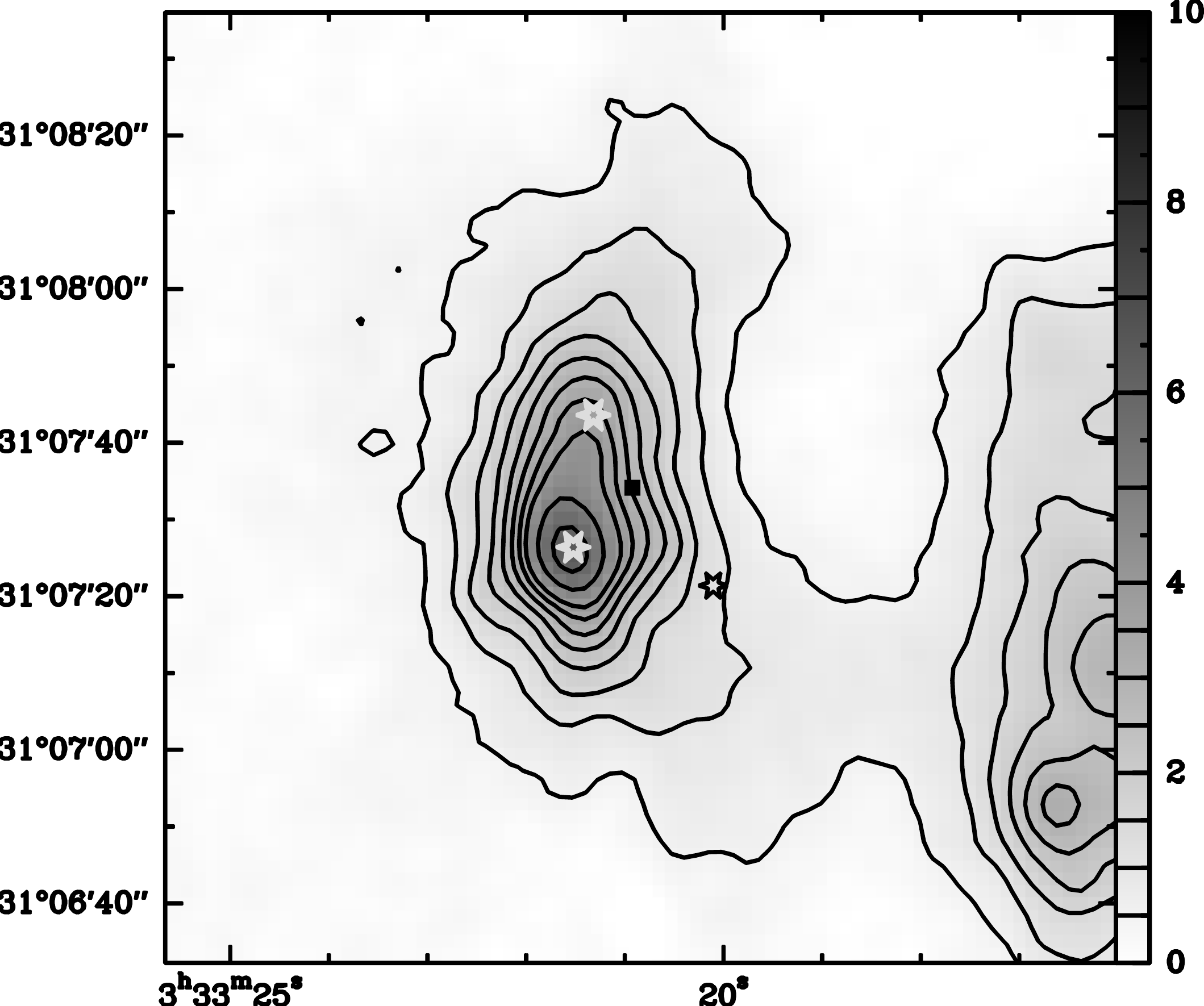}
\caption{Continuum map at 350 $\mu$m obtained at the CSO with the SHARC instrument.
The contours have a step of 0.5 Jy/beam from 0.5 to 4 Jy/beam
and 1 Jy/beam for larger fluxes, as drawn on the right. 
The white stars indicate the position of B1--bS and B1--bN and the black star corresponds
to the \textit{Spitzer} source [EDJ2009] 295. The black dot gives the position of our molecular survey.
} \label{fig:cont350} \vspace{-0.1cm}
\end{center}
\end{figure}

\begin{figure}
\begin{center}
\includegraphics[angle=0,scale=0.4]{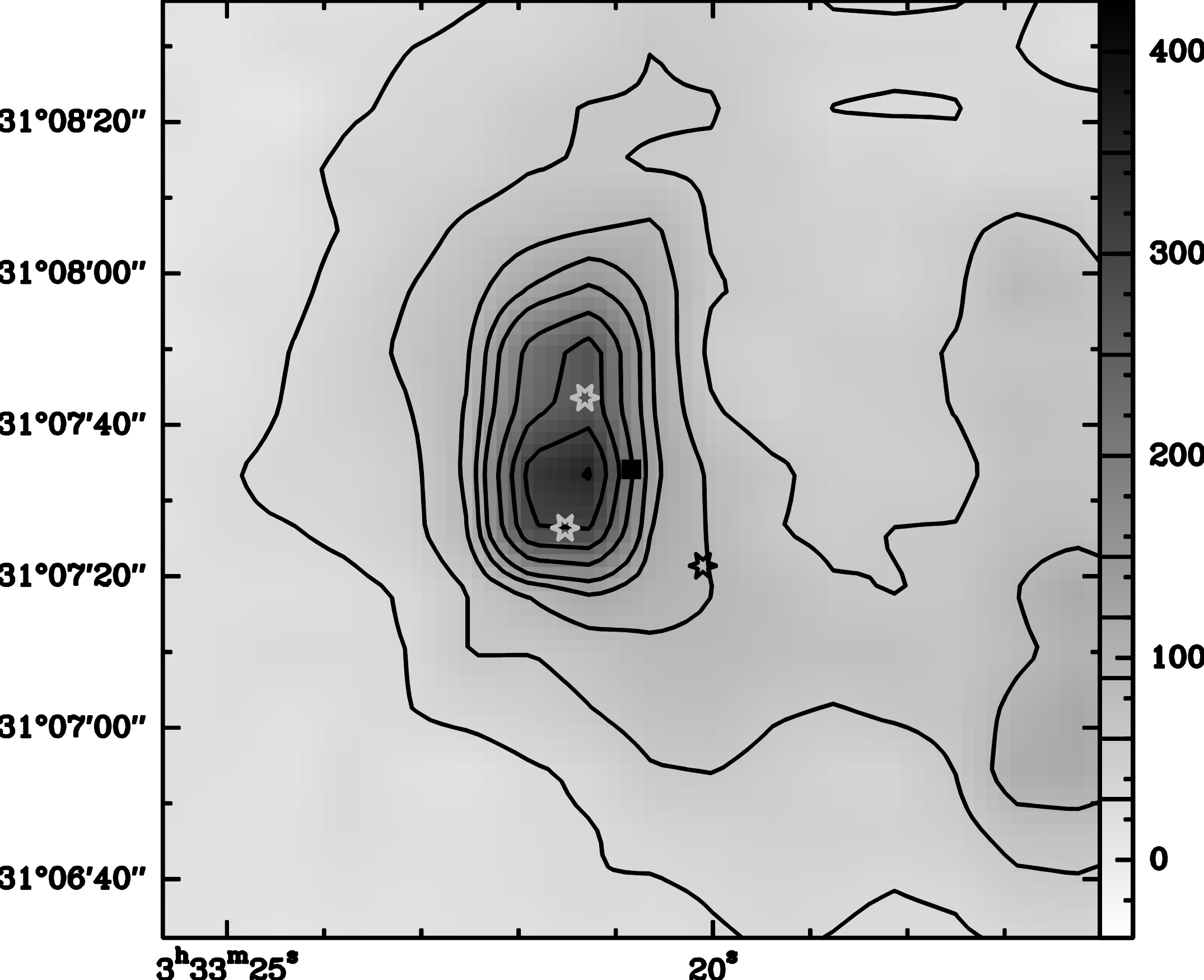}
\caption{Same as Fig. \ref{fig:cont1300} but for the 1.2 mm continuum observed at the IRAM telescope 
with the MAMBO instrument. The contours
have a step of 30 mJy/beam from 30 to 150 mJy/beam and 50 mJy/beam for higher fluxes, 
as drawn on the lookup table on the right.}
\label{fig:cont1300} \vspace{-0.1cm}
\end{center}
\end{figure}

In order to derive the gas temperature and H$_2$ density profiles,
we used continuum observations at $\lambda$ = 350 $\mu$m and $\lambda$ = 1.2 mm,
respectively obtained at the CSO and IRAM telescopes. The corresponding
continuum maps are reported in Fig. \ref{fig:cont350} and \ref{fig:cont1300}.
In these figures, we indicate the positions of the B1--bS and B1--bN cores identified by \citet{hirano1999} as well as the 
\textit{Spitzer} source [EDJ2009] 295 ($\alpha$ = 03$^h$33$^m$20.34$^s$ ; $\delta$ =  31$\degr$07$\arcmin$21.4$\arcsec$ (J2000)) referenced in \citet{jorgensen2006} and \citet{evans2009}.
The peak position of the 1.2 mm observations is found at 
$\alpha$ = 03$^h$33$^m$21.492$^s$ ; $\delta$ =  31$\degr$07$\arcmin$32.87$\arcsec$ (J2000),
which is offset by (+8$\arcsec$,-1$\arcsec$) with respect to the position of the molecular line observations 
given in Sect. \ref{section:observations}.
The 350 $\mu$m observations peak at the B1--bS position and is thus offset by $\sim$6$\arcsec$ 
to the south--south west, with respect to the 1.2 mm peak position. 
In the modeling, we assumed that the central position of the density distribution is half--way between the 350 $\mu$m 
and 1.2 mm peaks, which corresponds to the coordinates 
$\alpha$ = 03$^h$33$^m$21.416$^s$ ; $\delta$ =  31$\degr$07$\arcmin$29.79$\arcsec$ (J2000).
In order to characterize the continuum radial behaviour, we performed an annular average of the map around
this position. In the B1 region, there are two other cores, labelled B1--d and B1--c by \citet{matthews2002}, 
which are respectively offset by $\sim$80$\arcsec$ to the SW and $\sim$125$\arcsec$ to the NW from the central position. 
While performing the average, we discarded the points distant by less than 30$\arcsec$ from these two cores.
The fluxes defined in this way are represented in Fig. \ref{fig:SED} and the error bars represent 
one standard deviation, $\sigma$.

\subsubsection{Modelling}

In order to infer the radial structure of the source from these observations, we performed 
calculations using a ray--tracing code. As an input, we fix the
H$_2$ density and temperature profiles which gives, as an output, the intensity as a function
of the impact parameter. The model predictions are then compared to the observations by 
convolving with the telescopes beams, which are respectively approximated by Gaussians of 9$\arcsec$ and 15$\arcsec$
FWHM for the CSO and IRAM telescopes.

As a starting point, we performed a grid of models assuming that the source is isothermal. 
We fixed the dust temperature at $T_d$ = 12K, which is the typical
temperature of the region, according to the analysis of NH$_3$ observations reported by \citet{bachiller1990}
or \citet{rosolowsky2008}.
We parametrized the density profile as a series of power laws, i.e :
\begin{align}
n_0(r) & =  n_0                                                                                      &  \textrm{if } & \quad r < r_0 \\
n_i(r)  & =  n_{i-1}(r_{i-1}) \times \left(\frac{r_{i-1}}{r}\right)^{\alpha_i}  &  \textrm{if } & \quad r_{i-1} < r < r_i 
\end{align}
and we computed a grid of models with the following parameters kept fixed :
$r_0$ = 5$\arcsec$, $r_1$ = 35$\arcsec$, $r_2$ = 125$\arcsec$, 
$r_3$ = 450$\arcsec$ and $\alpha_3$ =  3.0.
The density profile is thus described by three free parameters, the central H$_2$ density $n_0$, 
and the slopes of the first and second regions, 
$\alpha_1$ and $\alpha_2$. This choice is based on some preliminary models, for which we found that these delimiting radii were 
accurate in order to reproduce the morphology of the SED. Moreover, the slope of region 3 was found to be poorly
constrained which motivated us to keep it fixed.

The dust absorption coefficient is parametrized according to
\begin{eqnarray}
\kappa_{abs}(\lambda) = \kappa_{1300} \left( \frac{1300}{\lambda} \right)^{\beta}
\end{eqnarray}
with $\lambda$ expressed in $\mu$m and $\kappa_{1300}$ being the absorption coefficient at 1.3 mm. 
Values for the dust absorption coefficient were determined by \citet{ossenkopf1994}, for the case of 
dust grains surrounded by ice mantles. The values were found to be in the range 
$0.5 < \kappa_{1300} < 1.1$ cm$^2$/g, where $\kappa_{1300}$ is given in function of the dust mass.
Hence, we fixed the gas--to--dust mass ratio to 100 and furthermore fixed the absorption coefficient to 
$\kappa_{1300}$ = 0.005 cm$^2$/g, where $\kappa_{1300}$ is this time refereed to the mass of gas. 
The results of the models depend only on the product 
$\kappa_{1300} \times n_0$.
The grid is then computed considering $\beta$ as an additional free parameter. 

\begin{figure}
\begin{center}
\includegraphics[angle=0,scale=0.5]{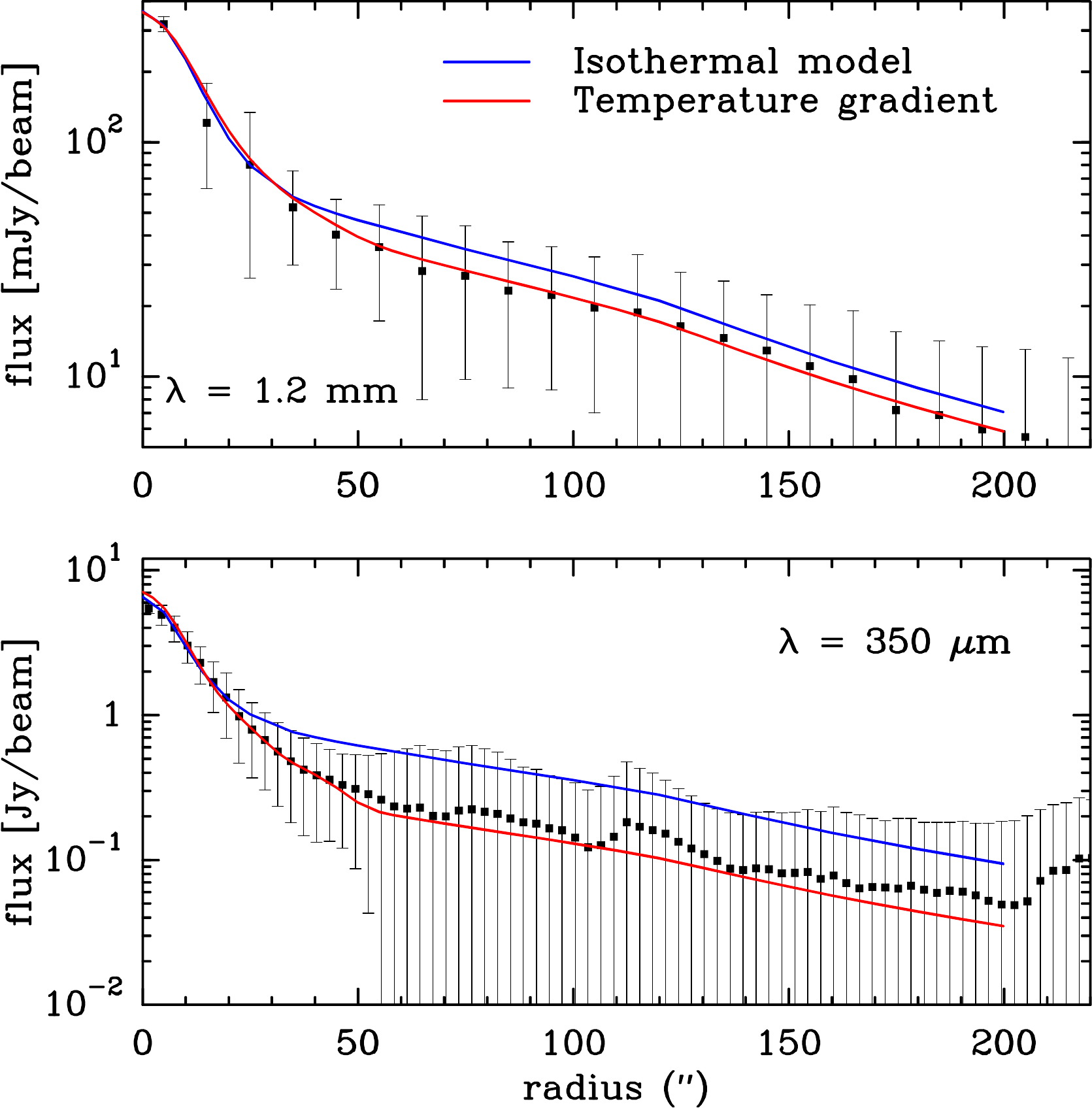}
\caption{Comparison of the observed and model fluxes as a function of radius at 350 $\mu$m (bottom panel) and 1.2 mm (upper panel). The blue curve corresponds to a model obtained assuming that the source is isothermal at T$_d$=12K. The red curve corresponds to a model where we introduced a temperature gradient.} \label{fig:SED} \vspace{-0.1cm}
\end{center}
\end{figure}

From the results of the grid, it appears that the best models are found for parameters in the range :
$3\, 10^6 < n_0 < 6 \, 10^6$ cm$^{-3}$, $ 2.0 < \alpha_1 < 2.4$, $1.2 < \alpha_2 < 1.5$ and $1.7 < \beta < 2.4 $.
However, we could not find a model that would fit satisfactorily the radial profiles at both wavelengths.
The origin of the problem is illustrated in Fig. \ref{fig:SED}. In this figure, the blue curve corresponds to a model
that satisfactorily reproduces (i.e. within 30\%) the 1.2 mm radial profile. It can be seen that this model
is able to reproduce the flux at 350 $\mu$m, for radii $r < 20 \arcsec$. However, for larger radii, 
the flux is overestimated by a factor 3. Obviously, such a radial
dependence cannot be accounted for by a modification of the $n_0$, $\kappa_{1300}$ or
$\beta$ parameters, since changes in these parameters would modify the relative fluxes at both wavelengths 
and independently of the radius. 

\begin{figure}
\begin{center}
\includegraphics[angle=270,scale=0.37]{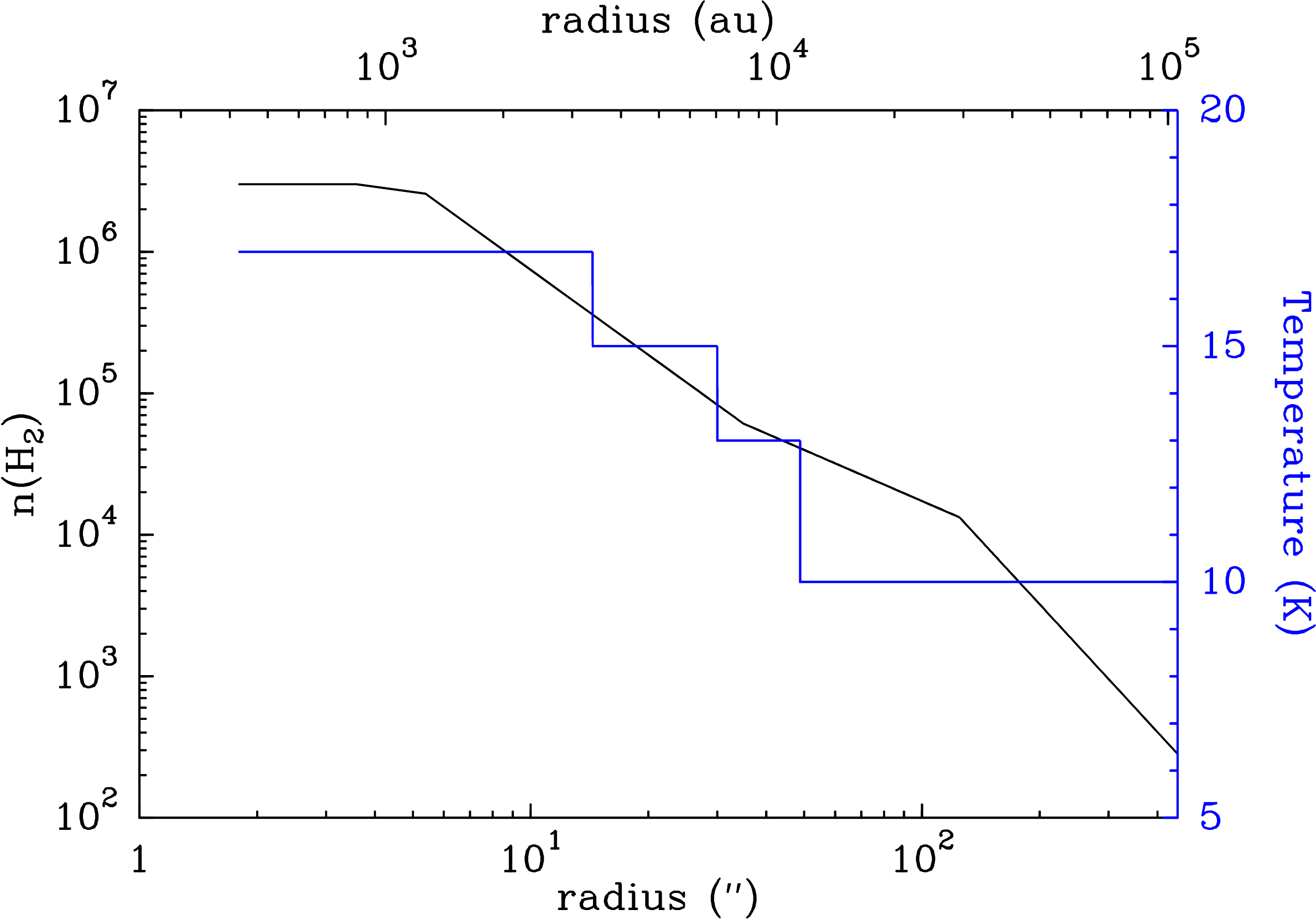}
\caption{Temperature (blue curve, right axis) and H$_2$ density (black curve, left axis) 
derived from the SED fitting.} \label{fig:structure} \vspace{-0.1cm}
\end{center}
\end{figure}

A solution to this problem is to introduce a temperature gradient 
in the model. Indeed, an increase of the dust temperature will affect differently the fluxes at
350 $\mu$m and 1.2 mm, producing a larger increase at the shorter wavelength.
We thus modified the best model obtained from the grid analysis
by introducing a gradient in the dust temperature, from 17K in the center 
to 10K in the outermost region. The resulting
fit is shown by the red curve in Fig. \ref{fig:SED}. The parameters of this model are $n_0 = 3\, 10^6$ cm$^{-3}$, 
$\alpha_1 = 2$,  $\alpha_2 = 1.2$ and $\beta = 1.9$. The dust temperature and H$_2$ 
density are represented in Fig. \ref{fig:structure}.
With the density profile defined this way, the peak value of the H$_2$ column density is
N(H$_2$) = $2.1 \, 10^{23}$ cm$^{-2}$. The average column density, within a 30$\arcsec$ FWHM beam,
is N(H$_2$) = $7.6 \, 10^{22}$ cm$^{-2}$, in good agreement with the estimate of \citet{johnstone2010}, i.e.
N(H$_2$) = $8.2 \, 10^{22}$ cm$^{-2}$, obtained assuming the same beam. The FHWM of the column
density distribution is $\sim$16$\arcsec$. Finally, the current density profile corresponds to enclosed masses of 
0.5, 1.8, 4.5 and 13.1 M$_{\sun}$ for delimiting radii of 10, 30, 60 and 120$\arcsec$, 
respectively ($\sim$ 0.01, 0.03, 0.07 and 0.14 pc, assuming a distance to B1 of 235 pc).

\subsubsection{Comparison with previous studies}

B1b has been the subject of many studies with both ground-- and space--based telescopes.
In particular, interferometric observations by \citet{hirano1999} lead to the identification of two cores, separated
by $\sim$20$\arcsec$ and designated as B1--bS and B1--bN. These objects are respectively offset  by (+6$\arcsec$,-7$\arcsec$)
and (+5$\arcsec$,+10$\arcsec$) with respect to the central position of our observations (see Sect. \ref{section:observations}) and are thus both at a projected distance $\sim$10$\arcsec$. By modelling the SED from 350 $\mu$m to 3.5 mm,
these authors found that both objects can be characterized by a dust temperature 
of T$_d$ = 18K, with a density distribution consistent with $n(r)$ $\propto$ r$^{-1.5}$ and without central flattening.   
They estimated the bolometric to submm luminosities to $L_{bol}/L_{submm}$ $\sim$ 10, well 
below the limit of 200 for class 0 objects \citep{andre1993}. Thus, these characteristics 
would suggest that the two sources are very young class 0 objects.

Recently, \citet{pezzuto2012} characterized B1--bS and B1--bN by combining \textit{Spitzer}, \textit{Herschel} and 
ground--based observations. The modelling of the SEDs showed that the two objects are more evolved than
prestellar cores, but have not yet formed class 0 objects. In particular, the non-detection of B1--bS at 24 $\mu$m 
and detection at 70 $\mu$m place this source as a candidate for the first hydrostatic core stage. 
However, the bolometric luminosity of this source, estimated as $L_{bol}$ $\sim$ 0.5 L$_{\sun}$ is above the limit 
of 0.1 L$_{\sun}$ that characterizes the maximum luminosity at this evolutionary stage \citep{omukai2007}.
The non--detection of B1--bN at these two wavelengths gives this source the status of a less evolved object. 
\citet{pezzuto2012} described these objects by using a combination of blackbody plus modified--blackbody
components, that respectively characterize the central object and the surrounding dust envelope. For the two
objects, the central temperature was found to be of the order of 30K and the envelope has a typical 
temperature $\sim$ 9K. However, as noted by these authors, in a realistic model, the emission of the central object 
will be absorbed by the dusty envelope which will lead to a stratification in the dust temperature. 
 
Even more recently, \citet{huang2013} presented 1.1 mm continuum interferometric maps of the B1b region where 
the B1-bS and B1-bN objects are spatially resolved. For those two objects, they quote the parameters from the upcoming work by 
\citet{hirano2013} that describe the fit of the SED. B1-bS and B1-bN 
are respectively described by indexes $\beta = 1.3 \pm 0.2$ and $1.8 \pm 0.4$ and T$_{d}$ = $18.6 \pm 1.6$ K
 and $15.6 \pm  2.2$ K.


\subsection{Molecular excitation modeling}\label{modelling:molecules}

The line radiative transfer (RT) is solved using the short--characteristics
method and includes line overlap for the molecules with
hyperfine structure. The underlying theory and numerical code are described in \cite{daniel2008}.
For most of the molecules, the spectroscopy is retrieved from the \textit{splatalogue} 
database\footnote{http://splatalogue.net/} which centralizes the spectroscopic 
data available in other databases like JPL\footnote{http://spec.jpl.nasa.gov/} 
\citep{pickett1998} or CDMS\footnote{http://www.astro.uni-koeln.de/cdms/} \citep{muller2001,muller2005}.
The rate coefficients are retrieved from the BASECOL\footnote{http://basecol.obspm.fr/}
database \citep{dubernet2012}. The references for both the spectroscopy and collisional rate coefficients 
are further described in what
follows when considering the individual molecules.
In the case of linear molecules that have a hyperfine 
structure resolved by the observations, but for which the collisional rate coefficients just 
take into account the rotational structure (e.g. the HNC isotopologues), we emulate hyperfine rate coefficients using the 
IOS--scaled method described in \citet{neufeld1994}. It has been shown by \citet{faure2012} that
this method of computing hyperfine rate coefficients leads to accurate results when compared 
to rate coefficients obtained with a more accurate recoupling scheme. 
In the case of the symmetric or asymmetric tops (i.e. NH$_3$ and NH$_2$D), the hyperfine rate 
coefficients are obtained using the M--random method.
Moreover, in the case of collisional
systems that consider He as collisional partner, we scale the rate coefficients in order 
to emulate rate coefficients with H$_2$, using a scaling factor that depends on the ratio of the reduced
masses of the two colliding systems. The same approximation is used to determine the rate coefficients
of the rarest isotopologues. 

\begin{figure*}
\begin{center}
  \subfigure[]{\includegraphics[angle=270,scale=0.3]{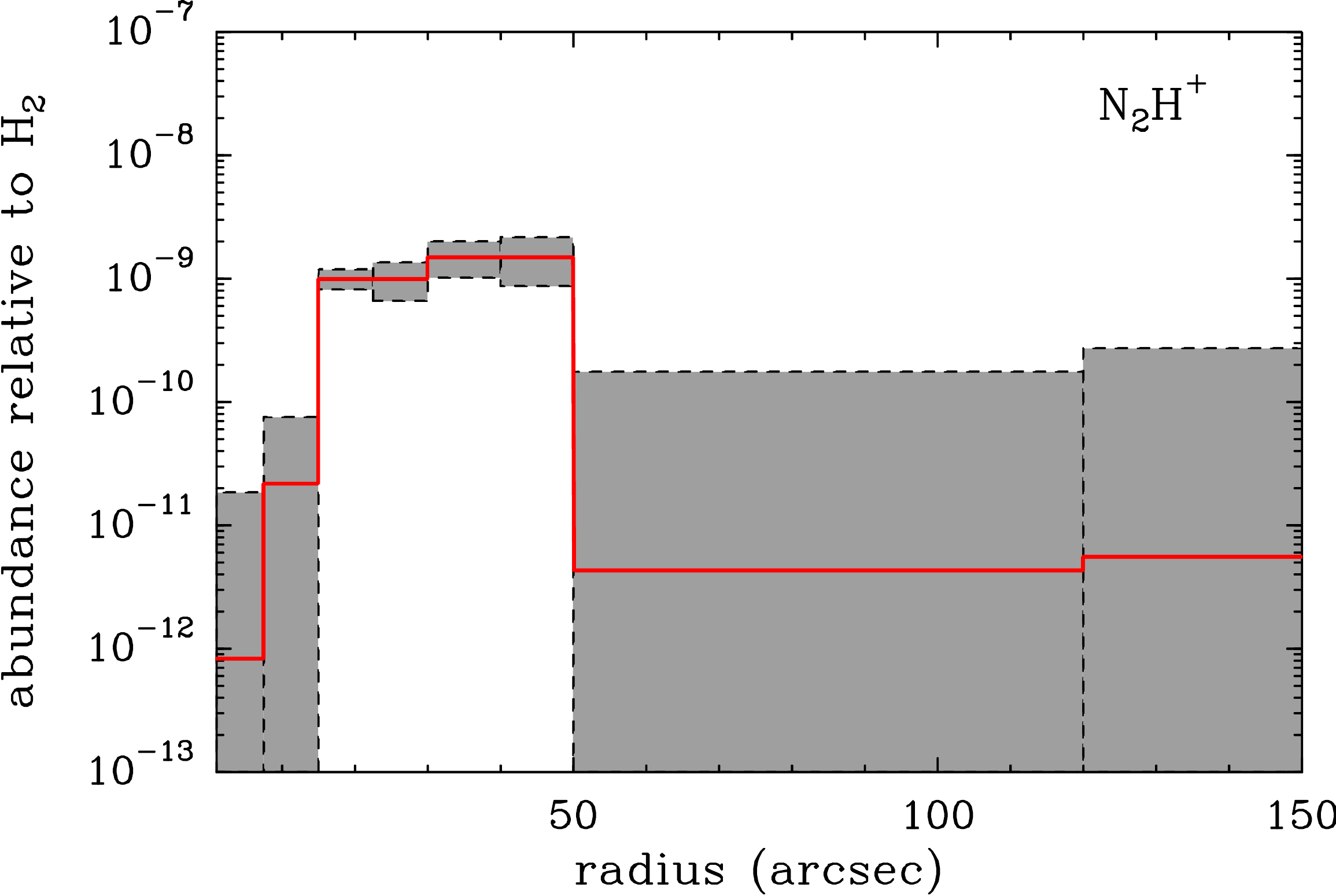}} \quad
  \subfigure[]{\includegraphics[angle=270,scale=0.3]{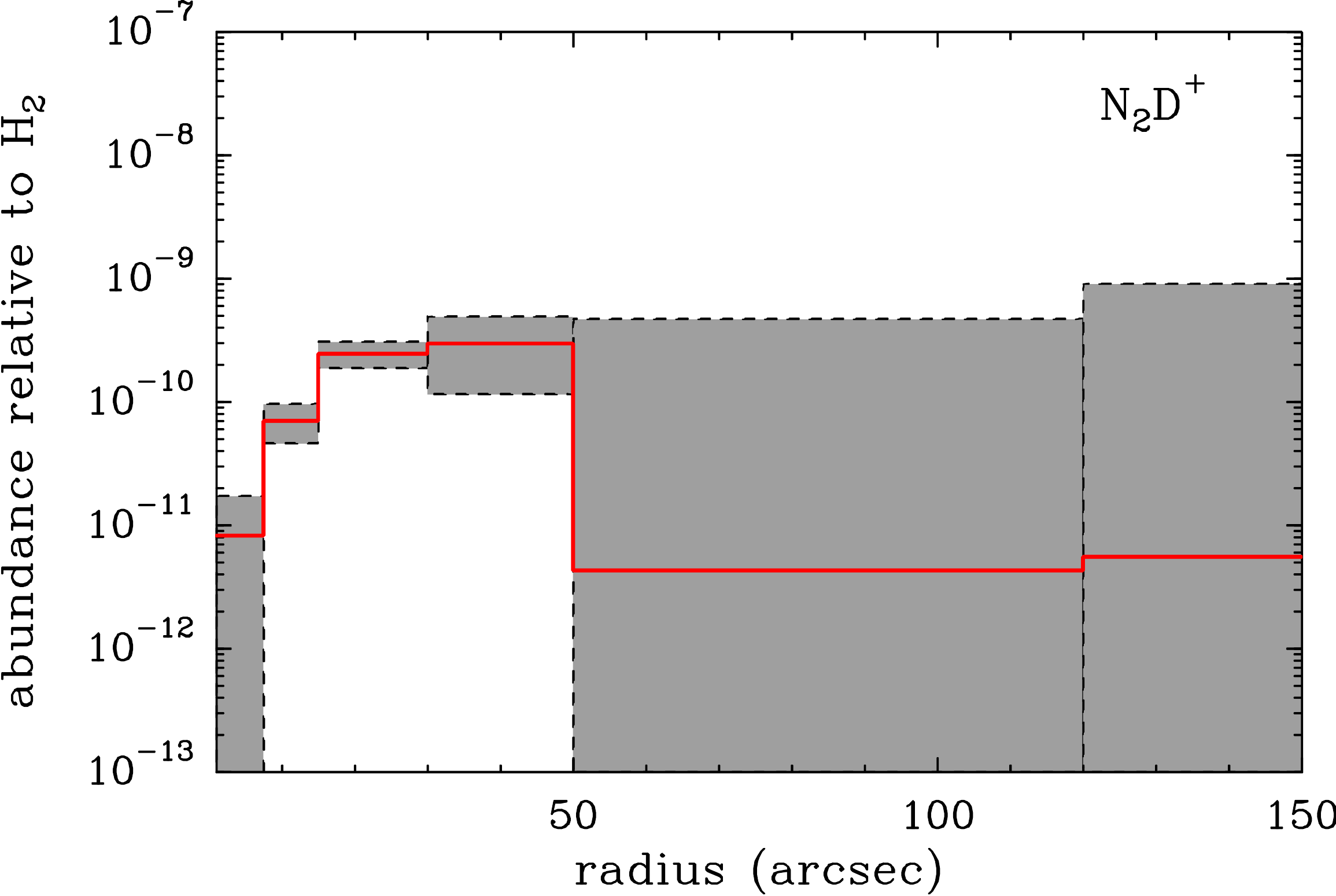}} \\
  \subfigure[]{\includegraphics[angle=270,scale=0.3]{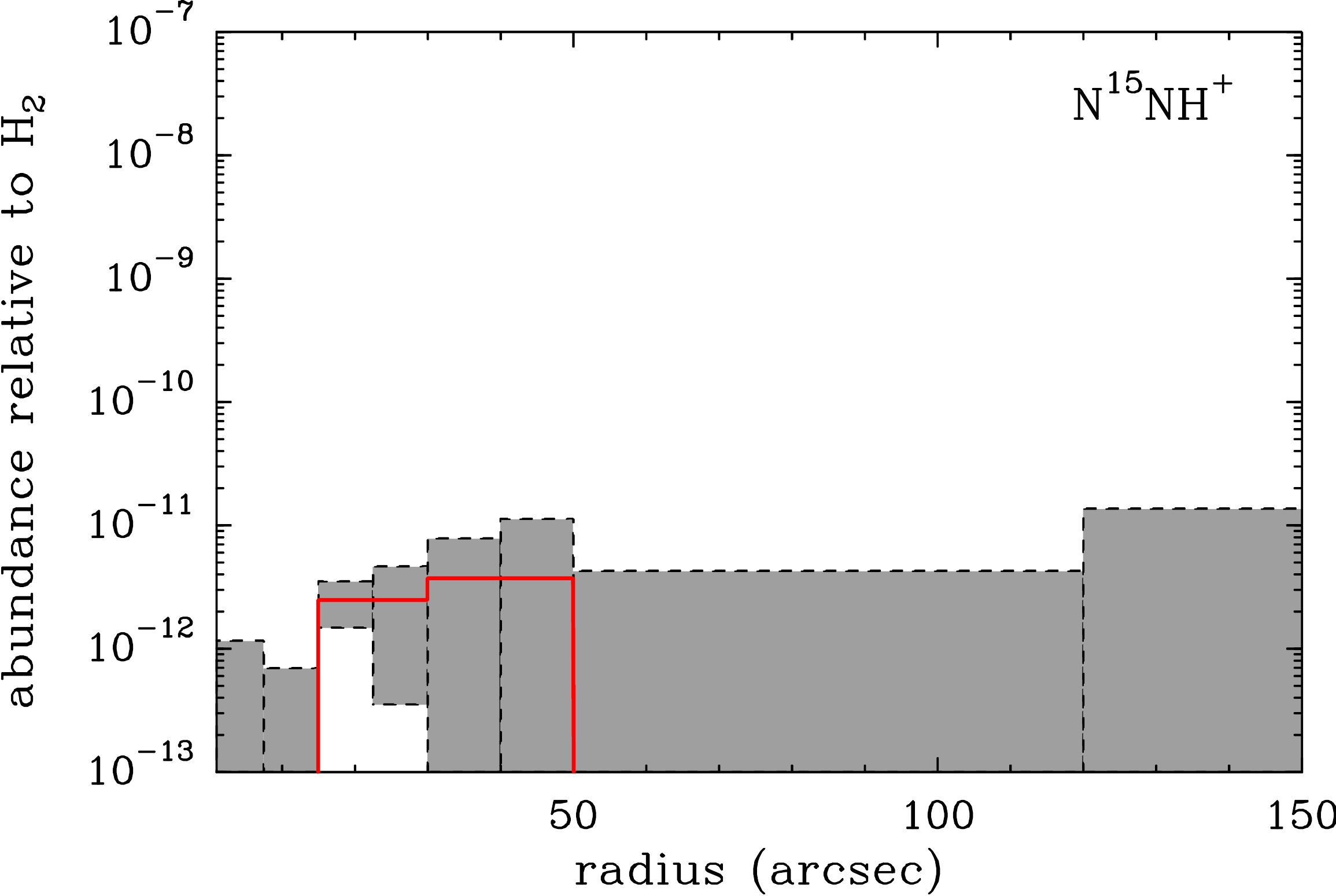}} \quad
\end{center}
\caption{Abundance profiles of the N$_2$H$^+$ isotopologues derived from the modeling (red lines). The shaded
areas give a confidence zone for the abundance as delimited by the lower and upper boundaries in each region (see Sect. \ref{modelling:molecules} for details).}
\label{profil_abondance-N2H+}
\end{figure*}

\begin{figure*}
\begin{center}
  \subfigure[]{\includegraphics[angle=270,scale=0.3]{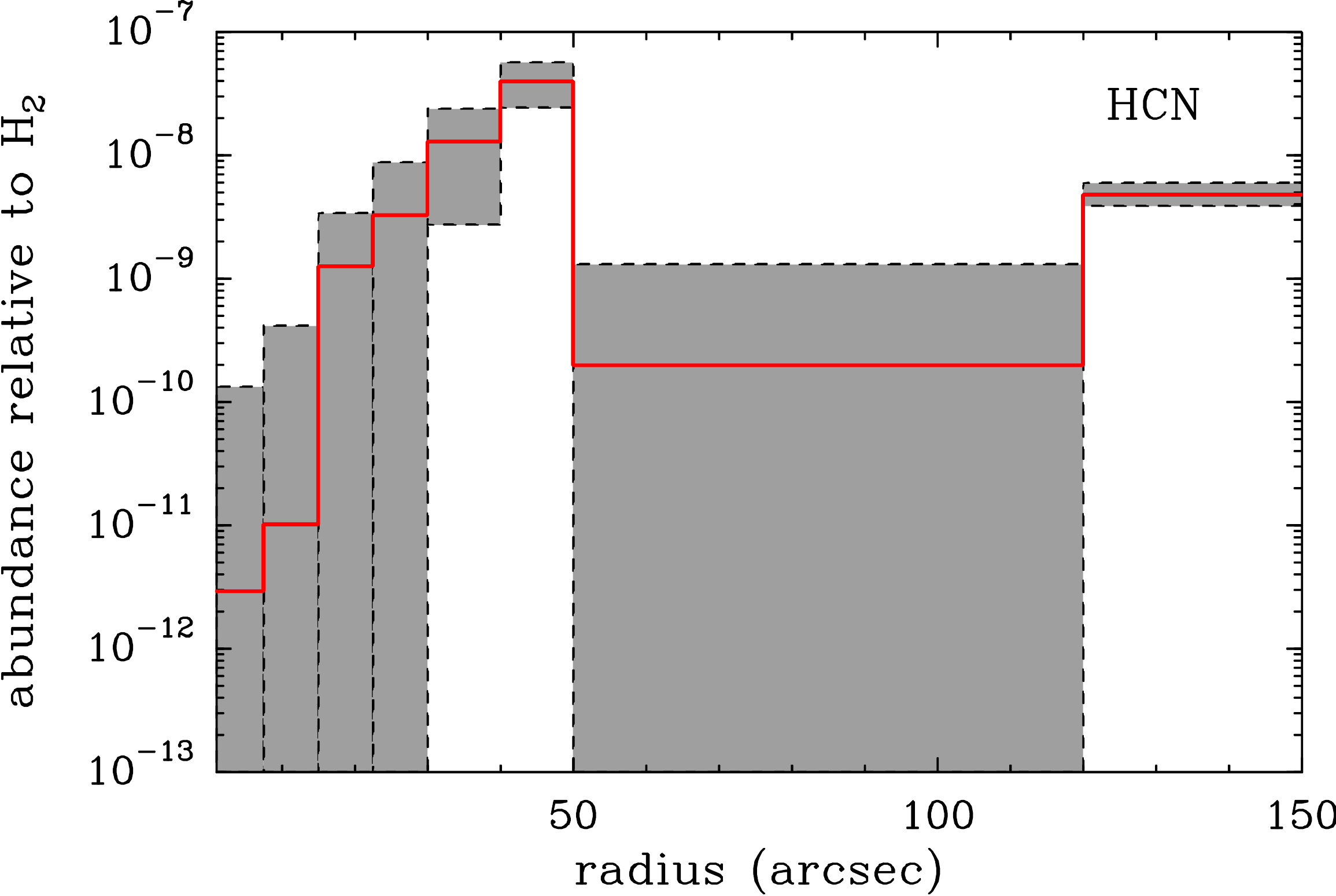}} \quad
  \subfigure[]{\includegraphics[angle=270,scale=0.3]{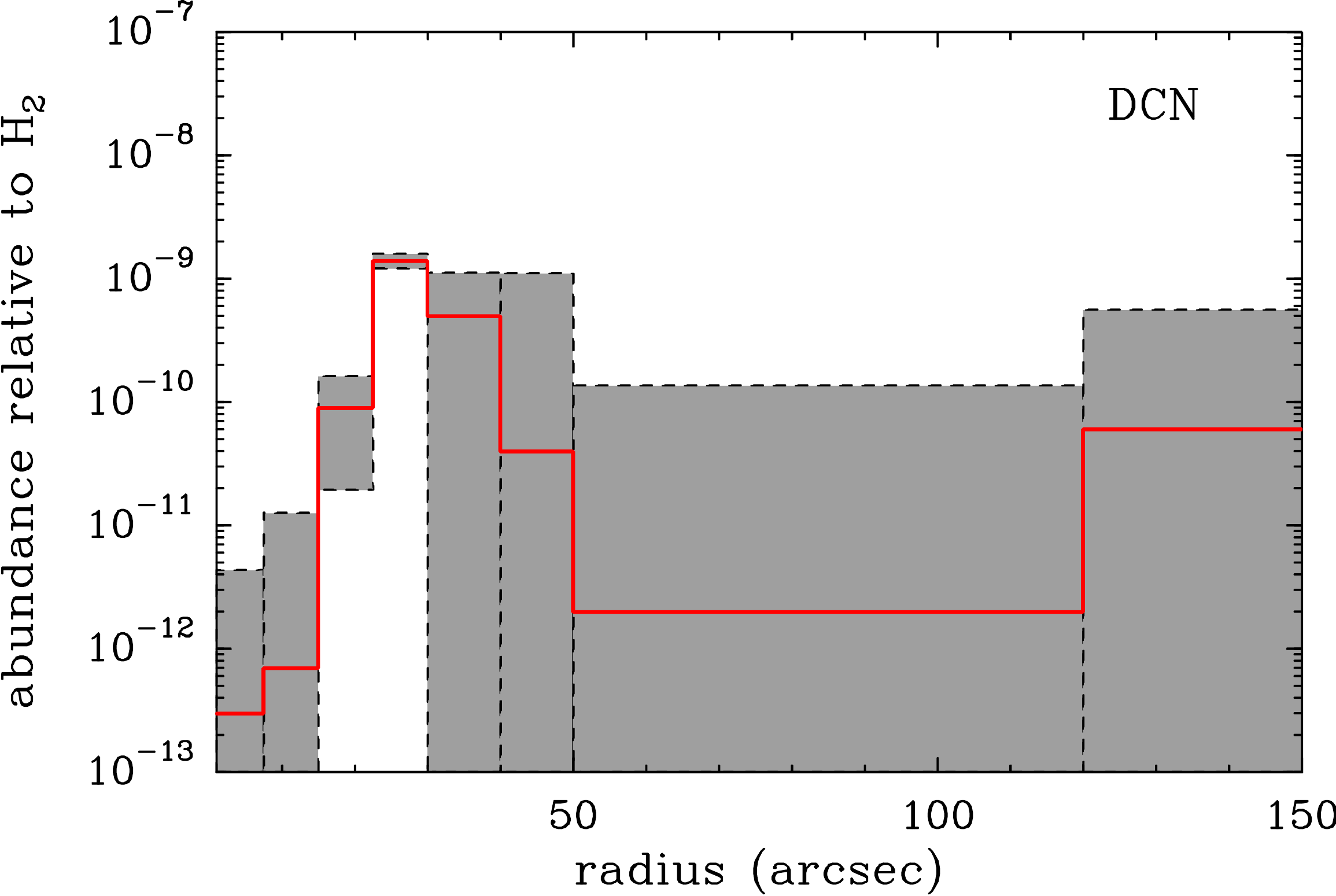}} \\
  \subfigure[]{\includegraphics[angle=270,scale=0.3]{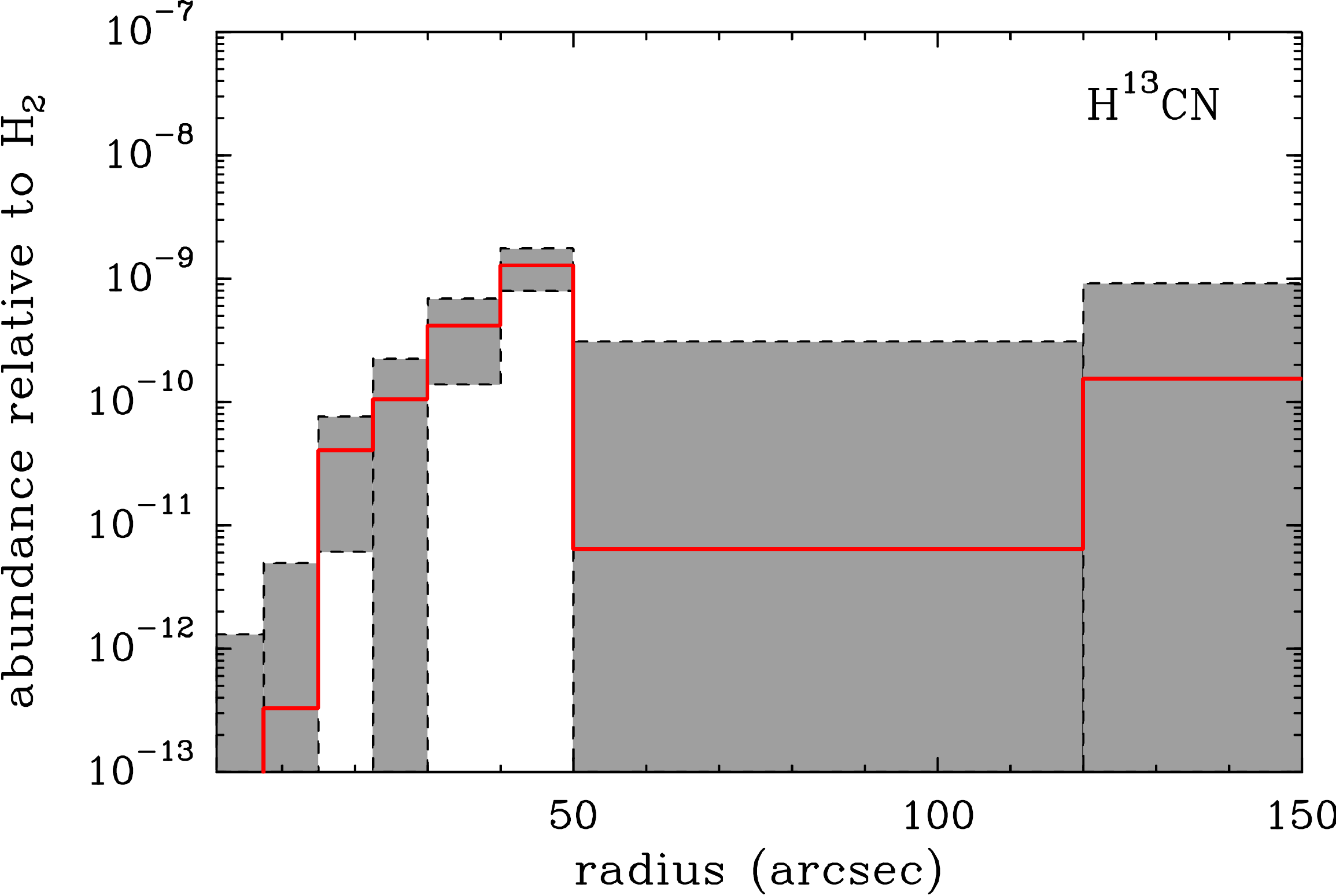}} \quad
  \subfigure[]{\includegraphics[angle=270,scale=0.3]{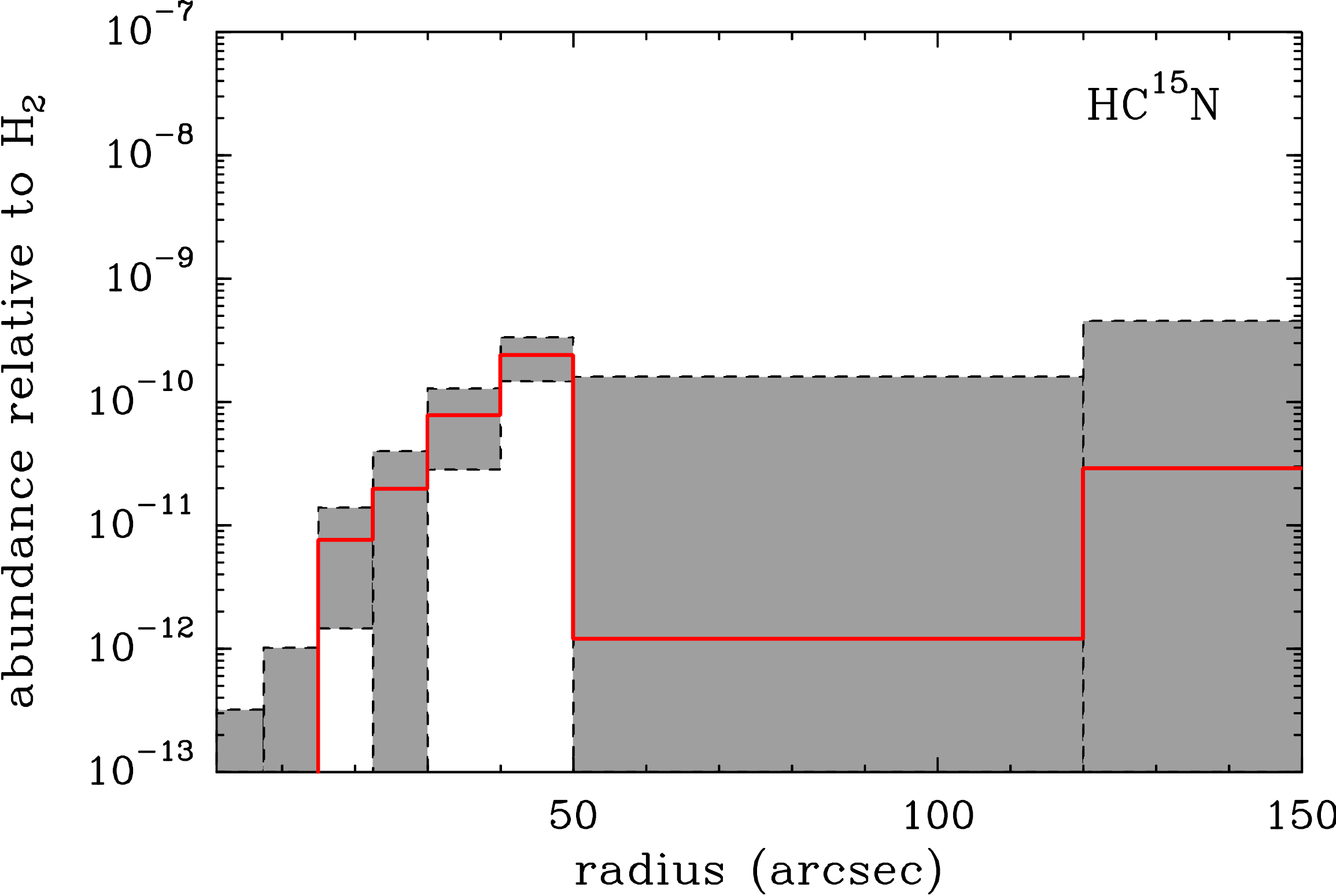}} \\
\end{center}
\caption{Abundance profiles of the HCN isotopologues derived from the modeling.}
\label{profil_abondance-HCN}
\end{figure*}

\begin{figure*}
\begin{center}
  \subfigure[]{\includegraphics[angle=270,scale=0.3]{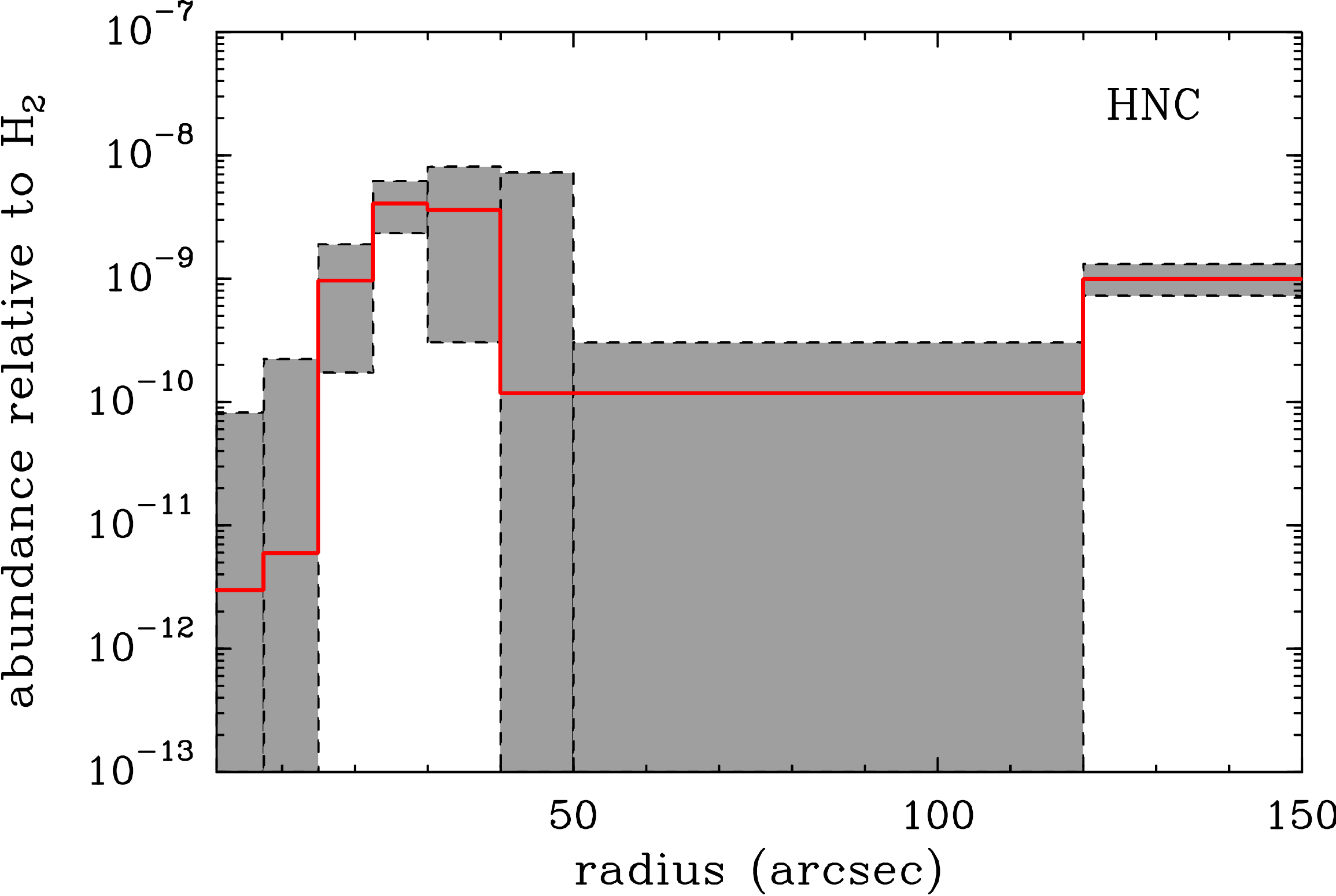}} \quad
  \subfigure[]{\includegraphics[angle=270,scale=0.3]{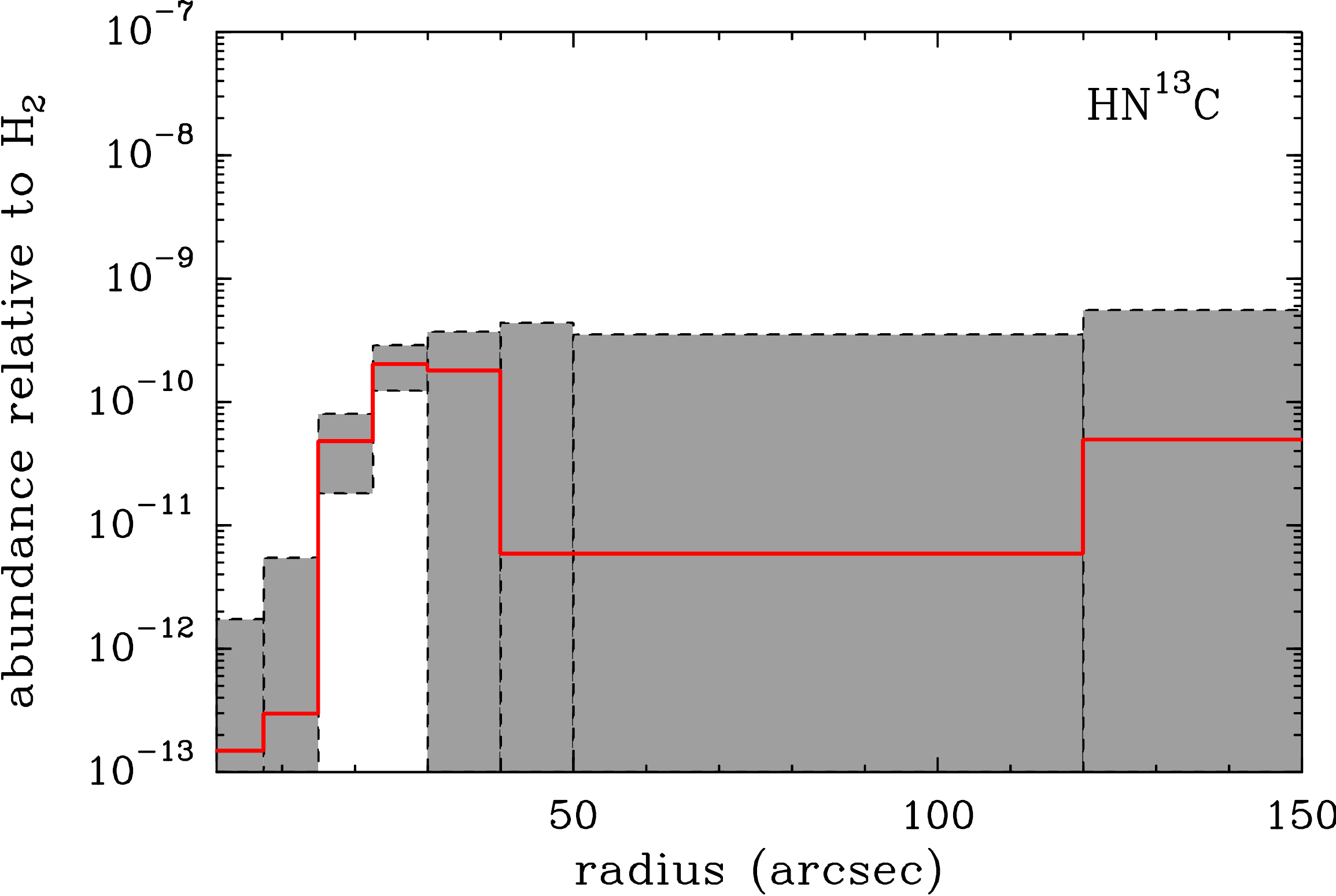}} \\
  \subfigure[]{\includegraphics[angle=270,scale=0.3]{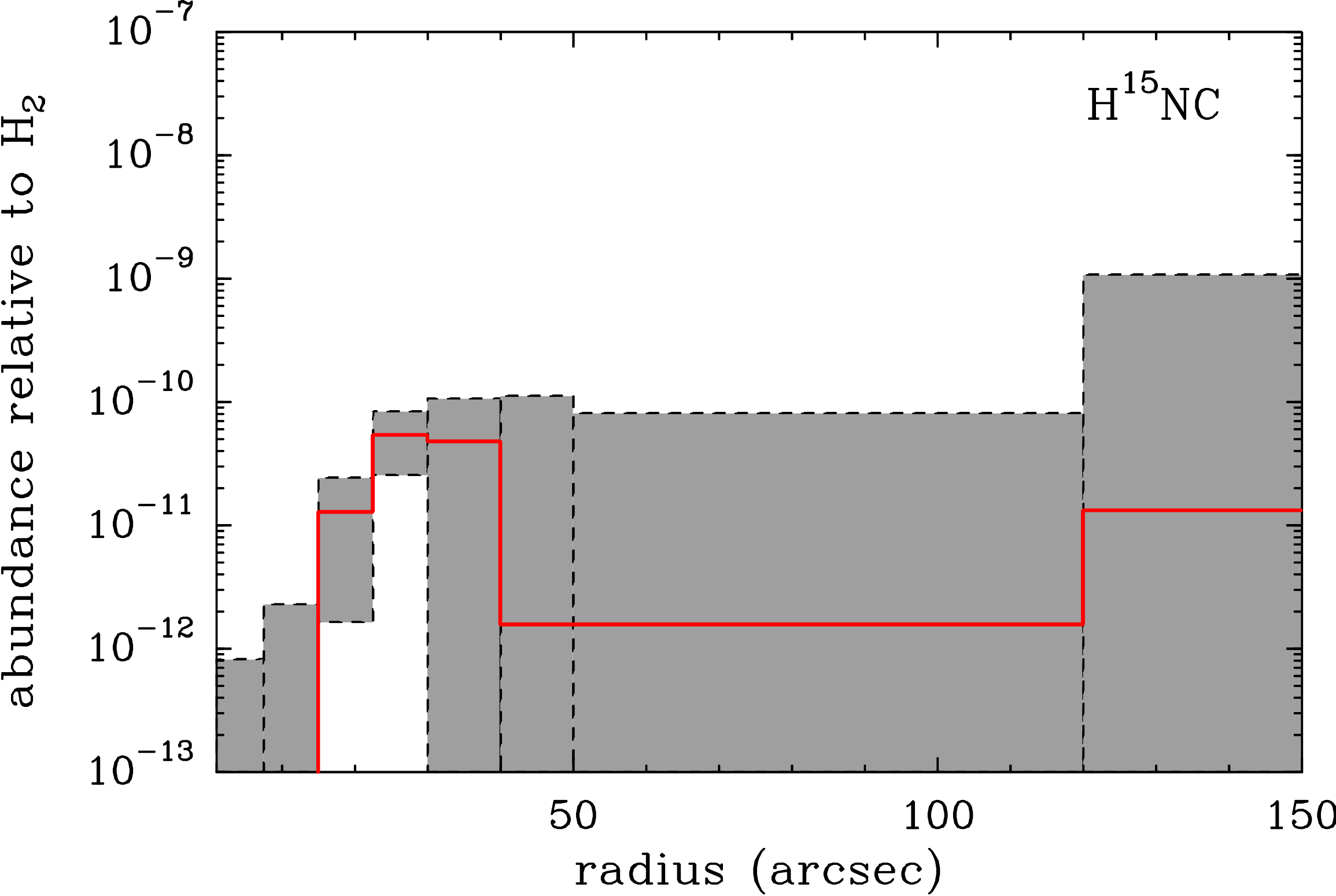}} \quad
  \subfigure[]{\includegraphics[angle=270,scale=0.3]{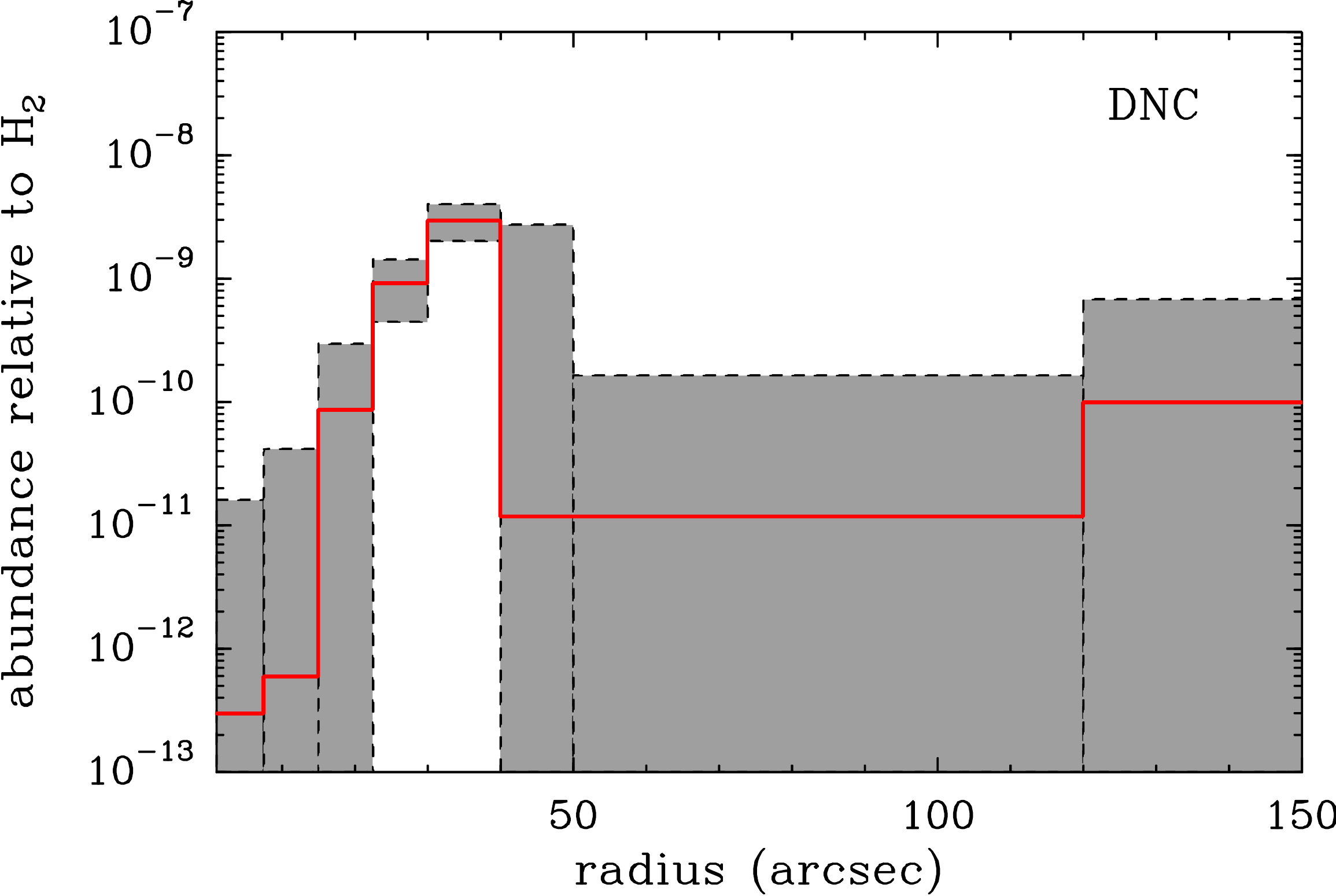}} \\
   \subfigure[]{\includegraphics[angle=270,scale=0.3]{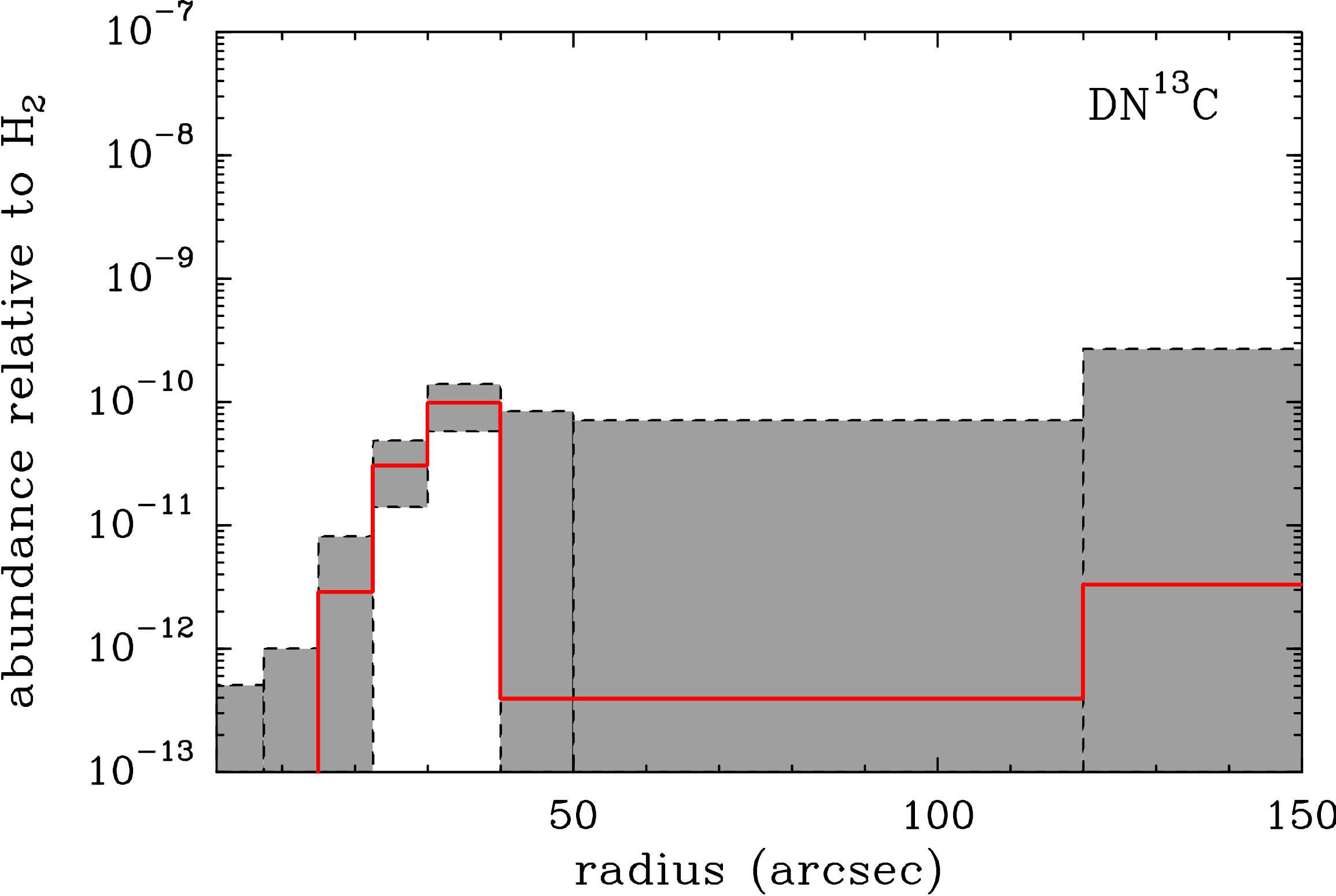}} 
\end{center}
\caption{Abundance profiles of the HNC isotopologues derived from the modeling.}
\label{profil_abondance-HNC}
\end{figure*}

\begin{figure*}
\begin{center}
  \subfigure[]{\includegraphics[angle=270,scale=0.3]{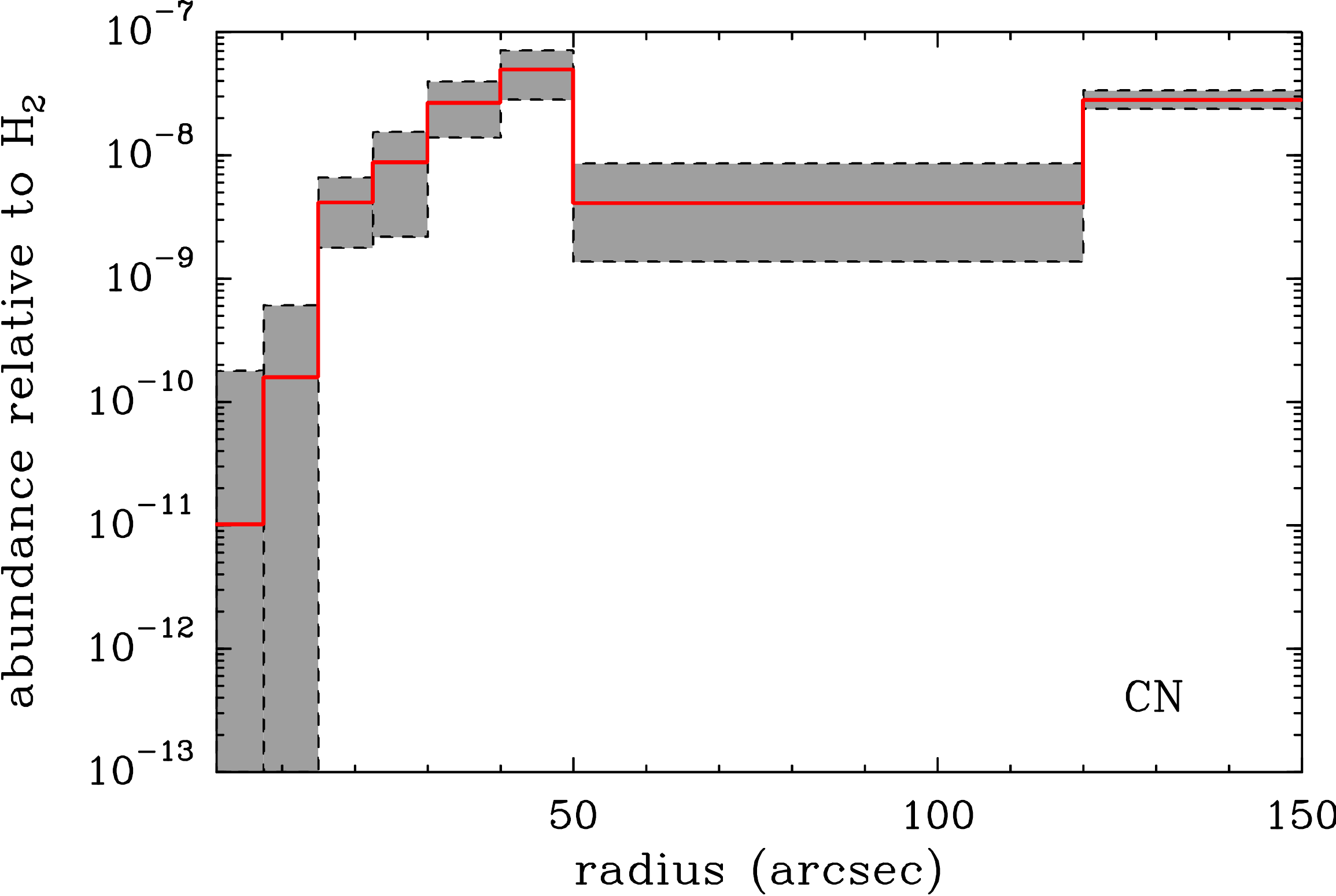}} \quad
 \subfigure[]{\includegraphics[angle=270,scale=0.3]{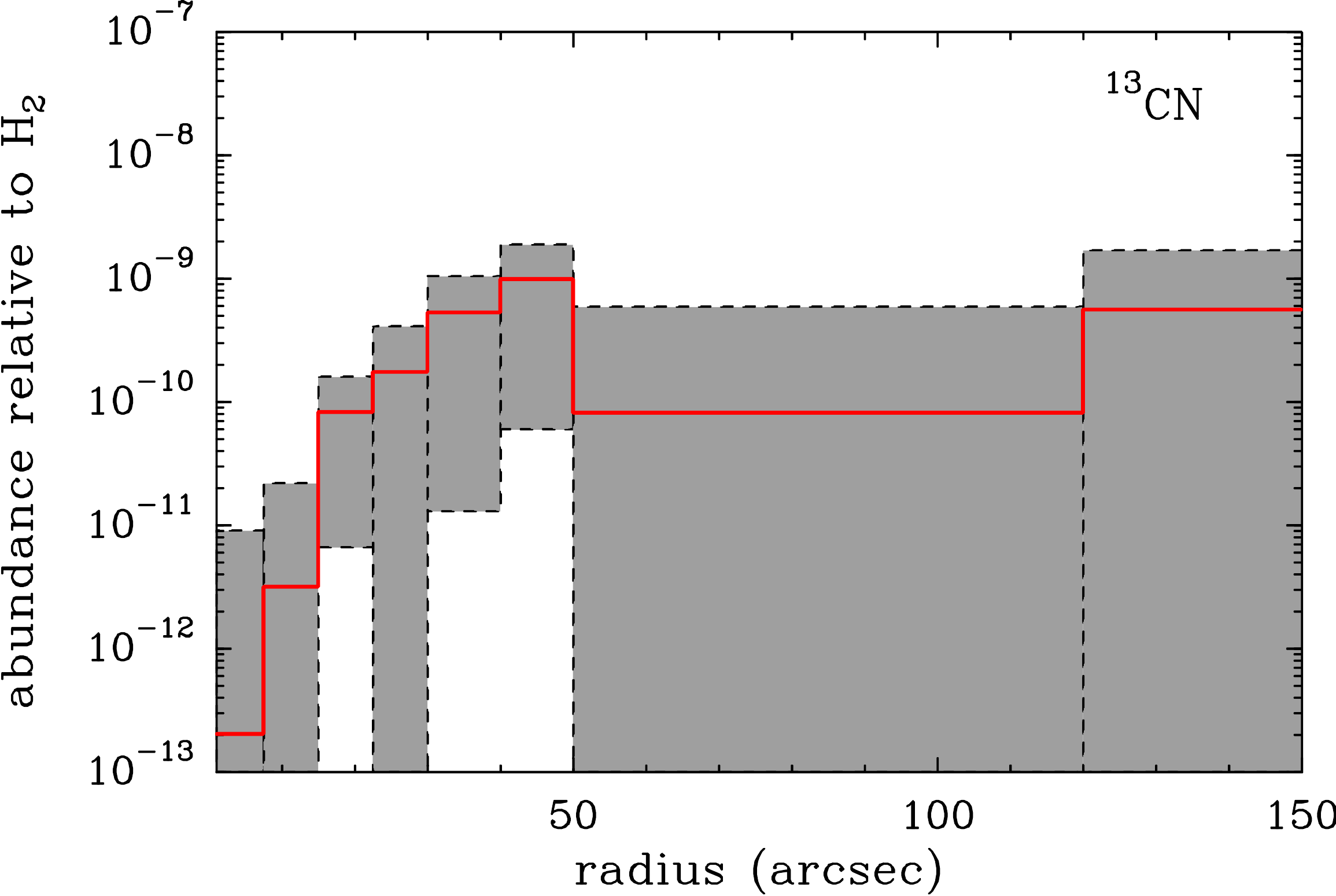}} \\
  \subfigure[]{\includegraphics[angle=270,scale=0.3]{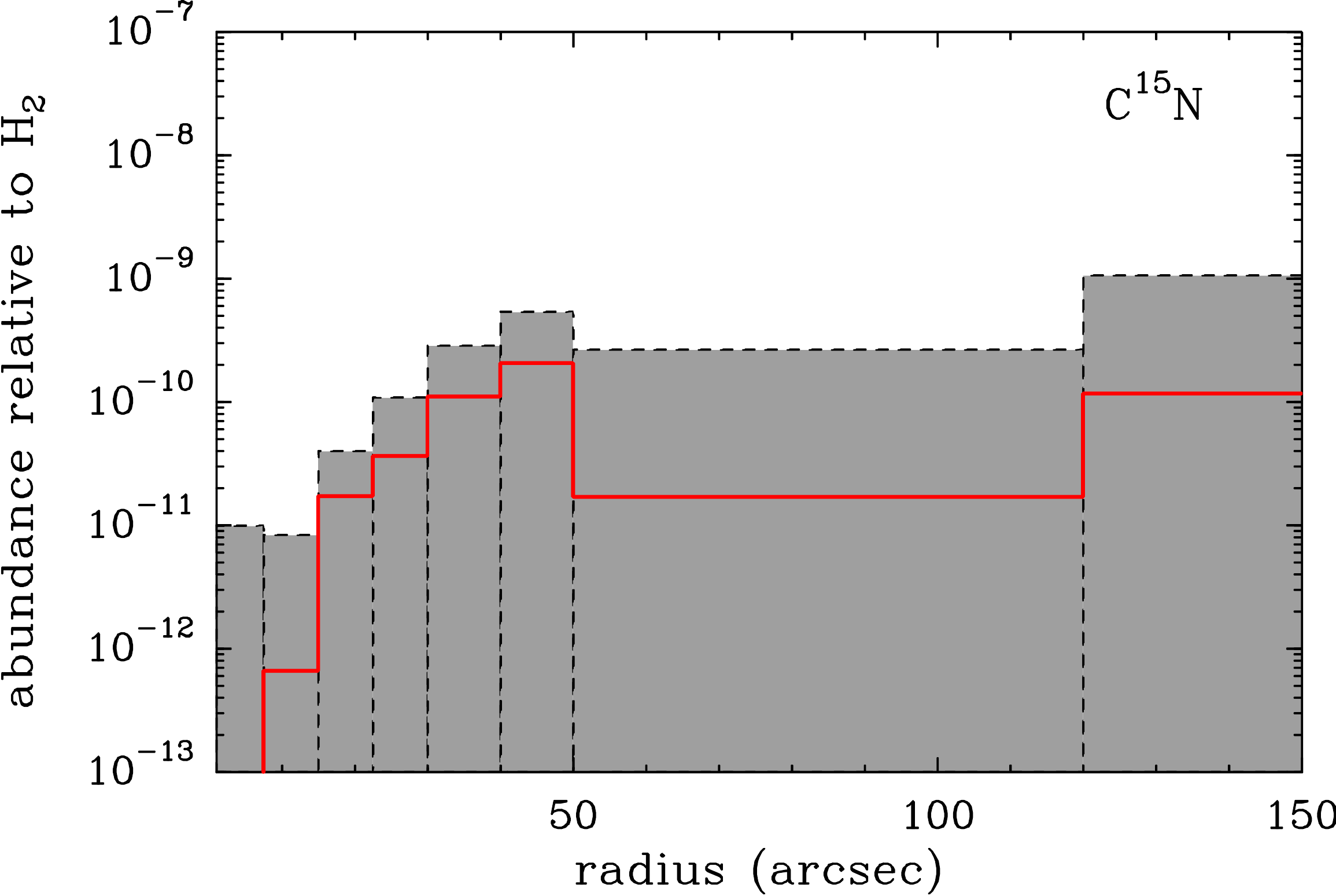}} \quad
\end{center}
\caption{Abundance profiles of the CN isotopologues derived from the modeling.}
\label{profil_abondance-CN}
\end{figure*}

\begin{figure*}
\begin{center}
   \subfigure[]{\includegraphics[angle=270,scale=0.3]{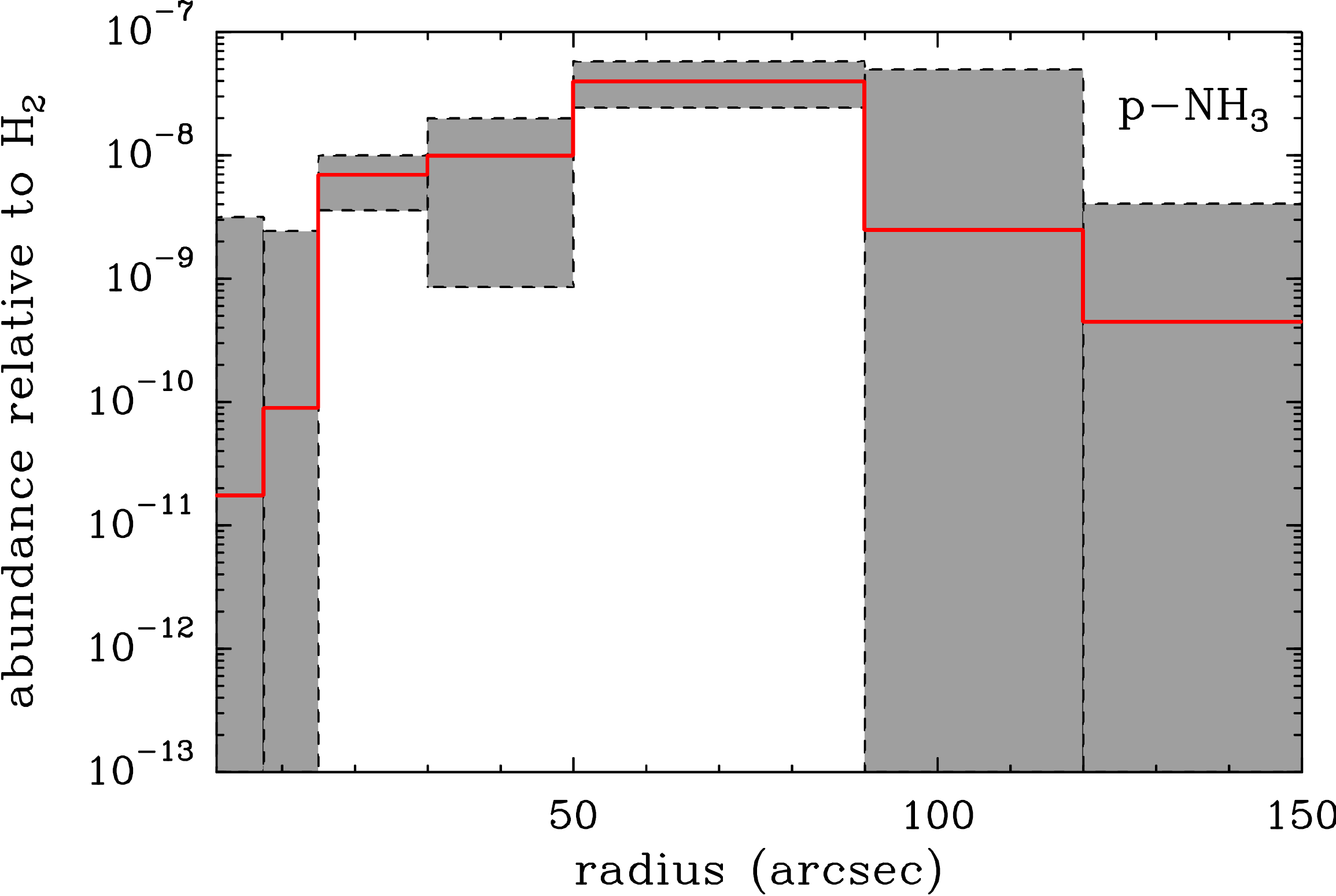}} \quad
   \subfigure[]{\includegraphics[angle=270,scale=0.3]{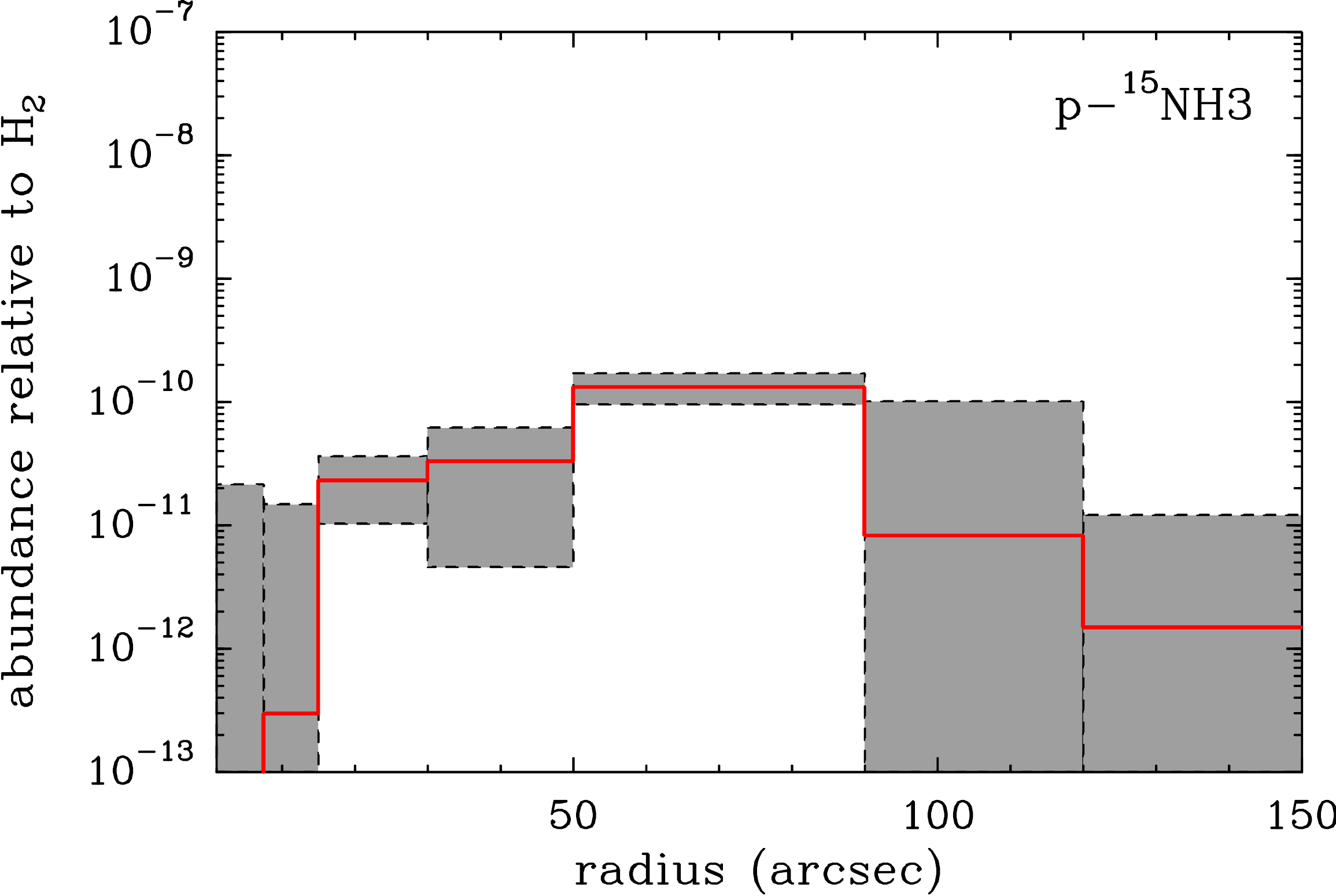}} \\
   \subfigure[]{\includegraphics[angle=270,scale=0.3]{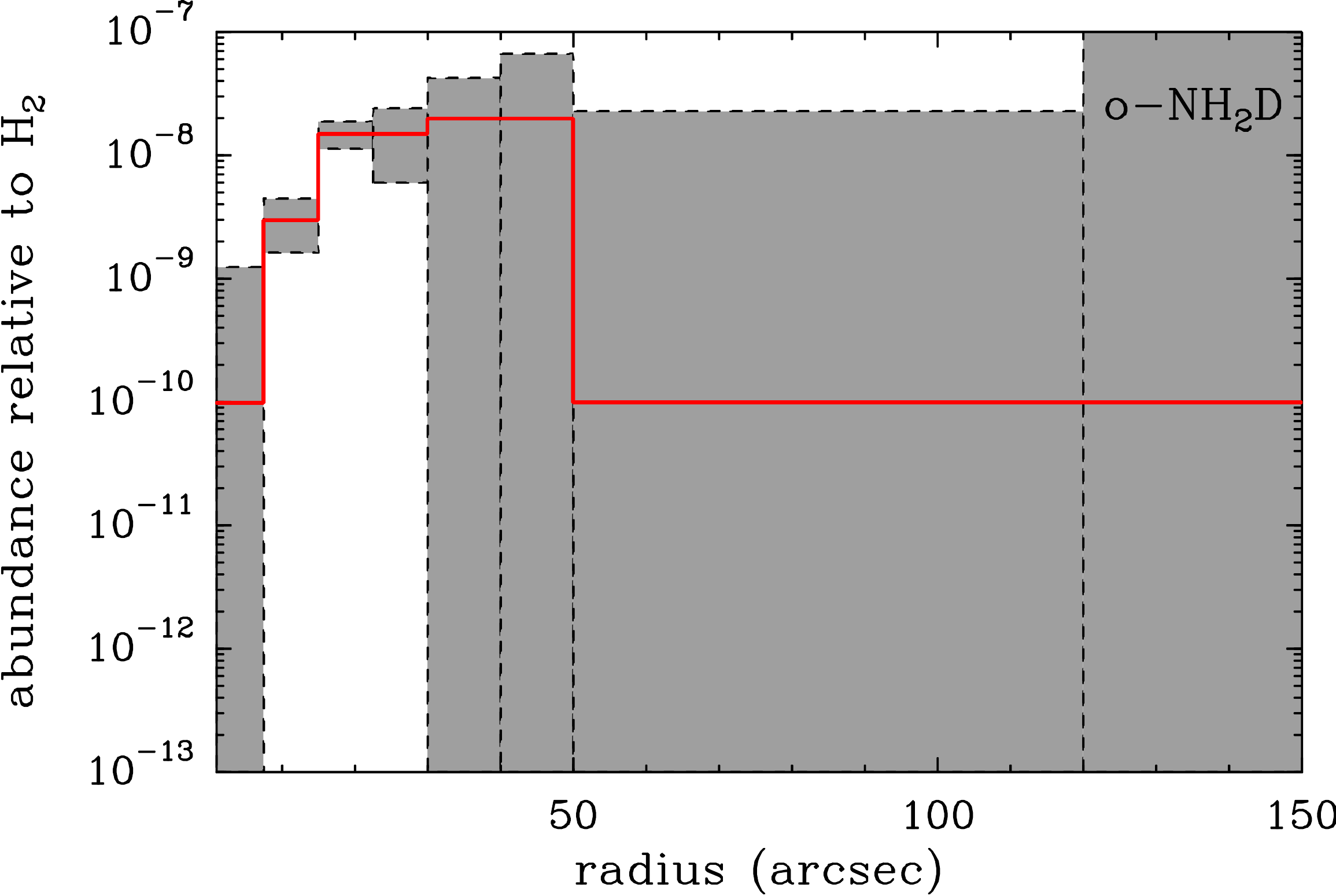}} \quad
   \subfigure[]{\includegraphics[angle=270,scale=0.33]{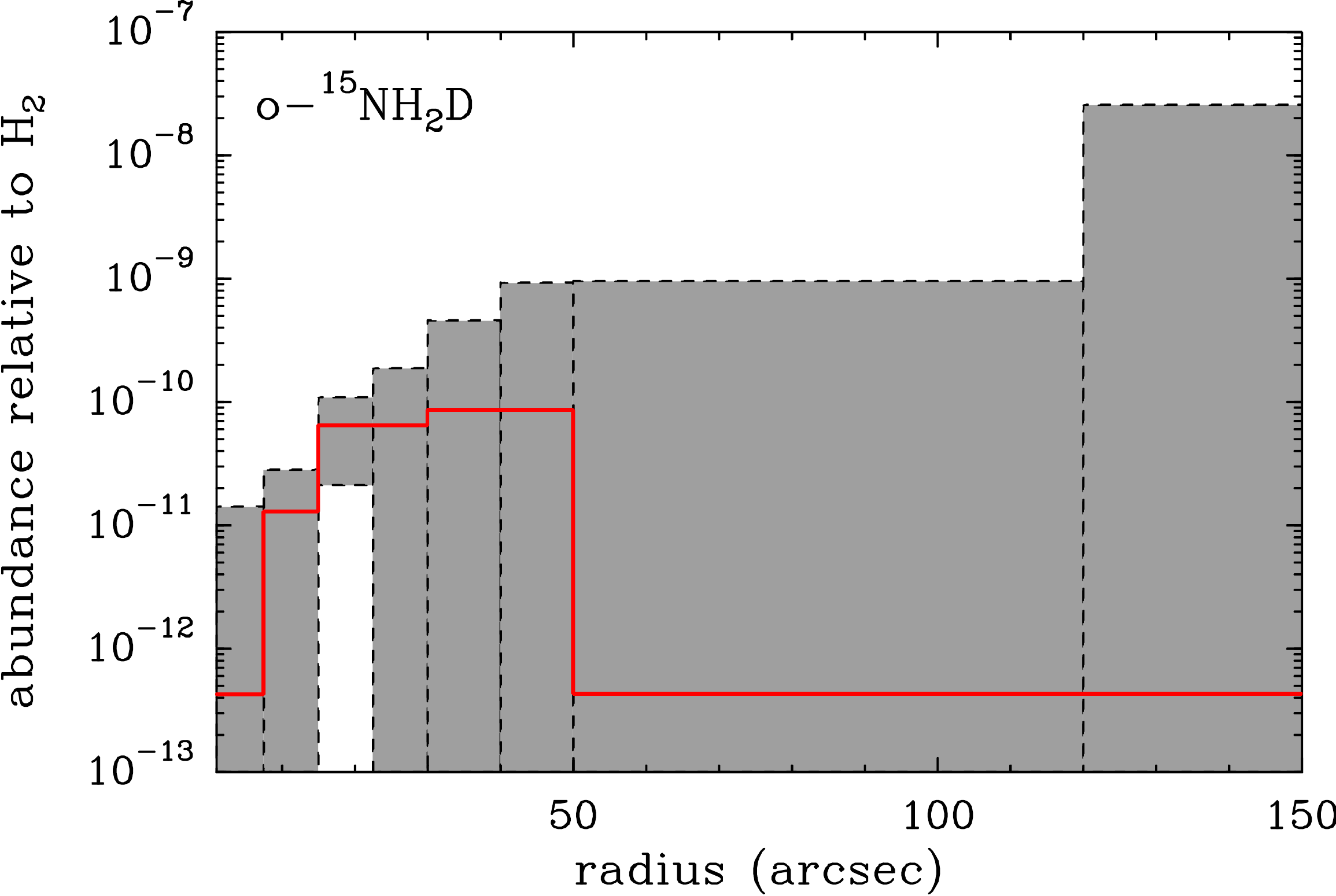}} \\
\end{center}
\caption{Abundance profiles of the NH$_3$ isotopologues derived from the modeling.}
\label{profil_abondance-NH3}
\end{figure*}

\begin{table*}
\caption{Column densities and isotopic ratios derived from the modelling}
\label{table-lineparameters} \centering
\begin{tabular}{l|c|rcc}
\hline \hline
\multicolumn{1}{c|}{}           &  \multicolumn{1}{c|}{log(N)}           &  \multicolumn{2}{c}{ratio to main related isotopologue \tablefootmark{a}} & ratio with $^{12}$C/$^{13}$C = 60 \\
\hline

\hline 
p--NH$_3$  &  14.74$^{+0.25}_{-0.13}$ & & ... \\ [0.1cm]
p--$^{15}$NH$_3$  & 12.25$^{+0.43}_{-0.08}$ & NH$_3$/$^{15}$NH$_3$ = & 300$^{+55}_{-40}$ \\ [0.1cm]

\hline
o--NH$_2$D  & 14.73$^{+0.12}_{-0.08}$ & & ... \\ [0.1cm]
o--$^{15}$NH$_2$D  & 12.37$^{+0.33}_{-0.11}$ & NH$_2$D/$^{15}$NH$_2$D = &  230$^{+105}_{-55}$ \\ [0.1cm]

\hline
N$_2$H$^+$  &  13.48$^{+0.04}_{-0.04}$ & & ... \\ [0.1cm]
N$_2$D$^+$  &  13.02$^{+0.18}_{-0.07}$ & N$_2$H$^+$/N$_2$D$^+$ = & 2.9$^{+0.5}_{-1.0}$ \\ [0.1cm]
$^{15}$NNH$^+$  & $<$ 10.69 & N$_2$H$^+$/$^{15}$NNH$^+$ $\,\,\,\,$ & $ > 600$ \\ [0.1cm]
N$^{15}$NH$^+$  & 10.88$^{+0.50}_{-0.11}$ & N$_2$H$^+$/N$^{15}$NH$^+$ =  & 400$^{+100}_{-65}$ \\ [0.1cm]

\hline
HCN & 14.40$^{+0.09}_{-0.10}$ & & ... \\ [0.1cm]
H$^{13}$CN  & 12.91$^{+0.19}_{-0.10}$ & HCN/H$^{13}$CN = & 30$^{+7}_{-4}$ \\ [0.1cm]
HC$^{15}$N  & 12.18$^{+0.42}_{-0.10}$ & HCN/HC$^{15}$N = & 165$^{+30}_{-25}$ & 330$^{+60}_{-50}$\\[0.1cm]
DCN  &  13.08$^{+0.31}_{-0.09}$ &  HCN/DCN = & 20$^{+6}_{-10}$ & 40$^{+12}_{-20}$ \\ [0.1cm]
D$^{13}$CN  & $< 12.38$ & DCN/D$^{13}$CN $\,\,\,\,$ & $>5$ \\ [0.1cm]
\hline

HNC & 13.77$^{+0.14}_{-0.13}$ & & ... \\ [0.1cm]
HN$^{13}$C  &  12.47$^{+0.37}_{-0.14}$ & HNC/HN$^{13}$C = & 20$^{+5}_{-4}$ \\ [0.1cm]
H$^{15}$NC  &  11.90$^{+0.33}_{-0.14}$ & HNC/H$^{15}$NC = & 75$^{+25}_{-15}$ & 225$^{+75}_{-45}$ \\ [0.1cm]
DNC  &  13.31$^{+0.16}_{-0.10}$ & HNC/DNC = & 2.9$^{+1.1}_{-0.9}$ & $9^{+3}_{-3}$ \\ [0.1cm]
DN$^{13}$C  & 11.84$^{+0.51}_{-0.14}$ & DNC/DN$^{13}$C = & 30$^{+8}_{-5}$ \\ [0.1cm]

\hline
CN  & 14.78$^{+0.05}_{-0.05}$ & & ... \\ [0.1cm]
$^{13}$CN  & 13.08$^{+0.19}_{-0.14}$ & CN/$^{13}$CN = & 50$^{+19}_{-11}$ \\ [0.1cm]
C$^{15}$N  & 12.40$^{+0.51}_{-0.14}$ & CN/C$^{15}$N = & 240$^{+135}_{-65}$ & 290$^{+160}_{-80}$ \\ [0.1cm]
\hline
\end{tabular}

\tablenotea{Unconvolved column densities obtained from the modeling of the various molecules. The ratio of the 
rarest isotopologues with respect to the main one are also given.}

\tablefoottext{a} For singly $^{13}$C-- or $^{15}$N-- or D-- substituted isotopologue, the column density ratio corresponds
to N(main) / N(singly subtituted). For doubly substituted isotopologues (wether $^{13}$C-- and D-- or $^{15}$N-- and D-- substituted), 
the ratio correspond to N(D-- substituted) / N(doubly substituted) 
\end{table*}

In the modeling, the center of the density distribution corresponds to the one adopted when 
modeling the continuum radial profiles and is given is Sect. \ref{modelling:continuum}. The offset with the position of the molecular
line observations is accounted for by computing the molecular spectra at a projected distance of 8$\arcsec$
from the center.
The abundance of a given molecule is obtained dividing the radial profile
in a maximum of eight regions. 
These regions are chosen as a sub--division of the regions chosen to describe the temperature profile.
Additionally, for each molecular species, the grid is defined to minimize the abundance variations between two consecutive shells.
The outermost radius considered in the model, for all the molecules,
is 450$\arcsec$. However, in Fig. \ref{profil_abondance-N2H+}--\ref{profil_abondance-NH3} where 
the abundances that result from the modelling are plotted, we truncate the radius at 150$\arcsec$ since
the abundance is constant outside 120$\arcsec$.
The abundance in each region
is then determined using a Levenberg-Marquardt algorithm, which searches for
the best value of the abundances on the base of a $\chi^2$ minimization.
Typically, a solution is found in 20--30 iterations.
In the case of the $^{13}$C and $^{15}$N isotopologues, the fit is performed simultaneously
with the main isotopologue. We force the abundance profiles of the various isotopologues 
to be identical within a scaling factor. In the case of the D--substituted isotopologues, we impose that the 
abundance in every region is in the range  $\chi$(HX)/1000 $<$ $\chi$(DX) $<$ $10 \times \chi$(HX). 
 
For N$_2$H$^+$ and o--NH$_2$D, the abundance profiles are better constrained because of 
the availability of emission maps, which is not the case for the other molecules. From the models computed
for o--NH$_2$D and N$_2$H$^+$, it appears, however, that the abundance profile derived just using the observations at the central
position should give reliable abundance profiles. The main uncertainty would concern
the balance of the molecular abundance between consecutive shells. However, the global shape of the abundance
profile would remain qualitatively unchanged. The fact that the abundance profile is sensitive to observations at a single position
comes from the steepness of the density/temperature profile, in conjunction with the availability  
of multiple lines for a single species with different critical densities. 

The error bars of the abundances are obtained by considering each region
separately. For each region, the upper and lower limit
of the abundance are set so that the $\chi^2$ does not exceed a given threshold. 
This way of computing the error bars does not account for the inter--dependence 
of the different regions, but gives indications 
of the influence of each region on the emergent intensities. For each molecule,
the best abundances are represented in Fig. \ref{profil_abondance-N2H+}, \ref{profil_abondance-HCN}, 
\ref{profil_abondance-HNC}, \ref{profil_abondance-CN} and \ref{profil_abondance-NH3} as red lines. The grey regions
correspond to the range of abundances delimited by the upper and lower limits. 
In the figures that compare the observed and modeled spectra, the grey regions correspond to line intensity uncertainties 
expected from the error bars set on the abundances.
From the results of the non--local radiative transfer modeling, we calculate for each molecular line an averaged excitation 
temperature $\overline{T}_{ex}$, as well as 
the line center opacity. These quantities are given in Table \ref{table-observations}.  The averaged excitation temperature 
is calculated by taking into account the regions of the cloud that contribute to the emerging line profile, the average being given
by Eq. (2) of \citet{daniel2012}. 

The N$_2$H$^+$ maps of \citet{huang2013} show that the B1--bS and B1--BN cores have
different V$_{LSR}$, B1--bS being at 6.3 km s$^{-1}$ and B1--bN at 7.2 km s$^{-1}$. 
In the present observations, the B1--bN component appears as a redshifted shoulder in the spectra of
some species. However, with the current angular resolutions, it is impossible to disentangle  
the exact contributions of the two emitting cores, since they spatially overlap. Additionally, our model is 1D spherical,
which does not allow us to describe in detail the complexity of this region.
We thus assumed a single V$_{LSR}$ $\sim$ 6.7 km s$^{-1}$. For some molecules, like N$_2$H$^+$ or NH$_3$,
the individual hyperfine component have symmetric shapes and with this assumption, it is possible to obtain
an overall good fit of the lines. However, for HCN or HNC, the emission at 7.2 km s$^{-1}$ is strong and 
this assumption is no longer valid. In such cases, we have chosen to calculate the $\chi^2$ just considering the blue 
part of the lines. The channels used to calculate the $\chi^2$ are shown 
as black thick histograms in the figures that compare the models to the observations.

In the model, we introduce two regions in terms of infall velocity and turbulent motion. The infall velocity in the region with
$r < 120\arcsec$ was set to 0.1 km s$^{-1}$ and the turbulent motion to 0.4 km s$^{-1}$. These values were chosen since they 
reproduce the widths of the molecular lines, as well as the asymmetry seen in some molecular lines, 
e.g. in the HNC or HCN spectra.
Outside 120$\arcsec$, the gas is considered
to be static and the turbulent motion is increased to 1.5 km s$^{-1}$
in order to reduce the emerging intensities of the HCN and HNC lines. Indeed, 
with such a high turbulence, 
the lines are absorbed but no self--absorption features are seen in the synthesized spectra. Thus, this enables 
to increase the amount of absorbing material to a larger extent than it would be possible by simply fitting the observed self--absorption
features. Consequently, this allows to increase 
the column density of the main isotopologues with respect to the 
$^{13}$C isotopologues since the latter are less affected by self--absorption effects. This is motivated by the fact that we derive low 
HCN/H$^{13}$CN and HNC/HN$^{13}$C column density ratios by compared to isotopic ratios in
local interstellar medium (see Sect. \ref{hcn}). Moreover, such an assumption is reasonable in view of the observations related to
the Perseus molecular cloud. Indeed, \citet{kirk2010} showed that the turbulent motion in this region follows 
the Larson's law \citep{larson1981}, 
with a turbulent motion that increases with the distance to the core center and typically reaches a value of 1 km s$^{-1}$ at a linear scale of 1 pc.
Finally, we emphasize that this outermost region of the model is poorly 
constrained by the data and its parameters should be treated with caution and seen as 
averaged values that would describe the overall foreground material.

Finally, in Table \ref{table-lineparameters}, we give the column densities of all the 
molecular species considered in this study, as well as the column density ratio of the rarest 
isotopologues with respect to the main ones. Since for a given species, the models are based
on various lines observed with different spatial resolutions, we give column densities which simply correspond
to an integration along the diameter of the sphere and which are not convolved with any telescope beam.

\subsection{N$_2$H$^+$}

\subsubsection{Spectroscopy and collisional rate coefficients}

The spectroscopic coupling constants for the hyperfine structure of N$_2$H$^+$ are taken 
from \citet{caselli1995}. We adopt the scaling of the rotational constants described in \citet{pagani2009},
where the way we calculate the line strengths is also described.
In the case of the $^{15}$N isotopologues, the hyperfine structure due to the $^{15}$N nucleus
is not resolved. In this case, we just introduce in the spectroscopy the spin of the $^{14}$N nucleus.
The coupling constants are from \citet{dore2009} and we use the same methodology as
\citet{pagani2009} in order to derive the line strengths of the two isotopologues.

The collisional rate coefficients are from \citet{daniel2005} and consider He as a collisional partner. 
In the case of the $^{15}$N 
isotopologues, these rate coefficients are also used but we apply the recoupling method to one
nuclear spin.

\subsubsection{Results}

In the case of N$_2$H$^+$, in addition to the observations of the $J$=1--0 and 3--2 lines at the central position, 
the abundance profile is constrained by taking into account the $J$=1--0 map shown in Fig. \ref{fig:N2H+_map}. To do so,
we compute the integrated intensity over the hyperfine components for each position of the map. These integrated intensities are then
averaged radially and the $\chi^2$ is calculated by taking into account both the central position and the integrated intensity radial
profile. 

\begin{figure*}
\begin{center}
  \subfigure[]{\includegraphics[angle=0,scale=0.35]{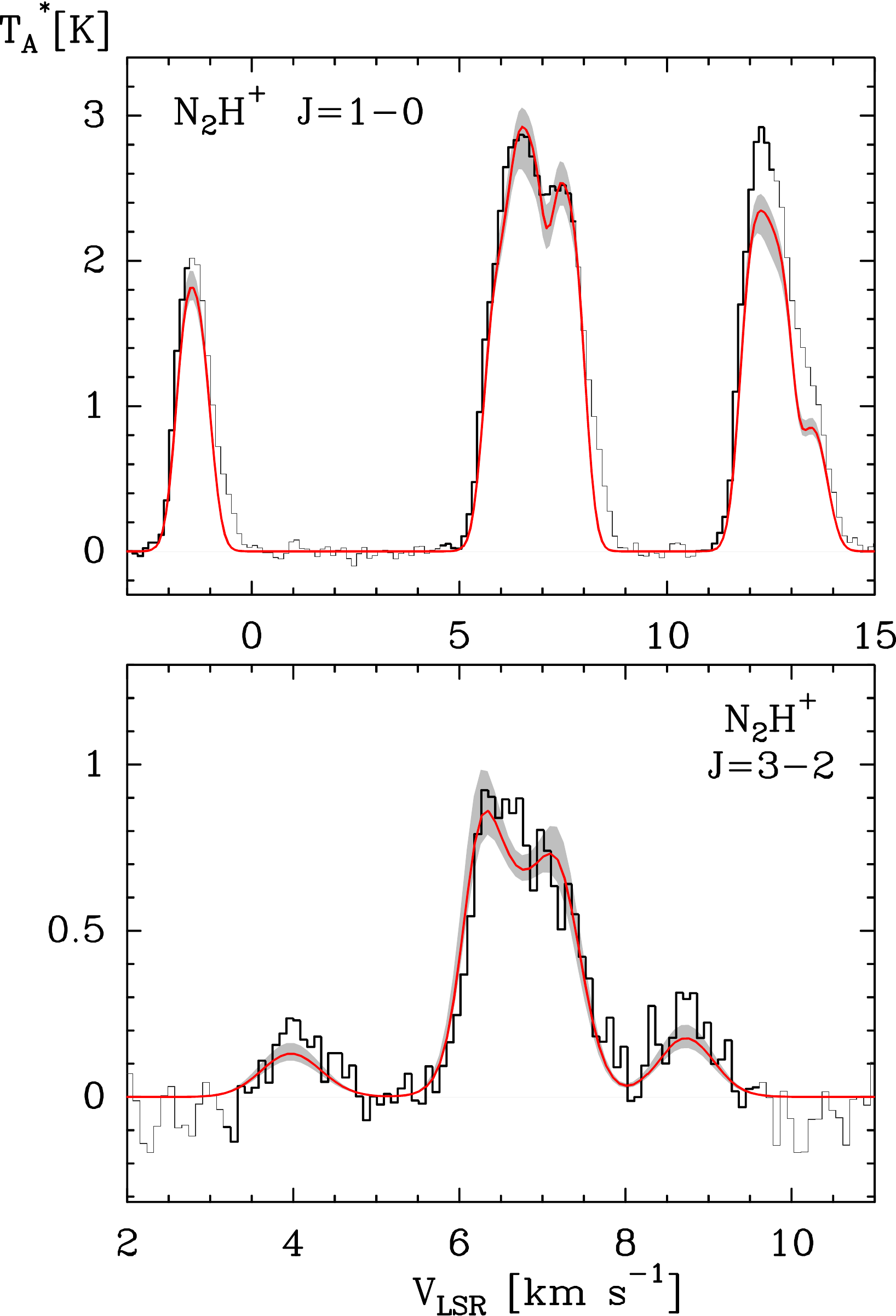}} \quad
  \subfigure[]{\includegraphics[angle=0,scale=0.35]{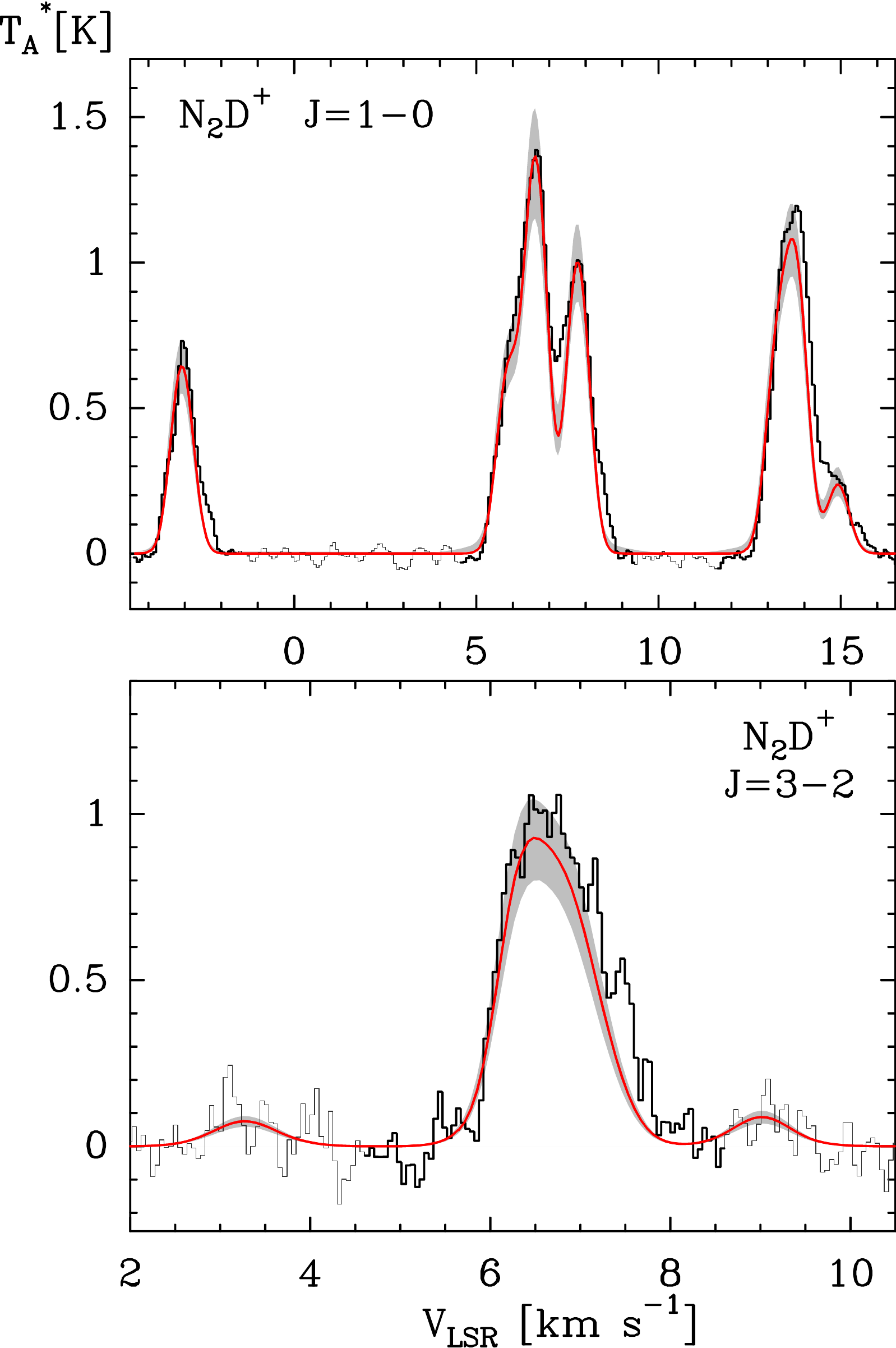}} \quad
  \subfigure[]{\includegraphics[angle=0,scale=0.35]{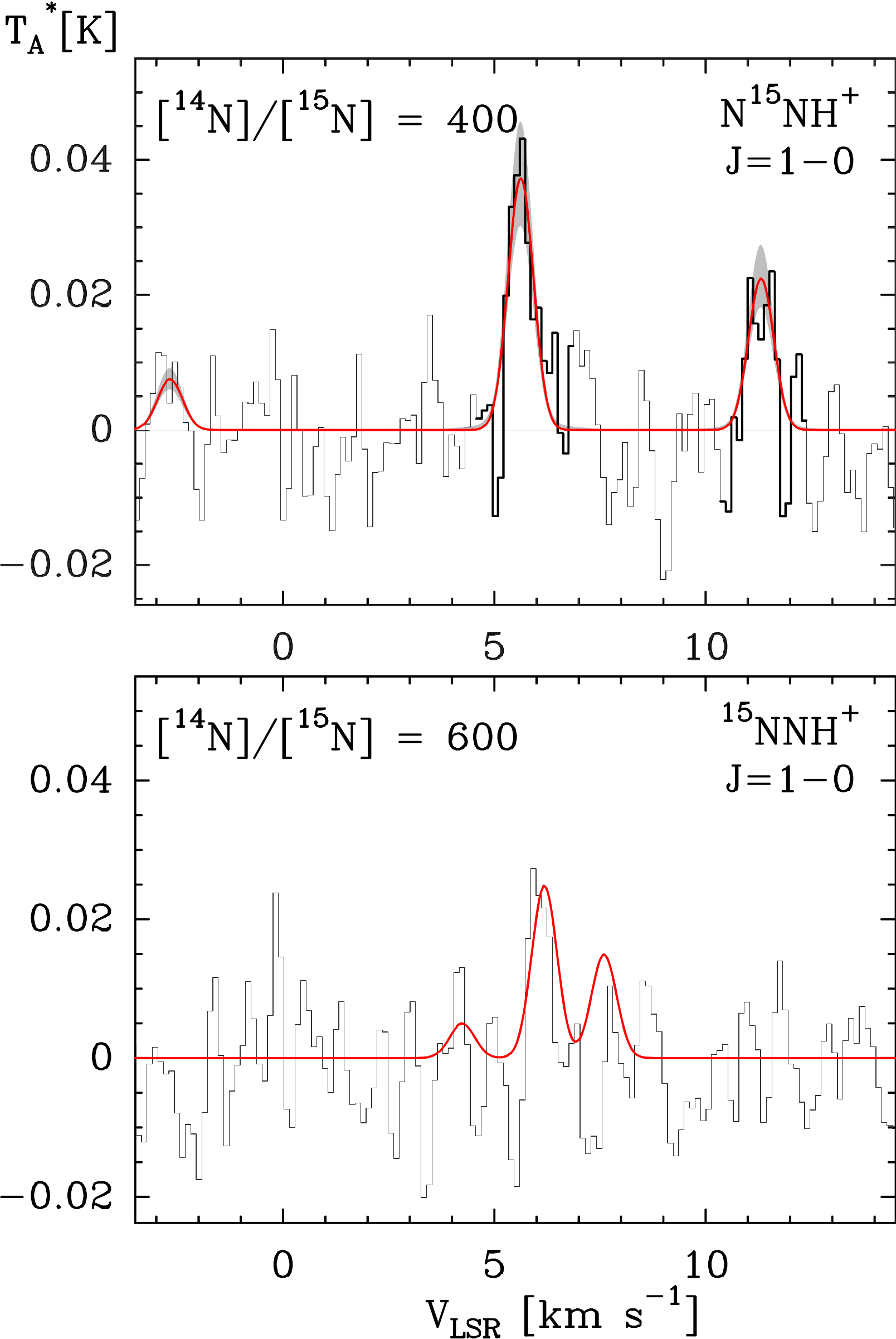}} 
\end{center}
\caption{Observed (histograms) and modeled (red lines) spectra for \textbf{(a)} N$_2$H$^+$ ($J$=1-0) (top panel) and 
N$_2$H$^+$ ($J$=3-2) (bottom panel) \textbf{(b)} N$_2$D$^+$ ($J$=1-0) (top panel) and 
N$_2$D$^+$ ($J$=3-2) (bottom panel) \textbf{(c)} N$^{15}$NH$^+$ ($J$=1-0) (top panel) and 
$^{15}$NNH$^+$ ($J$=1-0) (bottom panel). The shaded areas correspond to the variations expected
from the error-bars set on the abundance profiles and represented in Fig. \ref{profil_abondance-N2H+}.}
\label{fig:N2H+}
\end{figure*}

\begin{figure*}
\begin{center}
\includegraphics[angle=270,width=17cm]{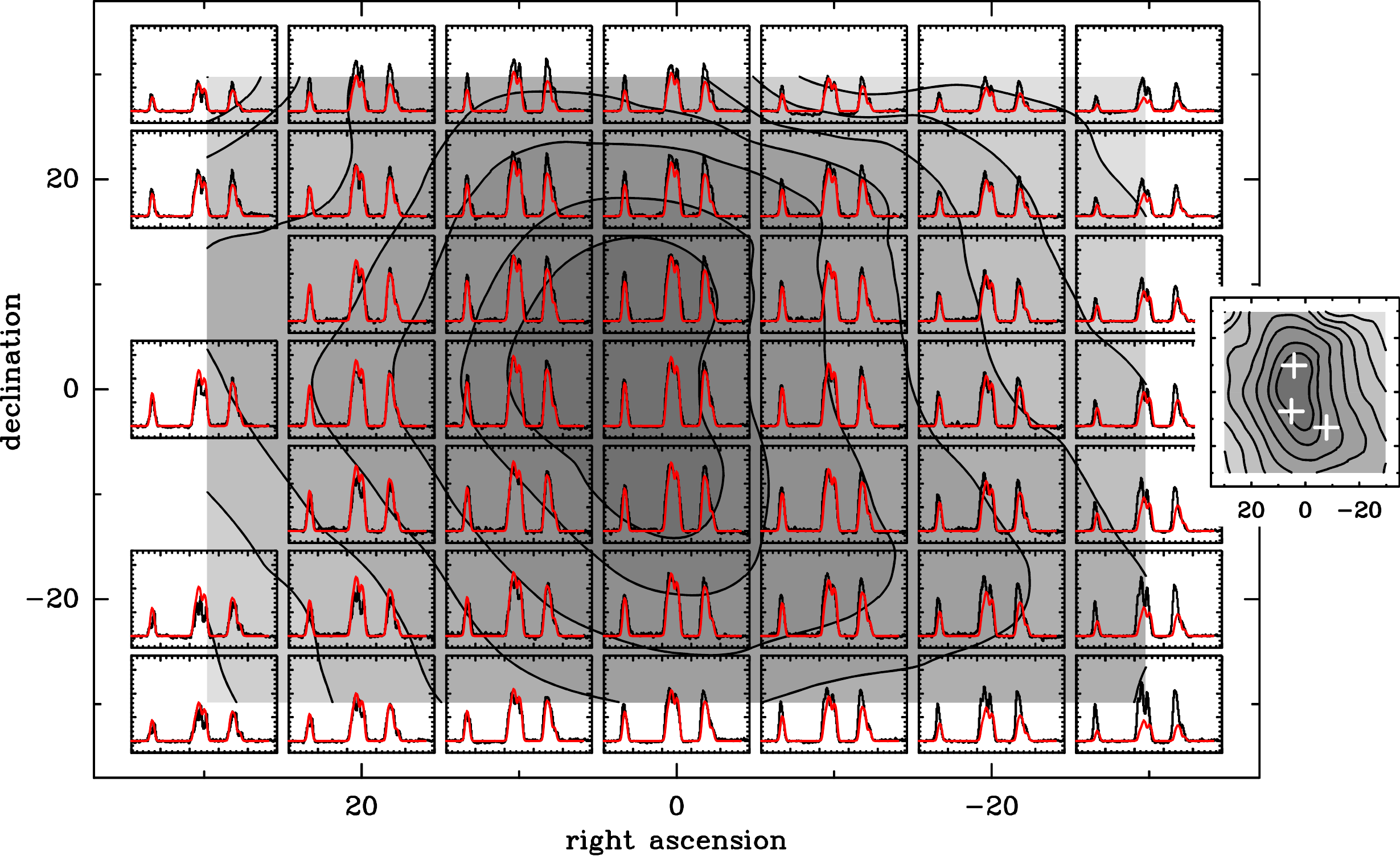}
\caption{Map of the observed (histograms) and modeled (red lines) spectra of N$_2$H$^+$ ($J$=1--0). 
The background corresponds to a map of the iscocontours of the intensity integrated over all the hyperfine components.
The map on the right side shows the isocontours with the same scale for the right ascension and declination. The white crosses indicate the position of the B1--bN and B1--bS sources identified by \citet{hirano1999}, as well as the \textit{Spitzer} source reported by \citet{jorgensen2006}.} \label{fig:N2H+_map} \vspace{-0.1cm}
\end{center}
\end{figure*}

The various abundance profiles derived from the modeling are shown in Fig. \ref{profil_abondance-N2H+}.
The comparison of the observations with the modeled line profiles is shown in 
Fig. \ref{fig:N2H+} and the map of the $J$=1-0 observations is compared with the 
model in Fig. \ref{fig:N2H+_map}. The column densities for each isotopologue, 
as well as the column density ratios, are given in Table \ref{table-lineparameters}.

\subsubsection{Comparison with previous studies}

%
%
%
%


\citet{kirk2007} have previously reported N$_2$H$^+$ maps of the whole Perseus molecular cloud
and various studies focused on a sample of prestellar or protostellar cores in this region \citep{roberts2007,emprechtinger2009,johnstone2010,friesen2013}.
Additionally, a few studies were specifically devoted to the B1b region \citep{gerin2001,huang2013}.
The column density estimates range from $\sim$$10^{13}$ to $4.5 \, 10^{13}$ cm$^{-2}$ and the current estimate
of $\sim$$3 \, 10^{13}$ cm$^{-2}$ thus falls between the previous ones. 

Moreover, the N$_2$D$^+$/N$_2$H$^+$ column density ratio was 
estimated in various studies. This ratio was found to be 0.15 by \citet{gerin2001},
0.25 by \citet{roberts2007}, 0.18 by \citet{emprechtinger2009} and 0.1 by \citet{friesen2013}. 
More recently, from interferometric observations, \citet{huang2013} estimated the ratio in the B1--bS 
and B1--bN cores and derived values of 0.13 and 0.27 towards those two positions. In the current study, we obtain 
a column density ratio $\sim$0.34 at the core center, a value higher than all the previous estimates. 

Part of the differences comes 
from the assumptions made in the respective analysis. Prior to this work, the analysis was carried 
out using the assumption of local thermodynamic equilibrium 
\citep[hereafter referred to as LTE approach and the details related to this method can be found in e.g.][]{caselli2002} 
and in most cases, 
the same excitation temperature was assumed for the two isotopologues. In the current model, by considering the excitation 
temperatures given in Table \ref{table-lineparameters},
we can see that this assumption is true within 20\% and 10\% for the $J$=1--0 and $J$=3--2 lines respectively. 
While this might not lead to substantial errors in the column density estimate, if based on the $J$=1--0 line, 
the error can however be larger for the $J$=3--2 line, since the $J$=3 energy level is at $\sim$26K. We refer the reader to the discussion performed in the 
o--NH$_2$D case, given in Sect. \ref{nh2d}, for further details.
Finally, from the current model, we infer that the average excitation temperature of N$_2$D$^+$ 
is higher than that of N$_2$H$^+$, which is a consequence of a more centrally peaked distribution of $\chi$(N$_2$D$^+$)
compared to $\chi$(N$_2$H$^+$). On the other hand, the excitation temperatures of the two $^{15}$N isotopologues of N$_2$H$^+$
are lower than that of N$_2$H$^+$. This comes from different line trapping effects. Thus, 
detailed radiative transfer models are required in order to interpret accurately the line intensities.

Additionally, regardless from the method of analysis, the models show that the respective abundances of 
N$_2$D$^+$ and N$_2$H$^+$ strongly vary with radius. Indeed, in the innermost region, i.e. $r < 15\arcsec$, we obtain
$\chi$(N$_2$D$^+$)/$\chi$(N$_2$H$^+$) larger than unity, while for greater radii, this ratio falls below 0.1. This radial variation 
of the two isotopologue abundances explains, to some extent, the dispersion in the column density ratio estimates. The reason 
is that the various studies were not all based on the same observational position and moreover, the telescope beams 
were not necessarily similar. The strong radial variations of $\chi$(N$_2$D$^+$)/$\chi$(N$_2$H$^+$) thus make 
beam dilution effects an important factor when estimating the column density ratio. 

\begin{figure}
\begin{center}
\includegraphics[angle=0,scale=0.5]{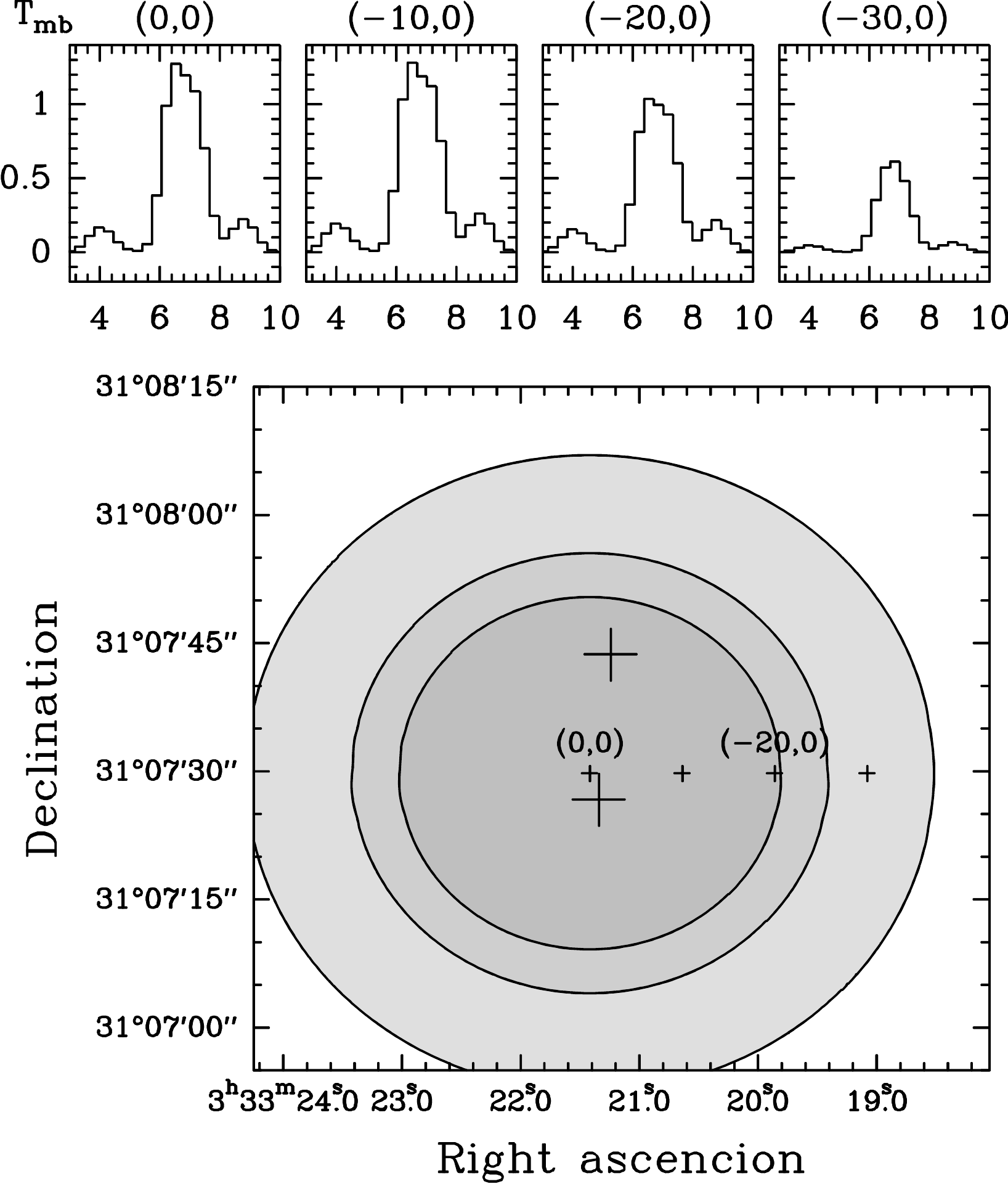}
\caption{Simulated map of the N$_2$H$^+$ $J$=3--2 integrated intensity as seen by a telescope of 
synthesized beam size of 4$\arcsec$ (bottom panel). This map can be directly compared to the map given in 
Fig. 2 (c) of \citet{huang2013}. The three isocontours shown are 0.51, 1.02 and 1.53 Jy km s$^{-1}$/beam
as in \citet{huang2013}. The large crosses indicate the positions of B1--bN and B1--bS. 
The upper panels shows line profiles as a function of the distance from the center.
The corresponding positions are indicated by small crosses on the map. 
The velocity resolution of the spectra is 0.3 km s$^{-1}$.} \label{fig:interfero_N2H+} \vspace{-0.1cm}
\end{center}
\end{figure}

The abundance profiles we derive for N$_2$H$^+$ and N$_2$D$^+$ (given in Fig. \ref{profil_abondance-N2H+}) 
show that for $r<15''$ (the region that encloses both B1--bS and B1--bN), the abundance of these species are 
reduced respectively by factors $\sim$100 and 10, with respect
to the maximum abundances which are found at radii $15\arcsec < r < 50\arcsec$. On the other hand, the interferometric maps of 
\citet{huang2013} show that the peak intensities of the N$_2$H$^+$($J$=3-2) and N$_2$D$^+$($J$=3-2) maps are found at the
positions of the B1--bN and B1--bS sources. Both findings may seem incompatible. It is thus important to 
emphasize the fact that even if the N$_2$H$^+$ and N$_2$D$^+$ abundance profiles show signs of depletion 
in the current analysis, the abundance is still high enough in this depleted region for the lines to be sensitive to the gas density, 
especially in the excited rotational lines. In other words, despite of the low abundances, the $J$=3-2 lines will still trace the density 
enhancements. This can be seen when considering the upper boundaries of the abundance represented in 
Fig. \ref{profil_abondance-N2H+}. Indeed, the upper limits of the N$_2$H$^+$ 
and N$_2$D$^+$ abundances in the region $r < 15\arcsec$ are still factors $>$10 and $>$5 below the maximum 
abundances found at $15\arcsec < r < 50\arcsec$. This is further illustrated by the fact that the synthetic
map of the N$_2$H$^+$ $J$=3-2 emission shown in Fig. \ref{fig:interfero_N2H+} has a
maximum in the inner $15\arcsec$ region.

Additionally, we recall that the description
of the innermost part of the B1b region is just a crude approximation of the exact geometry of the source, and a quantitative estimate
of the amount of depletion would require a more detailed analysis, preferentially based on a 3D modeling of the region. 
Indeed, this point is illustrated in Fig. \ref{fig:interfero_N2H+} where a synthetic map of the 
N$_2$H$^+$ $J$=3--2 emission is shown. This map is 
obtained assuming a FWHM telescope beam of 4$\arcsec$, which corresponds to the angular resolution of 
the map shown in Fig. 2 (c) of \citet{huang2013}. Compared to the interferometric map of
\citet{huang2013}, we see that the current model can reproduce, at least qualitatively, the radial variation of the $J$=3--2 
intensity for radii r $>$ 20$\arcsec$. However, the maximum integrated intensity predicted by the current model is 2 
Jy km s$^{-1}$/beam, and the integrated intensity is roughly constant within the central 15$\arcsec$. 
On the other hand, the interferometric
map shows that at the positions of B1--bS and B1--bN, the intensity can reach values of $\sim$3 Jy km s$^{-1}$/beam, and 
is above 2 Jy km s$^{-1}$/beam close to these sources, in a filament that connects the two objects. 
The current model is thus reliable enough to describe the envelope at large scales, but fails to mimic the 
complexity of the B1b region in the vicinity of the  B1--bS and B1--bN sources. 
To account for the geometry of the H$_2$ density distribution, it would be necessary to adopt a 3D model.

The N$_2$D$^+$/N$_2$H$^+$ column density ratio is mainly influenced by the abundance of the deuterated isotopologues
of H$_3^+$, whose abundances will be mainly governed by two processes. The first one is linked to the amount of CO that
remains in the gas phase. Indeed, if CO is abundant, it will preferentially react with the H$_3^+$ isotopologues and thus limits 
their abundances. As a consequence, the N$_2$H$^+$ deuteration will indirectly depend on the size 
of the dust grains, since smaller grains favor the depletion of CO \citep{flower2006a}. A second process that governs the N$_2$H$^+$
deuteration is linked to the ortho--to--para ratio of H$_2$ \citep{flower2006b,pagani2013}. Indeed, at low temperatures, 
the H$_3^+$ deuteration is initiated by the reaction
H$_3^+$ + HD $\to$ H$_2$D$^+$ + H$_2$ which has an exothermicity of $\sim$230K. The backward reaction is thus
negligible at low temperature ($T_k$ $\sim$10K), which leads to a high deuterium fractionation for H$_3^+$. However, once created, 
an efficient destruction of H$_2$D$^+$ is possible if it collides with an o--H$_2$ molecule, since the fundamental rotational level for this 
symmetry is $\sim$170K above the para--H$_2$ ground state. In other words, the o--H$_2$ internal energy helps to overcome 
the barrier of the backward reaction. A key parameter for the N$_2$H$^+$ deuteration is thus the H$_2$ 
ortho--to--para ratio. In that respect, as demonstrated by \citet{pagani2013}, the timescale at which the collapse 
proceeds plays an important role. Indeed, if the
collapse is fast, the H$_2$ ortho--to--para conversion induced by 
the collisions with H$^+$ or H$_3^+$ does not have time enough to reach the 
steady--state value. Thus, both the initial H$_2$ ortho--to--para ratio 
and the age of the cloud (i.e the time elapsed since the collapse started)
will be key parameters when deriving the final N$_2$D$^+$/N$_2$H$^+$ ratio. 

By considering the models of \citet{pagani2013} for the N$_2$H$^+$ deuteration, it appears that the N$_2$D$^+$/N$_2$H$^+$ 
abundance ratio will strongly vary in the innermost region of the cloud. In their Fig. 13, we can see that in the innermost
0.02 pc, this ratio will vary from a value in the range 1--10 in the center to around 0.1 at 0.02 pc. The corresponding models
have a central density of $\sim$$10^6$ cm$^{-3}$, which is of the same order as the central density obtained in 
the current modeling. However, the comparison is only qualitative, since 
these models are aimed at describing prestellar cores where 
the central gas temperature is much below the temperature that characterize the center of the B1b region.  
At the distance of B1, the size of 0.02 pc corresponds to an angular size of $\sim$17$\arcsec$. By examining 
Fig. \ref{profil_abondance-N2H+}, where the N$_2$H$^+$ and N$_2$D$^+$ abundance profiles are plotted, we indeed derive 
$\chi$(N$_2$D$^+$)/$\chi$(N$_2$H$^+$) $>1$ for $r < 15\arcsec$ and will on the order of 0.1 outside this radius. 
Thus the current results are consistent with what can be expected from \citet{pagani2013} model prediction, for a fast collapse model 
and for a cloud older than $5 \, 10^5$ years. 
This relatively old age is consistent with the presence of a more evolved YSO in the vicinity of the B1b region,
while the fast collapse may be related to the triggering effect of outflows from other YSOs in the B1 cloud.


\subsection{HCN and HNC} \label{hcn}

\subsubsection{Spectroscopy and collisional rate coefficients}

For the HCN isotopologues, we used the HCN / H$_2$ collisional rate coefficients of \citet{benabdallah2012} that take
into account the hyperfine structure. In the particular case of HC$^{15}$N, the $^{15}$N nucleus does not lead to a resolved 
hyperfine structure. Thus, we only considered the rotational energy structure and the rate coefficients 
are derived from the hyperfine rate coefficients of \citet{benabdallah2012} by summing over the hyperfine levels.
For the HNC isotopologues, we used the rotational collisional rate coefficients for HNC / p--H$_2$ by \citet{dumouchel2011}. 
The hyperfine rate coefficients are determined using the IOS--scaled method.

It was shown by \citet{sarassin2010} and \citet{dumouchel2010}, when considering the rate coefficients for the HCN and HNC molecules with
He, that accounting for the specificities of the two isotopomers leads to important differences in the collisional rate
coefficients. In particular, the HNC and HCN rate coefficients were found to have different propensity rules, the 
HNC rate coefficients being higher for odd $\Delta J$ values, while the HCN rate coefficients are higher for even $\Delta J$ values.
This conclusion still holds when considering the rate coefficients with H$_2$ by \citet{benabdallah2012}
and \citet{dumouchel2011}. The importance of the new HNC rate coefficients in astrophysical models was emphasized by 
\citet{sarassin2010} who showed that an abundance ratio $\chi$(HNC)/$\chi$(HCN) $> 1$
that would be derived assuming the same rate coefficients for both isomers would instead be compatible with an abundance ratio
$\chi$(HNC)/$\chi$(HCN) $\sim 1$, if one takes into account the specific rate coefficients for the two isomers. 

\subsubsection{Results}

\begin{figure*}
\begin{center}
  \subfigure[]{\includegraphics[angle=0,scale=0.35]{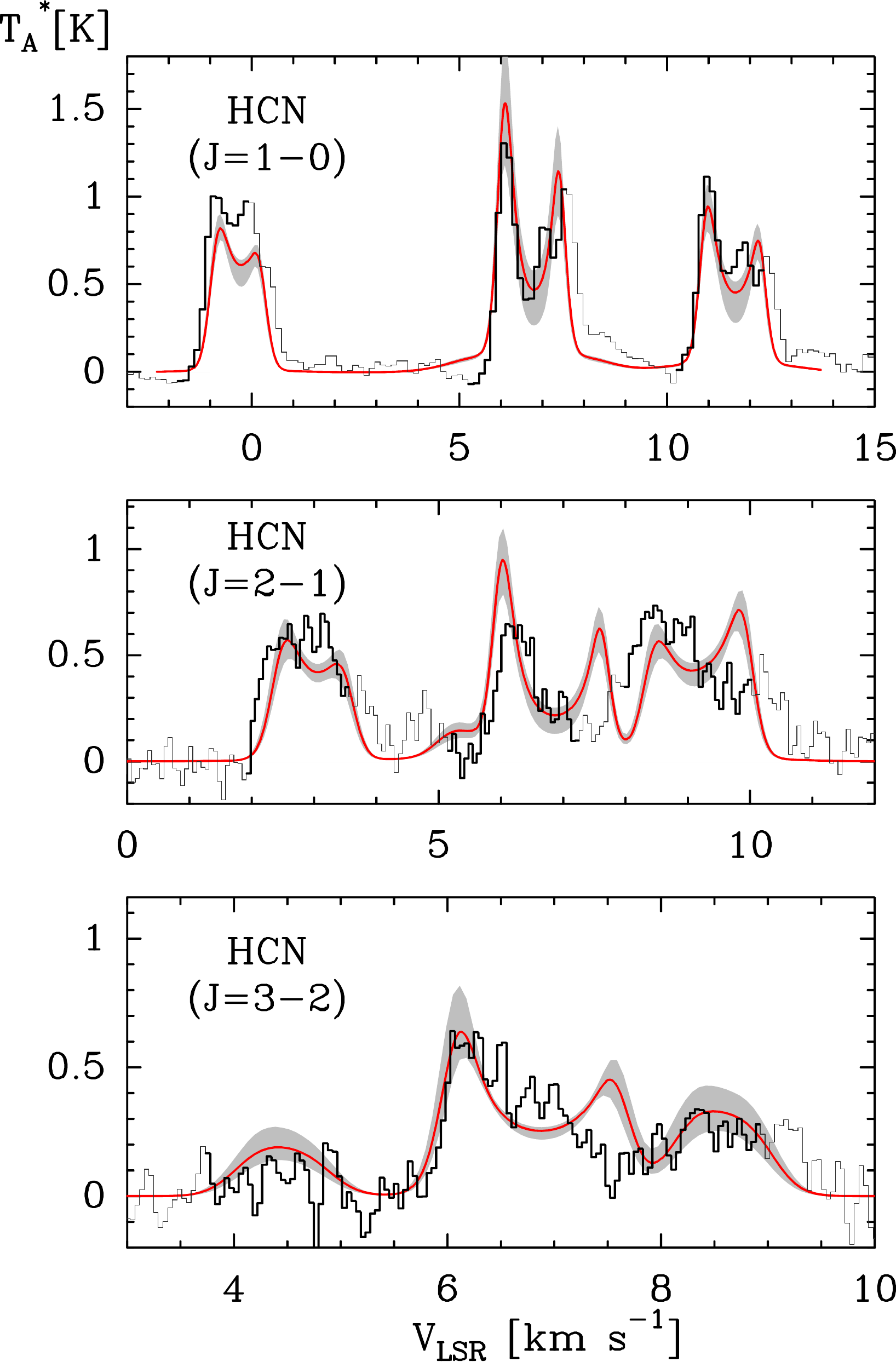}} \quad
  \subfigure[]{\includegraphics[angle=0,scale=0.35]{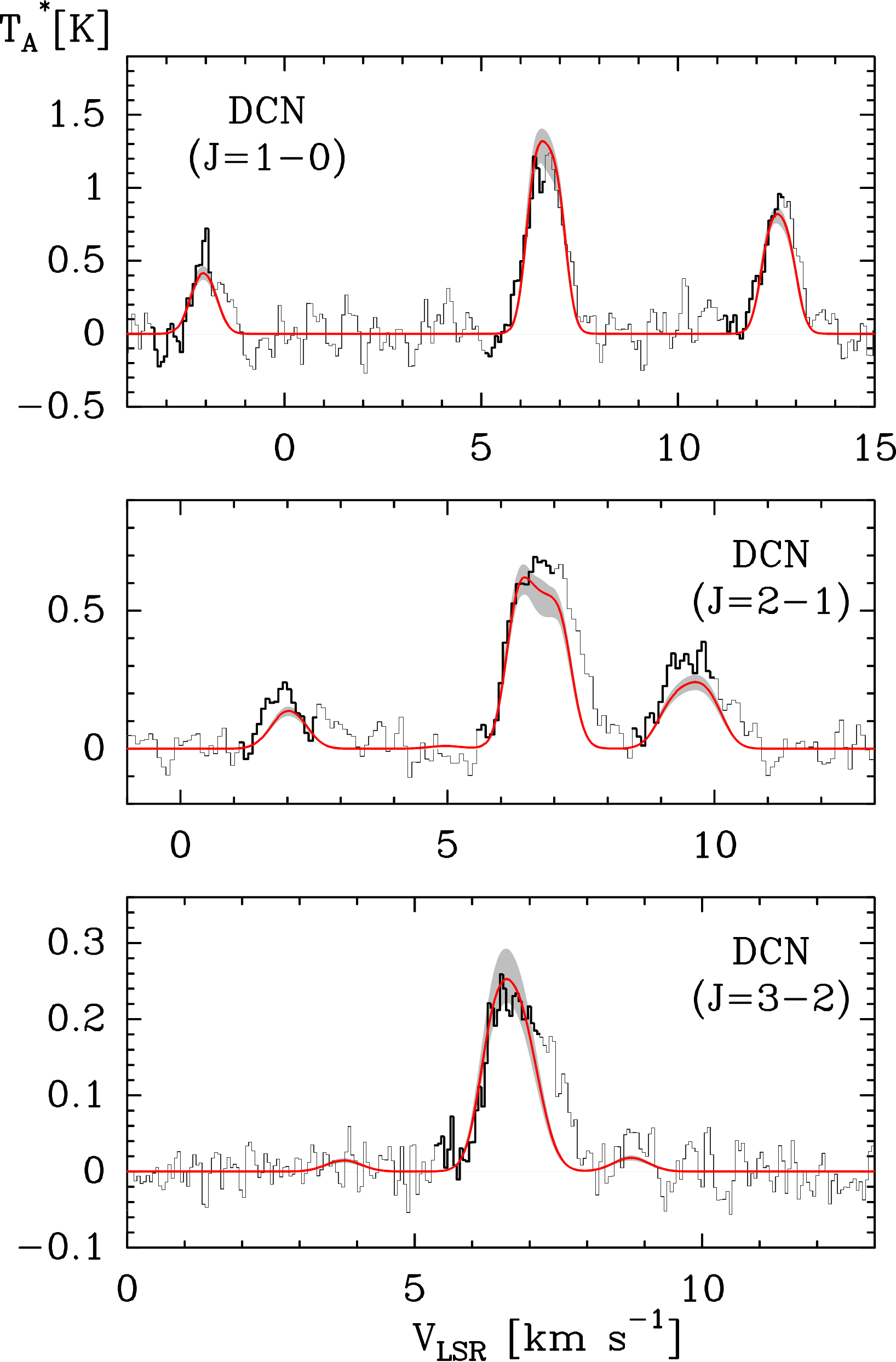}} \quad
  \subfigure[]{\includegraphics[angle=0,scale=0.35]{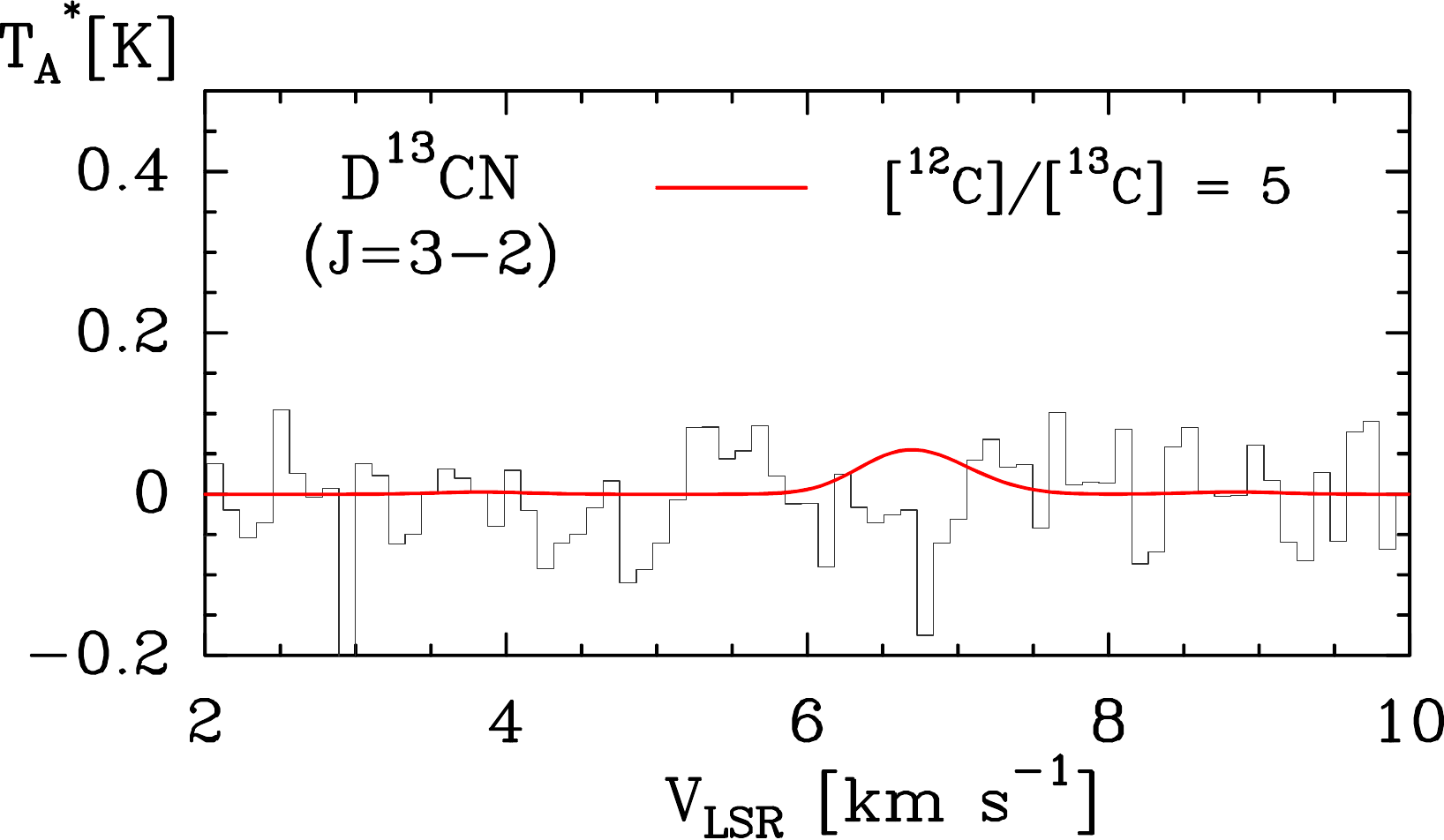}} \\
  \subfigure[]{\includegraphics[angle=0,scale=0.35]{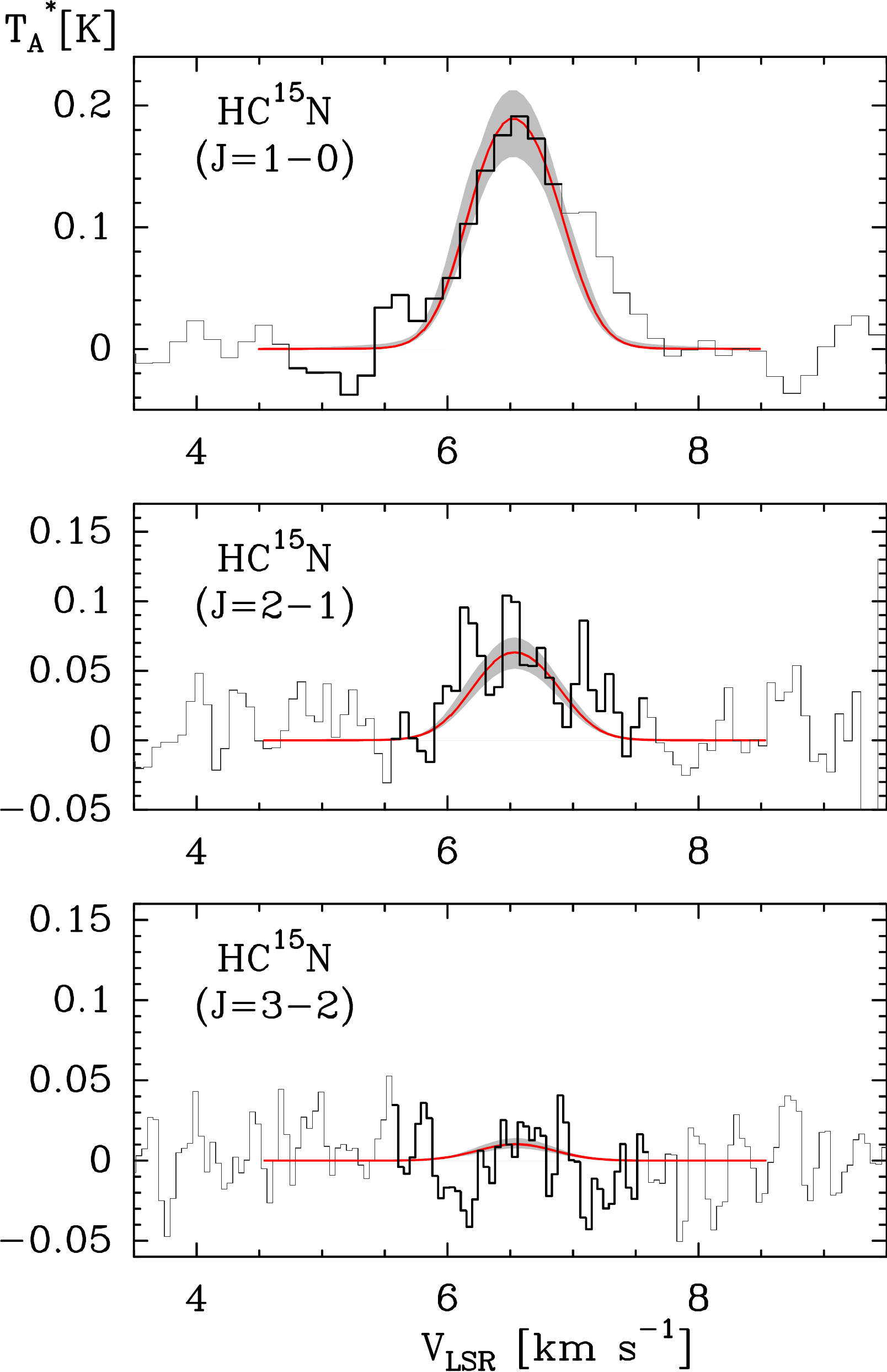}} \quad
  \subfigure[]{\includegraphics[angle=0,scale=0.35]{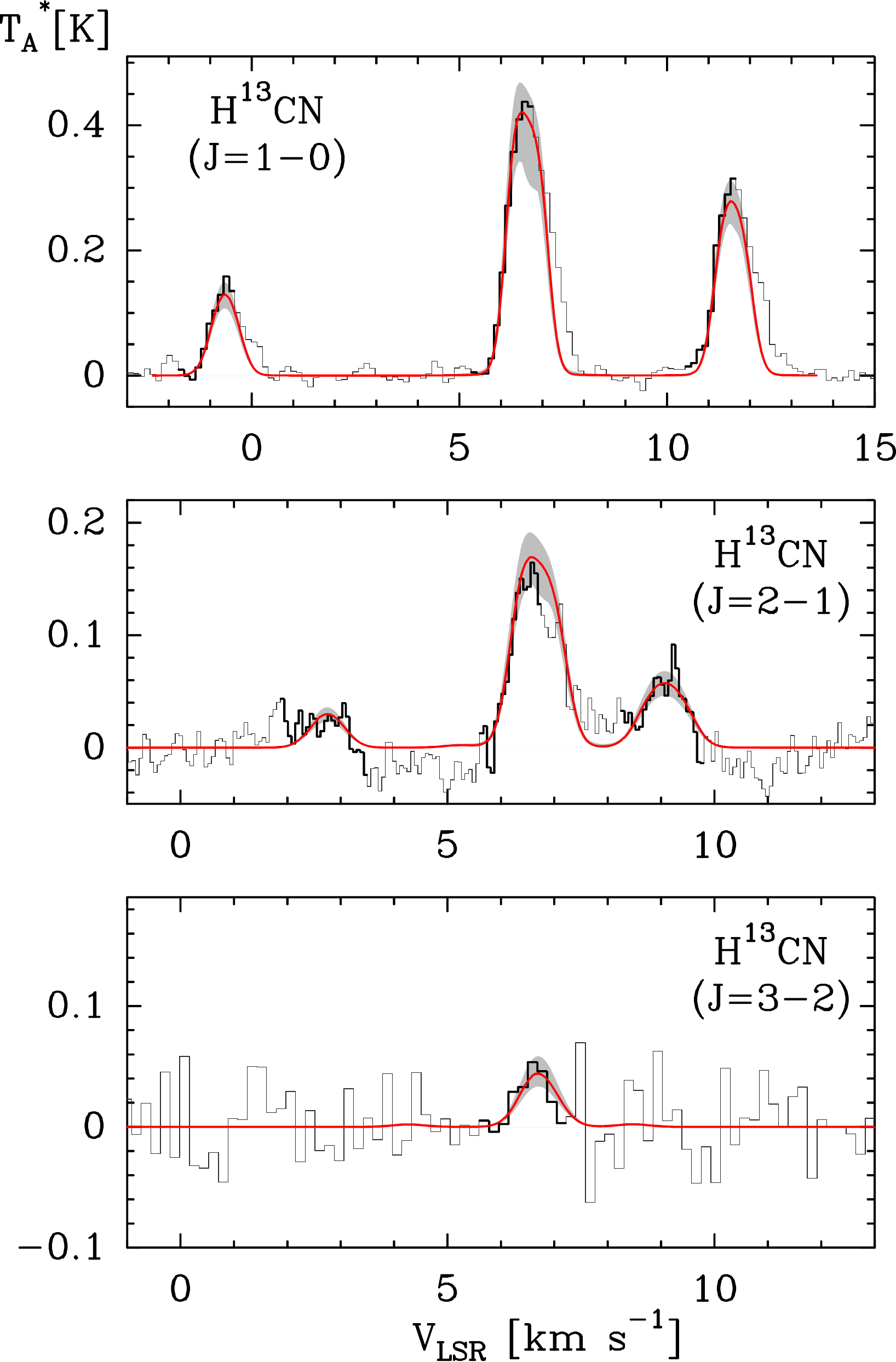}} \quad
\end{center}
\caption{
Observed (histograms) and modeled (red lines) spectra for 
\textbf{(a)} HCN ($J$=1-0) (top panel), HCN ($J$=2-1) (middle panel) and HCN ($J$=3-2) (bottom panel)
\textbf{(b)} DCN ($J$=1-0) (top panel), DCN ($J$=2-1) (top panel) and  DCN ($J$=3-2) (bottom panel)
\textbf{(c)} D$^{13}$CN ($J$ = 3-2) for a ratio of $\chi$(DCN) / $\chi$(D$^{13}$CN) = 5 (red line)
\textbf{(d)} HC$^{15}$N ($J$=1-0) (top panel), HC$^{15}$N ($J$=2-1) (middle panel) and HC$^{15}$N ($J$=3-2) (bottom panel)
\textbf{(e)} H$^{13}$CN ($J$=1-0) (top panel), H$^{13}$CN ($J$=2-1) (middle panel) and H$^{13}$CN ($J$=3-2) (bottom panel)
.}
\label{fig:HCN}
\end{figure*}

\begin{figure*}
\begin{center}
  \subfigure[]{\includegraphics[angle=0,scale=0.35]{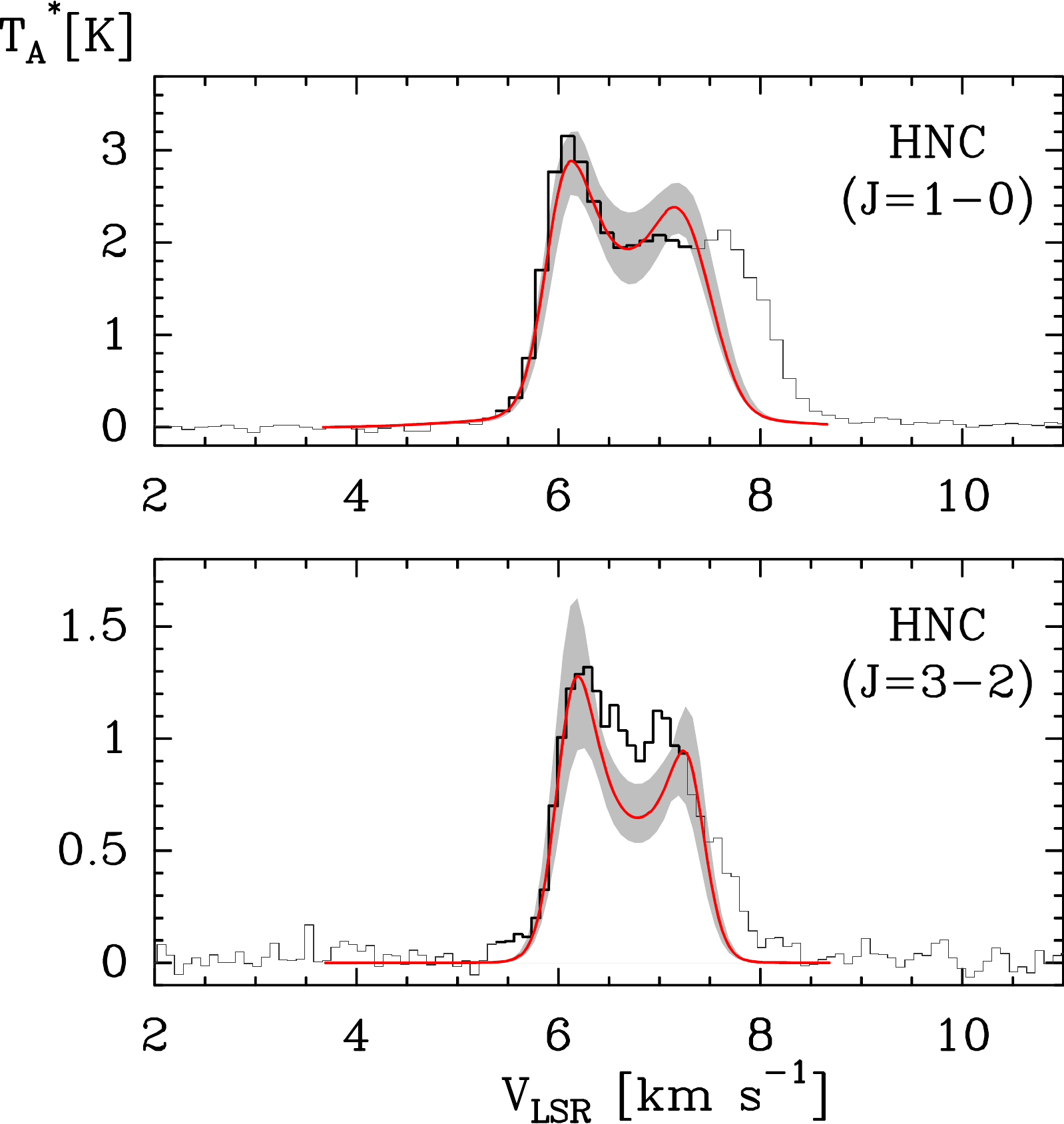}} \quad
  \subfigure[]{\includegraphics[angle=0,scale=0.35]{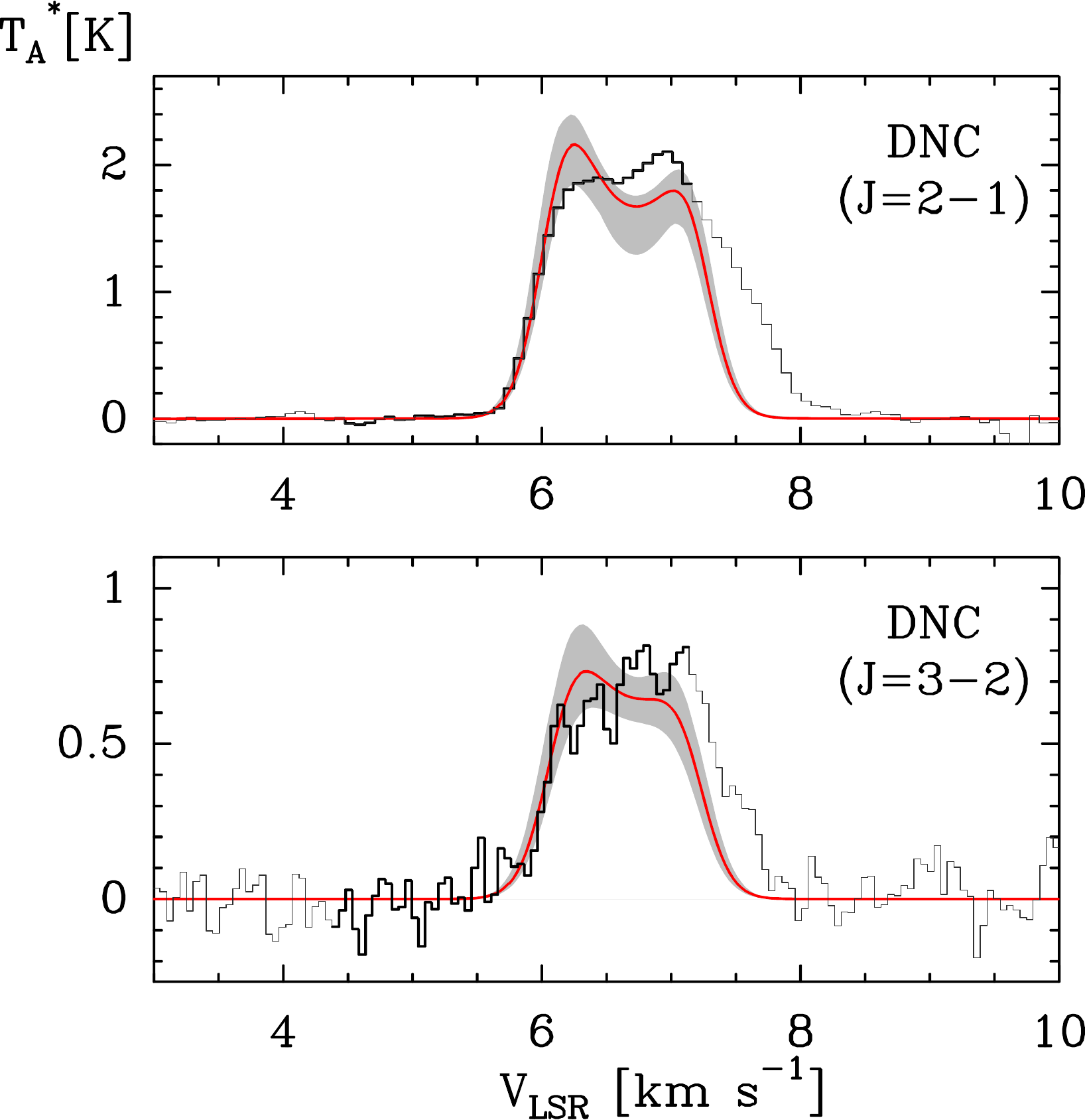}} \quad
  \subfigure[]{\includegraphics[angle=0,scale=0.35]{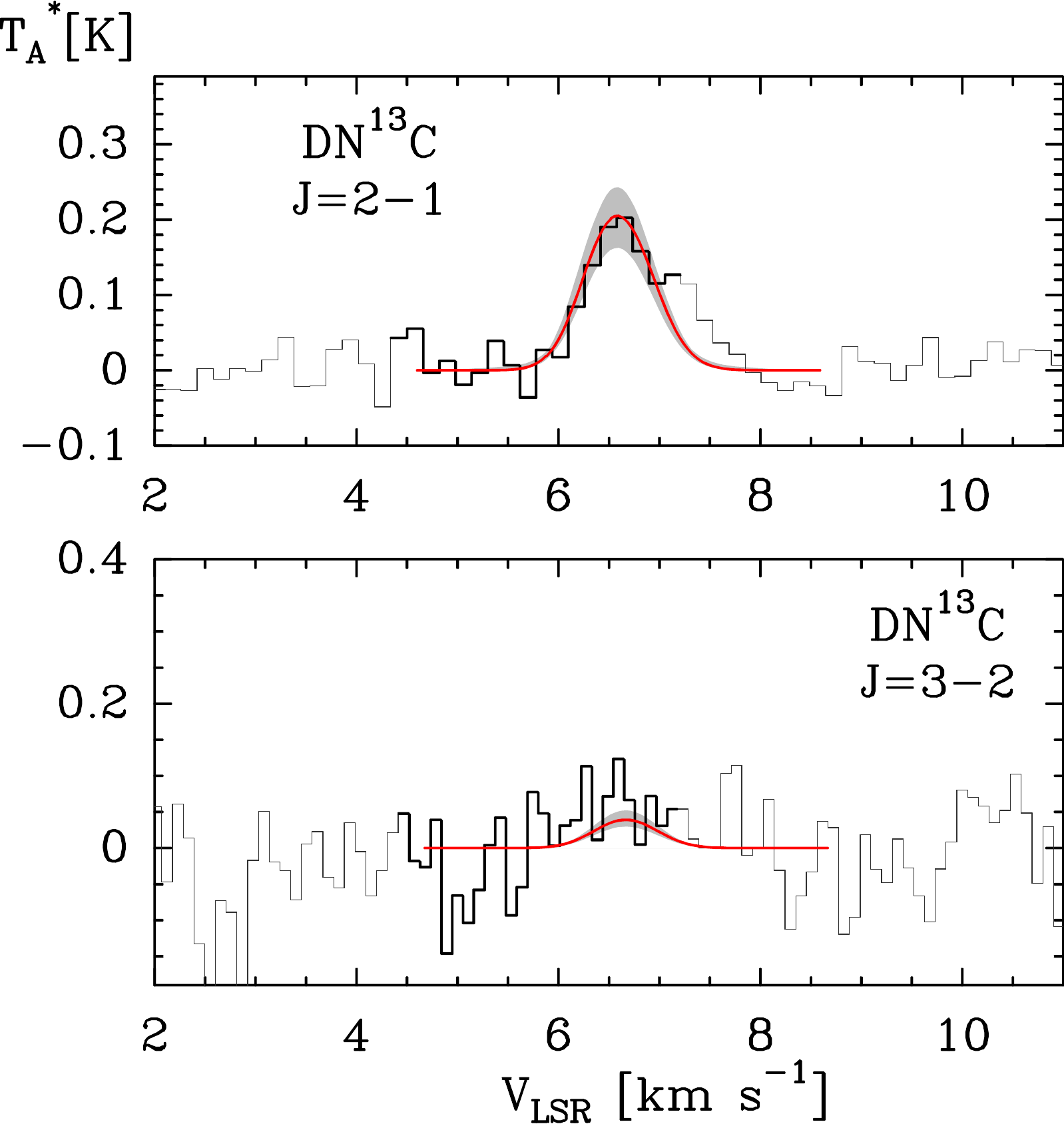}} \\
  \subfigure[]{\includegraphics[angle=0,scale=0.35]{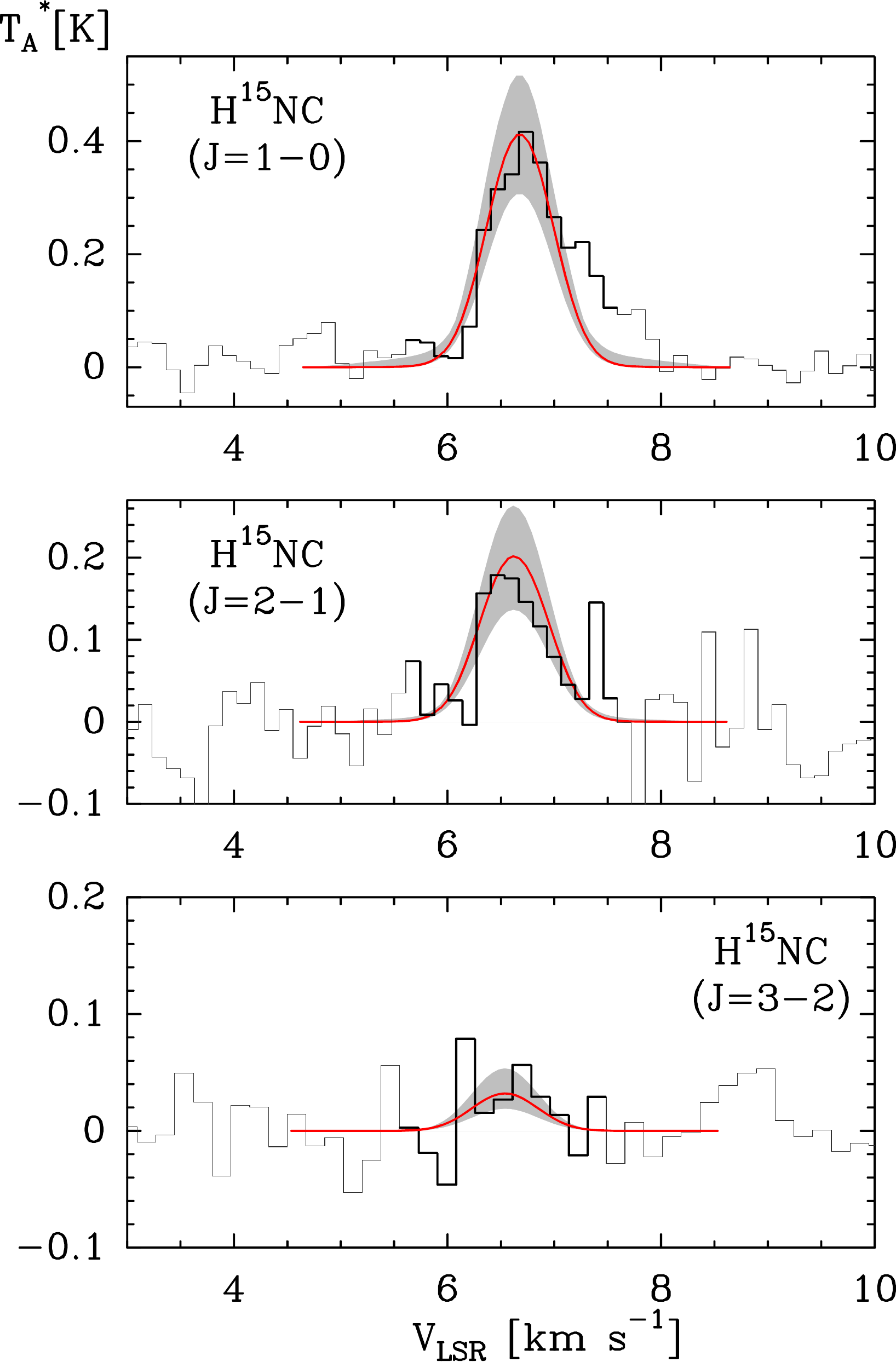}} \quad
  \subfigure[]{\includegraphics[angle=0,scale=0.35]{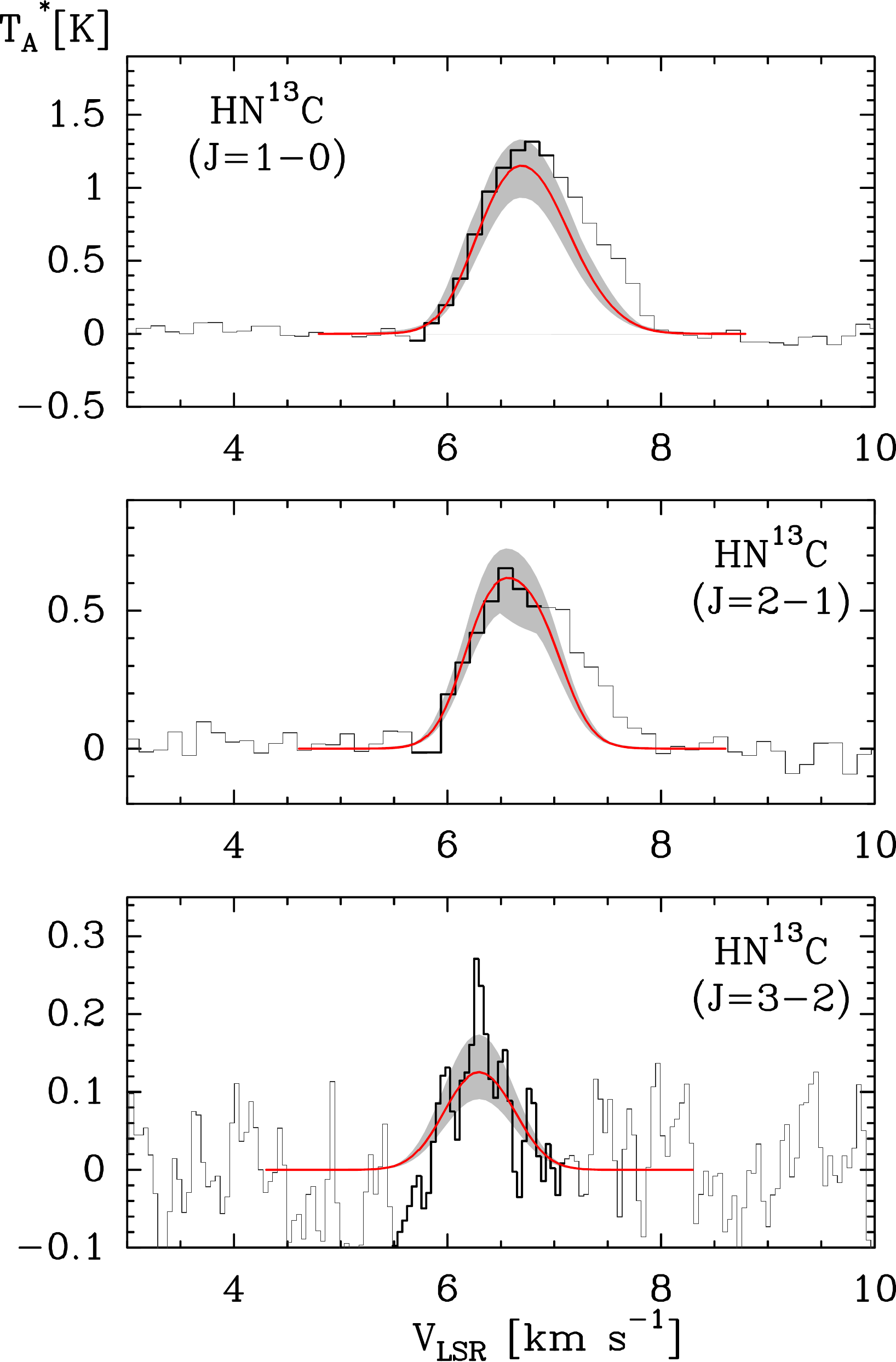}} \quad
\end{center}
\caption{
Observed (histograms) and modeled (red lines) spectra for 
\textbf{(a)} HNC ($J$=1-0) (top panel), and HNC (J=3-2) (bottom panel)
\textbf{(b)} DNC ($J$=2-1) (top panel), and  DNC ($J$=3-2) (bottom panel)
\textbf{(c)} DN$^{13}$C ($J$=2-1) (top panel), and  DN$^{13}$C ($J$=3-2) (bottom panel)
\textbf{(d)} H$^{15}$NC ($J$=1-0) (top panel), H$^{15}$NC ($J$=2-1) (middle panel) and H$^{15}$NC ($J$=3-2) (bottom panel)
\textbf{(e)} HN$^{13}$C ($J$=1-0) (top panel), HN$^{13}$C ($J$=2-1) (middle panel) and HN$^{13}$C ($J$=3-2) (bottom panel)
.}
\label{fig:HNC}
\end{figure*}

The various isotopologues abundance profiles derived from the modeling is shown in 
Fig. \ref{profil_abondance-HCN} for HCN and in Fig. \ref{profil_abondance-HNC} for HNC.
The comparison of the observations with the modeled line profiles are shown in 
Fig. \ref{fig:HCN} for the HCN isotopologues
and in Fig. \ref{fig:HNC} for the HNC ones.
In the case of D$^{13}$CN, we could only derive an upper limit for the molecular abundance.
The column densities for each isotopologue, as well as the column density ratios, are given in 
Table \ref{table-lineparameters}.

The current results show that the HCN/HNC column density ratio ranges from 2 to 4, depending
on the isotopologue considered (omitting the D-substituted isotopologues). 
If we consider the $^{13}$C and $^{15}$N isotopologues, we obtain a mean
ratio HCN/HNC $\sim$ 2.3.
Additionally, we found that HCN shows the lowest degree of deuteration 
fractionation of all the molecules considered in this work and, in particular, the degree of deuteration
is lower than for the HNC isomer. We thus obtained a ratio DCN/DNC $\sim$ 0.6.
All of these findings are in good agreement with the 
predictions of chemical models \citep[see e.g. Fig. 2 of][]{gerin2009}.

A main uncertainty in the modeling of HCN and HNC comes from the fact that we obtained $^{12}$C/$^{13}$C ratios 
which are respectively factors of 2 and 3 lower than expected for the local ISM \citep{lucas1998}. A main
issue while modeling HCN is that the lines have large opacities. Indeed, as we can see in Table \ref{table-lineparameters}, 
the opacities we
obtained for the $J$=1--0, 2--1 and 3--2 lines are $\sim$56, 90 and 37, respectively. With such high opacities, we can expect that part
of the HCN molecules will be hidden since their emission will be subsequently absorbed by the outermost layers 
of the cloud.

\begin{figure}
\begin{center}
\includegraphics[angle=270,scale=0.35]{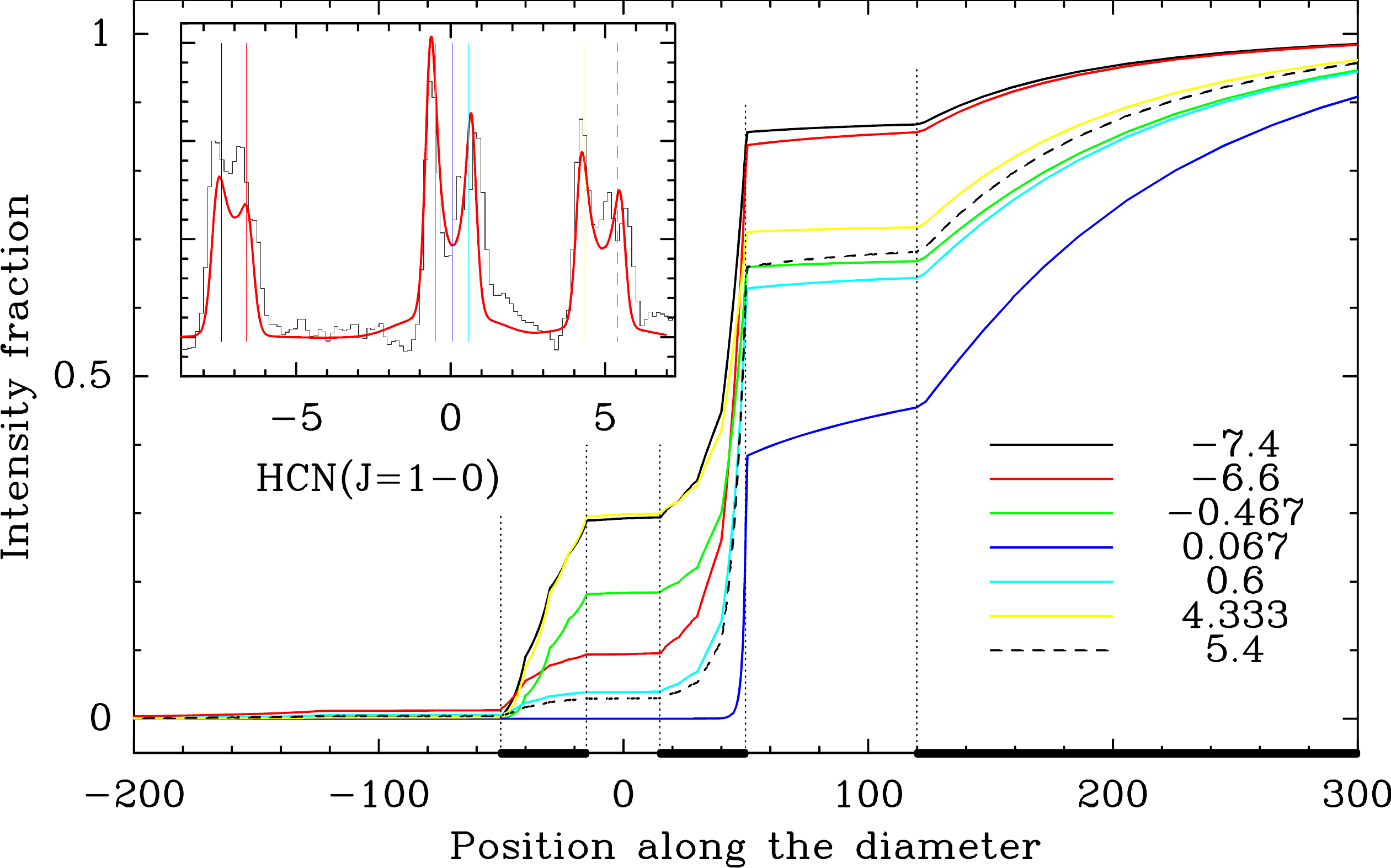} \\[0.5cm]
\includegraphics[angle=270,scale=0.35]{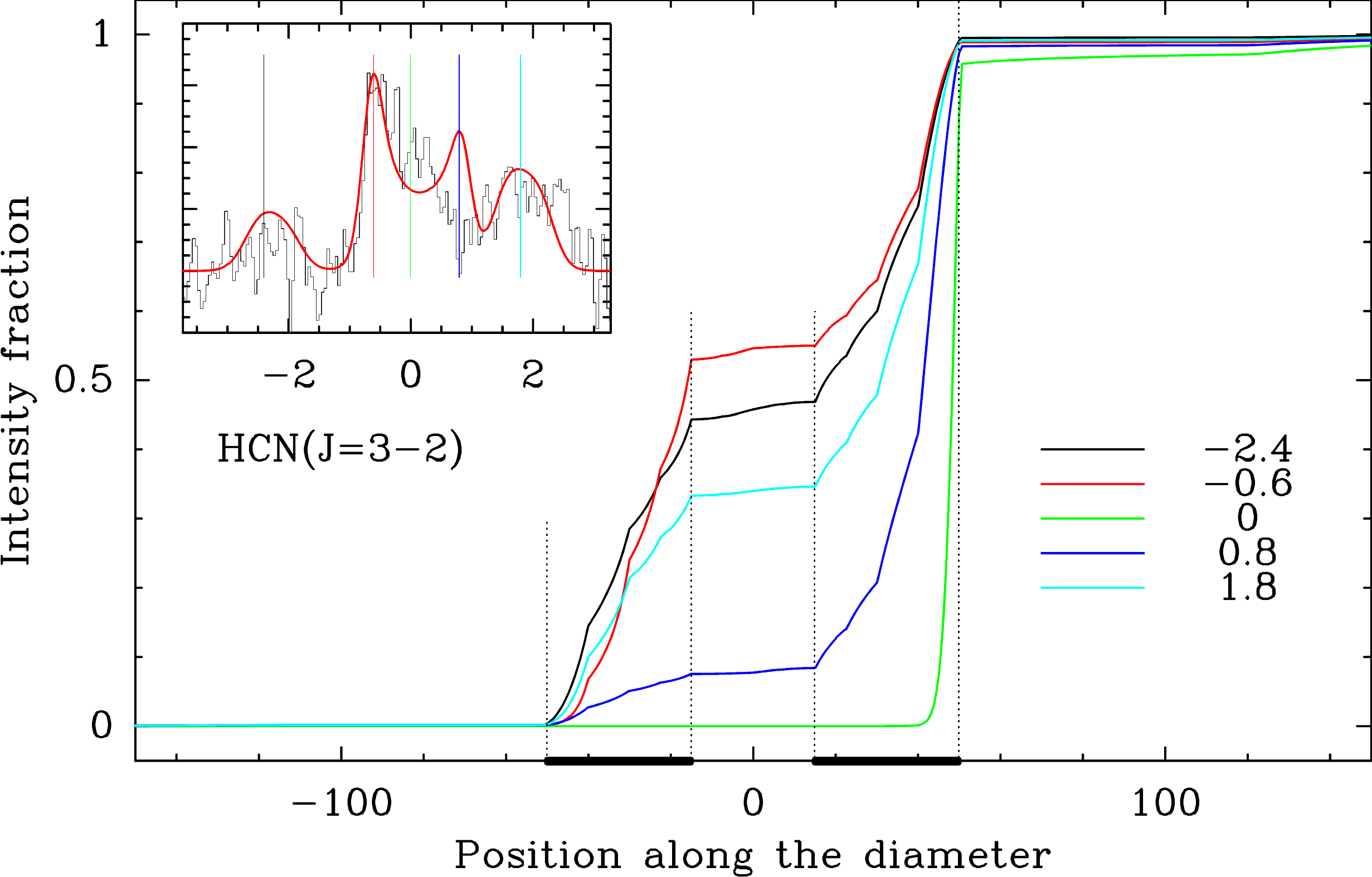}
\caption{Contribution to the emergent HCN $J$=1--0 (upper panel) and 3--2 (lower panel) intensities 
by the different radii of the sphere, along the diameter. The dotted vertical lines indicate the radii 
$r = 15\arcsec$, $50\arcsec$ and $120\arcsec$ that delimitate the regions where most of the 
photons are created. The main contribution to the emergent flux comes from the regions indicated by the 
bold segments in the $x$--axis. In each panel, the inset shows the emergent flux towards the centre of the model,
and the vertical lines indicate the velocities for which we considered the propagation along the diameter of the sphere.
} \label{fig:HCN_vel} \vspace{-0.1cm}
\end{center}
\end{figure}

In Fig. \ref{fig:HCN_vel}, we give the contribution of the various layers of the model to the emerging intensity,
using the quantity:
\begin{eqnarray}
I_v(x) =  \int_0^x S_v(y) \, e^{-\tau_v(y)} \, dy  
\end{eqnarray}
which gives the intensity at a velocity offset $v$, for an impact parameter which 
corresponds to the diameter of the cloud, $D$. The origin of the $y$--axis corresponds to the outer edge of the cloud
where the photons escape. The integration from $y=0$ to $y=x$ thus indicates the number of photons
created along that path and
accounts for the subsequent absorption of the photons by the outermost layers and 
until the photons escape from the cloud at $y=0$. 
The emerging intensity is thus $I_v(D)$ and in Fig. \ref{fig:HCN_vel}, we show the quantity $(I_v(D)-I_v(x))/I_v(D)$,
which indicates the percentage contribution of each radial point to the emerging intensity. 
In this figure, we show the contribution of the various regions to the emergent intensity, for the $J$=1--0 and $J$=3--2 lines
and at various velocity offsets. When considering the abundance of HCN given in Fig. \ref{profil_abondance-HCN}, we see that
regions depleted
in HCN, i.e.  $0'' < r < 15''$ and to a lesser extent $50'' < r < 120''$, have no significant contribution to the emergent intensity.
However, the whole region $15'' < r < 50''$ contributes to the line profile, and despite the large opacities, the innermost
part of this region is still visible. The contribution at a given radius, however, depends on the frequency considered. For example, if we 
consider the velocity offset $v=0.07$ km s$^{-1}$ for the $J,F$=1,2--0,1 line, we see that 40\% 
of the emergent flux is created in a thin layer, 
which is 1$\arcsec$ or 2$\arcsec$ wide, at a distance $r$$\sim$50$\arcsec$. When the velocity offset with respect to the hyperfine transition increases, the layer that contributes to 
the emergent flux becomes
wider. This can be seen considering, for example, the velocity offsets $v = 0.6$ km s$^{-1}$ for the 
$J,F$=1,2--0,1 line or
$v = 5.4$ km s$^{-1}$ for the $J,F$=1,1--0,1  line. Additionally, we can see that
the contribution of the backward hemisphere is more important for the blue part of the spectra, while the red part is 
more significantly influenced by the front hemisphere. This effect, which can be seen considering the velocity offsets 
$v = -7.4$ km s$^{-1}$ and -6.6 km s$^{-1}$ for the $J,F$=1,0--0,1 line,
is due to the infall motion we have introduced in the model. If we consider the relative contribution of 
various layers to the $J$=1--0 and $J$=3--2 lines,
we can see that the region  $120'' < r < 450''$ affects the $J$=1--0, line while the $J$=3--2 line remains mostly unaffected. 
The above conclusions are reached by considering a single impact parameter that crosses the center of the sphere.
When comparing the models to the observations, various impact parameters have to be considered in order 
to perform the convolution with the telescope beam. However, the conclusions reached above remain true for other impact
parameters. This explains why in Fig. \ref{profil_abondance-HCN}, the abundance in every region has an upper boundary when
we fully treat the coupling between the telescope beam and the emission that emerges from the cloud.
Finally, given
that all the regions where HCN is abundant contribute to some extent to the emerging profile, it is not possible, with the current model,
to increase the HCN abundance in a given region without modifying the emerging intensity. Hence, with the current assumptions, it is impossible to 
increase the $^{12}$C/$^{13}$C ratio to obtain a value that would be consistent with the local ISM value. 
To perform such a calculation, a multi--dimensional model would be needed.

\subsection{NH$_3$}

\subsubsection{spectroscopy and  collisional rate coefficients}

We used the collisional rate coefficients of \citet{maret2009} which are calculated considering
p--H$_2$($J$=0) as a collisional partner.
For both NH$_3$ and $^{15}$NH$_3$, accurate rest frequencies for the hyperfine transitions 
have been measured by \citet{kukolich1967,kukolich1968}. The hamiltonian was subsequently 
refined by \citet{hougen1972}. More recently, \citet{coudert2009} determined a more extensive
set of frequencies and line strengths using ab--initio calculations. These latter values are 
reported in the \textit{splatalogue} database. In the NH$_3$ modelling, we used the values from 
\citet{coudert2009}, except for the (1,1) and (2,2) lines. For these lines, the spectroscopy was retrieved
from the CLASS software since the rest frequencies are adapted from the experimental measurements
and are of greater accuracy.

When computing excitation of $^{15}$NH$_3$, the RT is done
including the unsplitted rotational levels of the molecule.
The spectrum for the hyperfine structure is then obtained using the formula:
\begin{eqnarray}
T_A^{hyp}(v) = \sum_i T_A^{rot}(v-v_i) \times \frac { 1 - \textrm{exp} \left[-s_i \times \tau^{rot}(v-v_i)  \right] }{1 - \textrm{exp} \left[-\tau^{rot}(v-v_i) \right]}
\end{eqnarray}
where $T_A^{hyp}(v)$ is the antenna temperature of the hyperfine lines at a velocity offset $v$,
$T_A^{rot}(v)$ the one of the rotational line, $v_i$ is the velocity offset of the $i^{th}$ hyperfine 
component, $s_i$ is its associated line strength and $\tau^{rot}(v)$ is the opacity of the rotational line.
The rest frequencies and line strengths of the (1,1) line corresponds to the values 
given by \citet{kukolich1967,kukolich1968}.

\subsubsection{Results}

\begin{figure}
\begin{center}
  \subfigure[]{\includegraphics[angle=0,scale=0.35]{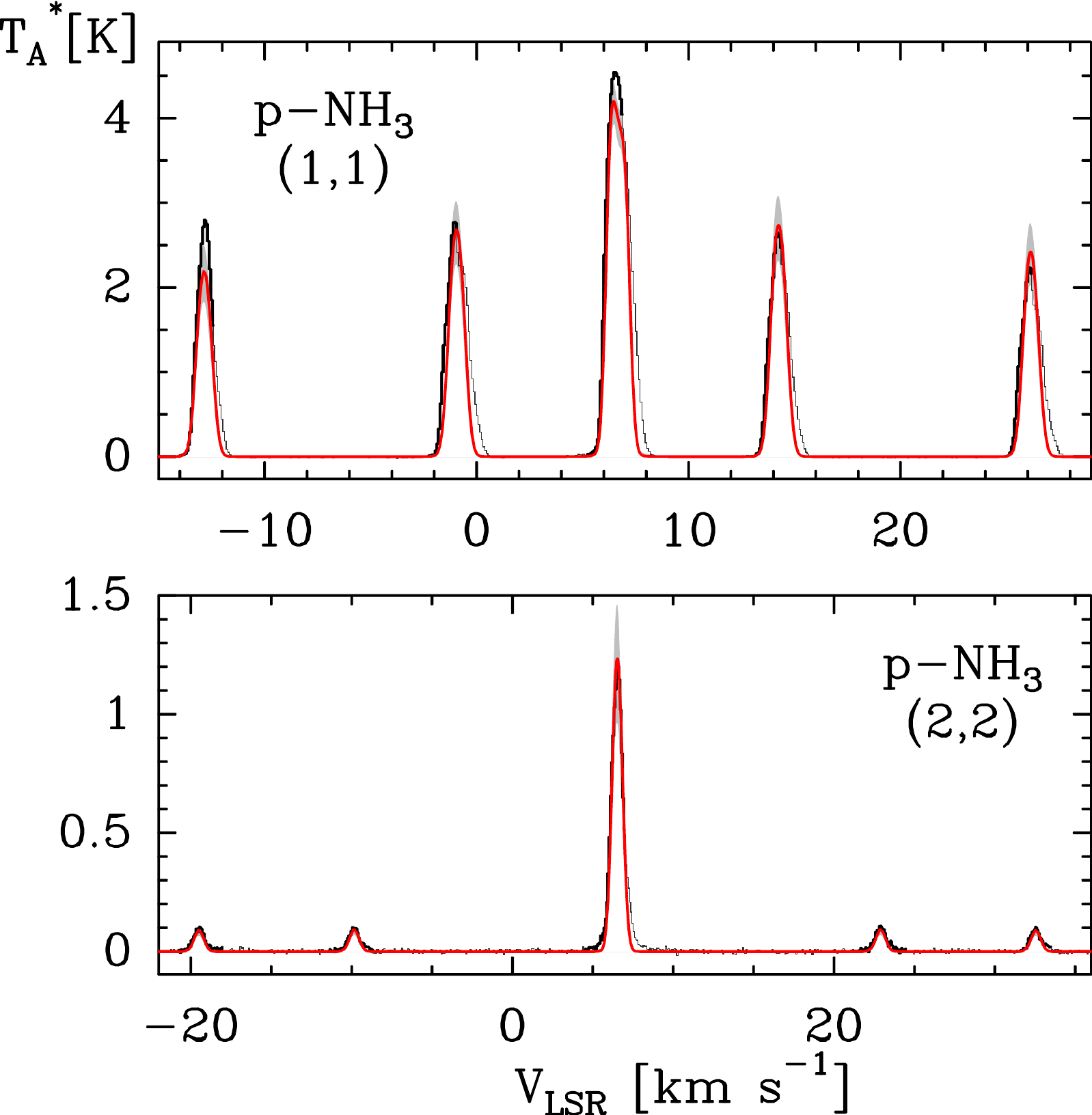}} \\
  \subfigure[]{\includegraphics[angle=0,scale=0.35]{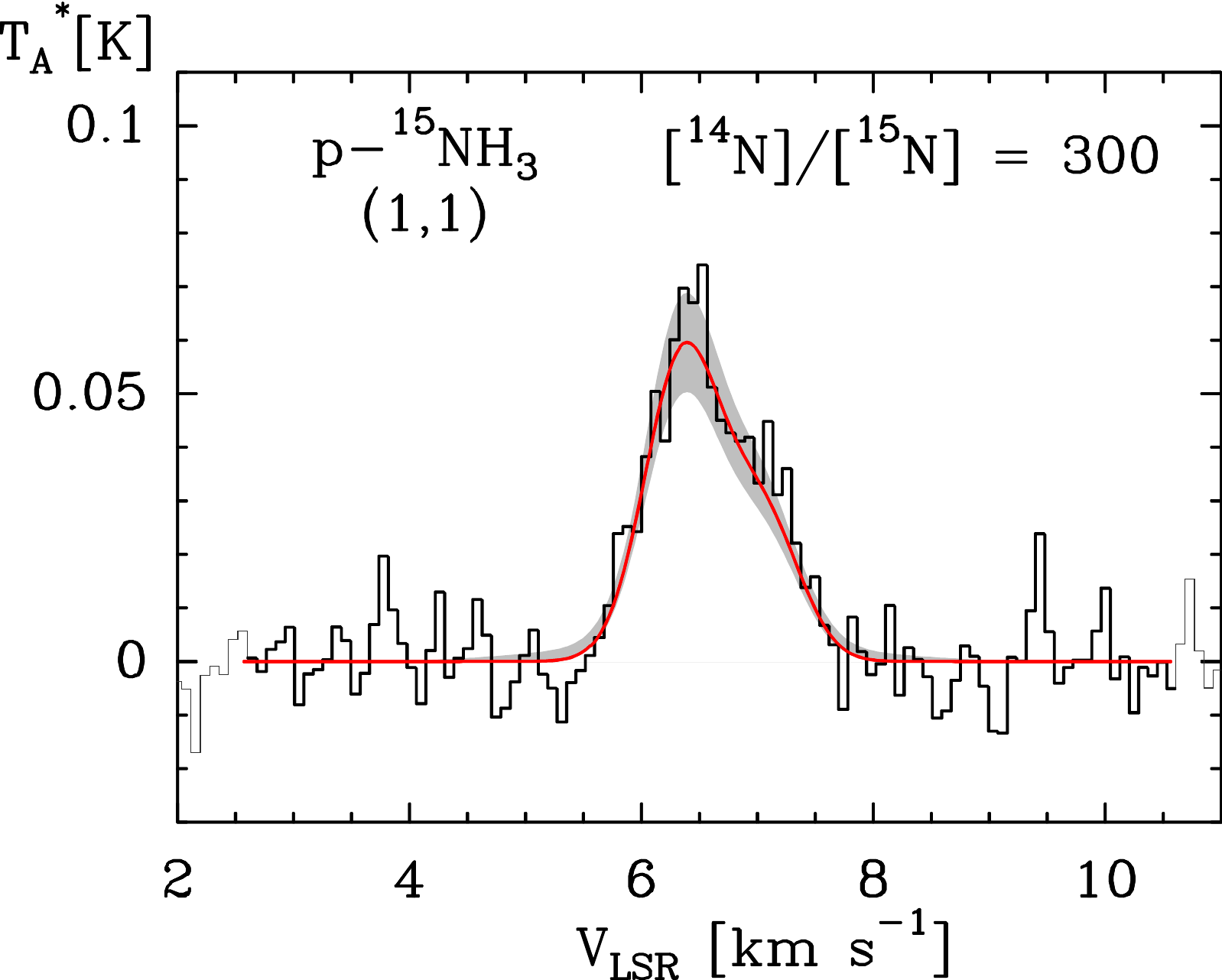}} 
\end{center}
\caption{
Observed (histograms) and modeled (red lines) spectra for 
\textbf{(a)} p--NH$_3$ (1,1) (top panel) and p--NH$_3$ (2,2) (bottom panel)
\textbf{(b)} p--$^{15}$NH$_3$ (1,1)
.}
\label{fig:NH3}
\end{figure}

Fig. \ref{profil_abondance-NH3} shows the isotopologue abundance profiles derived from the modeling.
The comparison of the observations with the modeled line profiles is shown in 
Fig. \ref{fig:NH3}.
The column density for each isotopologue, as well as the column density ratios, are given in 
Table \ref{table-lineparameters}.

\subsubsection{comparison with previous studies}

%
%


%
%



The p--NH$_3$ emission toward 
B1b was studied by \citet{bachiller1990,rosolowsky2008,johnstone2010,lis2010}. The current column density estimate, 
N(p--NH$_3$) =  $5.5 \, 10^{14}$ cm$^{-2}$, is in reasonable agreement 
with the previous works. The p--NH$_3$ column density was estimated to
$1.25 \, 10^{15}$ cm$^{-2}$ by \citet{bachiller1990}, 
$ 5.6 \, 10^{14}$ cm$^{-2}$ by \citet{rosolowsky2008},
$1.4 \, 10^{15}$ cm$^{-2}$ by \citet{lis2010} and 
$6.5 \, 10^{14}$ cm$^{-2}$ by \citet{johnstone2010}. 
Note that except in the latter study, the 
authors reported the ortho+para NH$_3$ column density, with an ortho--to--para ratio of 1. The values reported
here thus correspond to half of the column densities reported in these works. The column density 
estimates from the various studies agree within a factor $\sim$2, the current estimate being at the low--end 
of the reported values.

The NH$_3$ and $^{15}$NH$_3$ data used in the current study were previously analyzed 
by \citet{lis2010}. The column density ratio we obtained, i.e.  
N(NH$_3$)/N($^{15}$NH$_3$) = 300$^{+55}_{-40}$ agree, within the error bars, with the 
previous estimate of 334$^{+50}_{-50}$.


\subsection{NH$_2$D}\label{nh2d}

\subsubsection{Collisional rate coefficients}

For o--NH$_2$D, we used the collisional rate coefficients with He of \citet{machin2006}. Recently,
rate coefficients for ND$_2$H with H$_2$ have been determined by \citet{wiesenfeld2011} and it
was shown that for this molecule, the rate coefficients with H$_2$ are, on the average, a factor 10 
higher than the rate coefficients with He. We thus expect similar  
differences for NH$_2$D and, as a first guess, applied a scaling factor of 10 to the rate 
coefficients of \citet{machin2006}. However, with such an approach, the models fail to reproduce the spectral line 
shape of the $1_{1,1}s$--$1_{0,1}a$ line\footnote{In what follows, for simplicity, we omit the symmetry 
labels $a$ and $s$ when refereeing to the o--NH$_2$D energy levels since we only discuss the 
ortho symmetry}.
In particular, in order to reproduce the overall intensities of the 
hyperfine lines, it is necessary to adopt an abundance that would produce a large opacity for this line 
(i.e. $\tau > 10$). Subsequently, since the excitation temperature decreases with radius, this would entail
that the main hyperfine component (i.e. $F$=2--2) would show a self--absorption feature which is not seen
in the observed spectra. 
Consequently, in order to reproduce the observed profiles, it would be necessary to increase the density
to decrease the opacity needed to fit the hyperfine multiplets.  

Since all the other molecules are satisfactorily fitted with the current density profile, we examined in 
detail the impact of the assumption made on the collisional rate coefficients on the 
resulting line intensities. To do so, we used the Pearson's correlation coefficient $r$ defined as : 
\begin{eqnarray}
r = \frac{\sigma_{xy}}{\sigma_x \, \sigma_y}
\end{eqnarray}
where $\sigma_{xy}$ is the covariance between variables $x$ and $y$, defined as :
\begin{eqnarray}
\sigma_{xy} = \frac{1}{N} \sum_{i=1}^N \left(x_i -\bar{x}\right)\left(y_i - \bar{y} \right)
\end{eqnarray}
and $\sigma_x$ and $\sigma_y$ are the standard deviations with, for example :
\begin{eqnarray}
\sigma_{x} = \sqrt{\frac{1}{N} \sum_{i=1}^N \left(x_i -\bar{x}\right)^2}
\end{eqnarray}
In the above expressions, $\bar{x}$ and $\bar{y}$ stand for the mean values of variables $x$ and $y$.
The Pearson's coefficient defines the correlation between two variables. The closest $|r|$ is to 1, the higher
the correlation. When $r = 0$, the two variables are independent.

\begin{figure}
\begin{center}
\includegraphics[angle=270,scale=0.35]{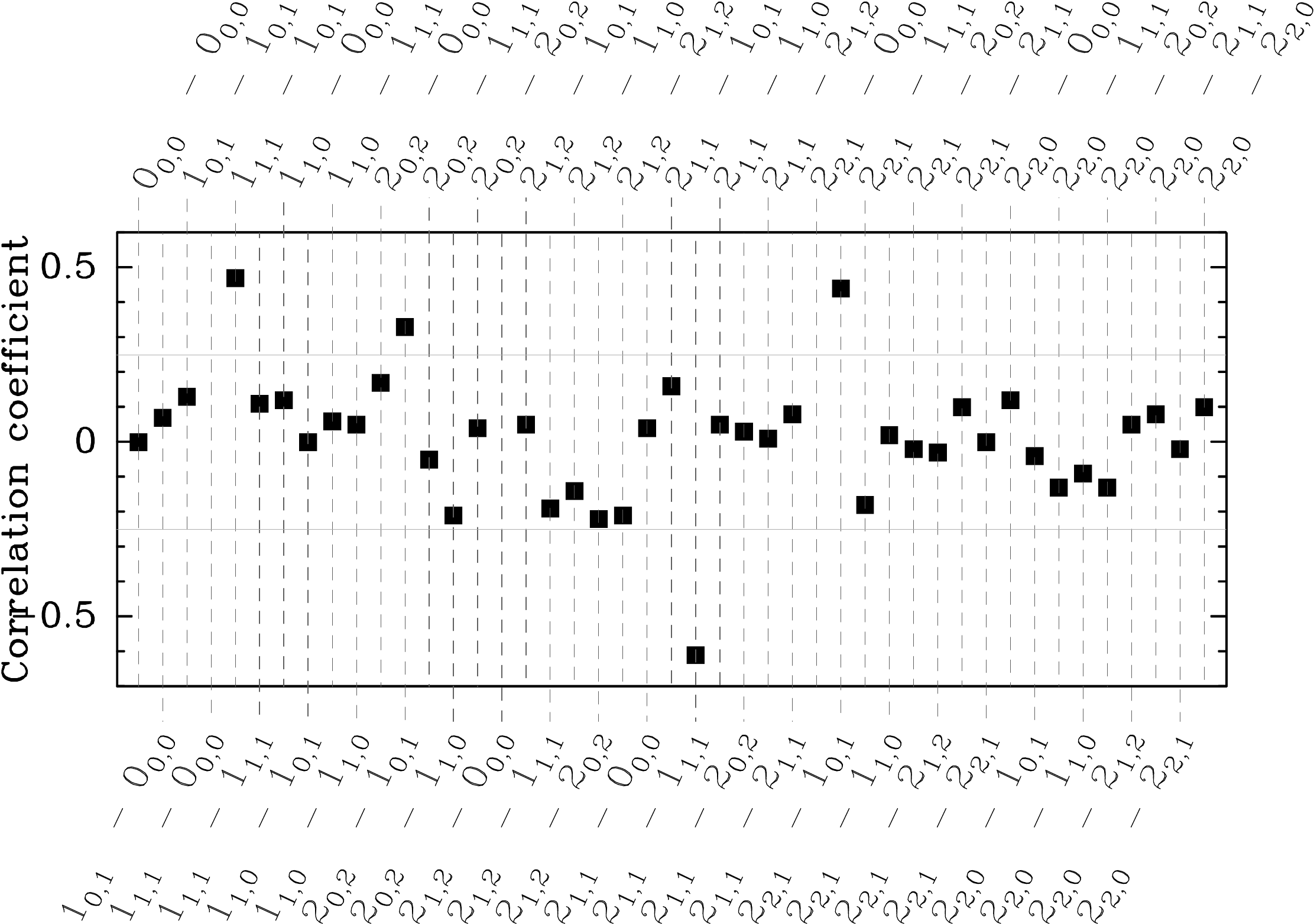}
\caption{Correlation between the magnitude of the individual collisional rate coefficients for o--NH$_2$D with the integrated intensity 
of the $1_{1,1}s--1_{0,1}a$ line. For each transition, the Pearson's correlation coefficient $r$ is given.} \label{fig:correlation} \vspace{-0.1cm}
\end{center}
\end{figure}

As previously said, the ND$_2$H / He and ND$_2$H / H$_2$ rate coefficients differ, on the average, by 
a factor 10. But, when looking closely to the differences, it appears that there is a large scatter in the 
ratio, with values typically in the range 3--30. We thus performed RT calculations where we randomly 
multiplied the NH$_2$D / He rate coefficients by a factor in the range 5--20. The statistics is performed on 
a sample of 100 calculations. In order to assess the influence
of a particular collisional rate coefficient on the resulting line intensity, we used the Pearson's coefficient previously defined 
and we identified $x$ to be the integrated line intensity and $y$ the rate coefficient of a particular transition.
The result of such a calculation is given in Figure \ref{fig:correlation}. 
In this figure, it can be seen that most of 
the transitions have a correlation coefficient such that $|r| < 0.25$. These transitions are thus only weakly correlated 
with the $1_{1,1}-1_{0,1}$ line intensity. Additionally, no transition appear to be strongly correlated with
 the line intensity, i.e. with $|r| \to 1$. This is due to the fact that since NH$_2$D is an asymmetric rotor, 
the pumping scheme is complex, which implies that the line intensity variations are due to various collisional transitions. 
The transitions that show the strongest correlation with the line intensity are the $1_{1,1}-1_{0,1}$, $2_{0,2}-1_{0,1}$,
$2_{1,1}-1_{1,1}$ and $2_{2,1}-1_{0,1}$ transitions. Except for the $2_{1,1}-1_{1,1}$ transition, an increase 
of the rate coefficient is correlated with an increase of the line intensities.
If we define a correlation coefficient that considers the sum of the rate coefficients for these transitions, weighted  
by $r/|r|$, we obtain a correlation coefficient $r = 0.88$. This shows that most of the pumping scheme is related 
to these four transitions. 

\begin{figure}
\begin{center}
\includegraphics[angle=0,scale=0.35]{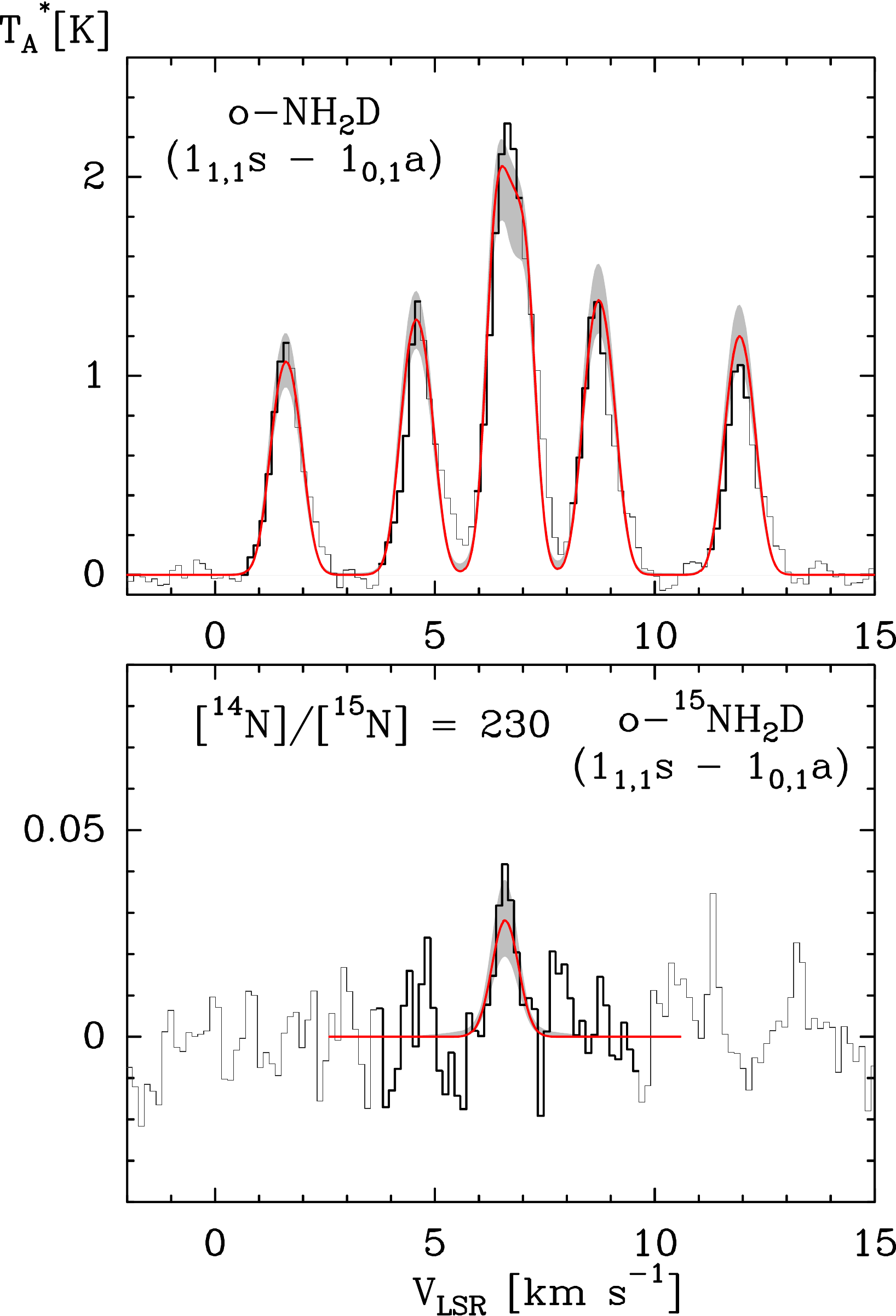}
\caption{Observed (histograms) and modeled (red lines) spectra of o--NH$_2$D ($1_{1,1}s-1_{0,1}a$) (top panel) and 
o--$^{15}$NH$_2$D ($1_{1,1}s-1_{0,1}a$) (bottom panel)} \label{fig:oNH2D} \vspace{-0.1cm}
\end{center}
\end{figure}

By comparing the rate coefficients for ND$_2$H with He and H$_2$, we can see that for the 
$1_{1,1}-1_{0,1}$, $2_{0,2}-1_{0,1}$, $2_{1,1}-1_{1,1}$ and $2_{2,1}-1_{0,1}$ transitions, the ratio
are respectively $\sim$ 52, 12, 10 and 13 at 10K and $\sim$ 35, 10, 10 and 13 at 25K. We thus refine the scaling
of the NH$_2$D / He rate coefficients, still applying an overall factor of 10 to the rates except for the 
$1_{1,1}-1_{0,1}$, $2_{0,2}-1_{0,1}$ and $2_{2,1}-1_{0,1}$ for which we apply factors of 45, 12 and 13, independently
of the temperature. With rate coefficients defined in this way, we are able to reproduce more satisfactorily the 
o--NH$_2$D observations shown in Figure \ref{fig:oNH2D}.

In order to test the influence of the above assumptions on the estimate 
of the NH$_2$D isotopologue column densities, we performed a few test calculations. More precisely, we 
ran models where the collisional rate coefficients
of the various transitions were multiplied by a random factor between 1/1.5 and 1.5. 
From one set to the other, we found that the best models were obtained introducing variations 
in the $\chi$(NH$_2$D) radial distribution. These variations introduce changes in the column 
density estimates at most of the order of 50\%. On the other hand, the changes in the 
column density ratios were found to be modest, 
always less than 10\%, with respect to the value of N(NH$_2$D)/N($^{15}$NH$_2$D) = 230.

\begin{figure*}
\begin{center}
\includegraphics[angle=270,scale=0.7]{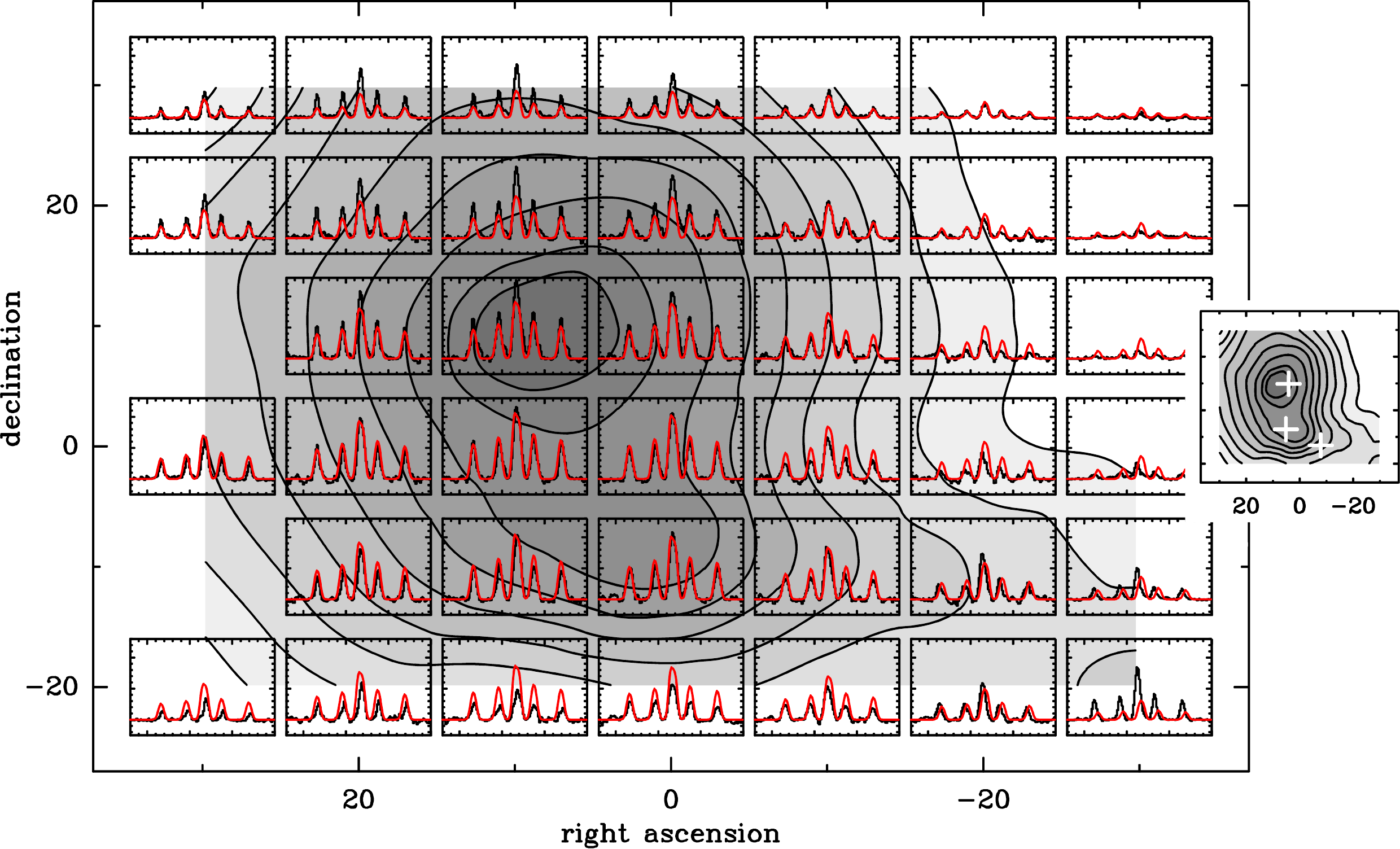}
\caption{Map of the observed (histograms) and modeled (red lines) spectra of o--NH$_2$D ($1_{1,1}s-1_{0,1}a$).
The background correspond to a map of the iscocontours of the intensity integrated of all the hyperfine components.
The map on the right--hand side shows the isocontours with the same scale for the right ascension and declination. The white crosses indicate the position of the B1-bN and B1-bS sources identified by \citet{hirano1999}, as well as the \textit{Spitzer} source reported by \citet{jorgensen2006}.} \label{fig:oNH2D_map} \vspace{-0.1cm}
\end{center}
\end{figure*}

\subsubsection{Results}

The isotopologue abundance profiles derived from the modeling are shown in 
Fig. \ref{profil_abondance-NH3}.
The comparison of the observations with the modeled line profiles are shown in 
Fig. \ref{fig:oNH2D} for both isotopologues
and in Fig. \ref{fig:oNH2D_map} we compare the $1_{11}-1_{01}$ NH$_2$D map
with the model.

\subsubsection{Comparison with previous studies}

The NH$_2$D and $^{15}$NH$_2$D observations considered in this work were previously
analyzed by \citet{gerin2009}. In that study, the column densities for both isotopologues 
were retrieved using LTE, which assumes a uniform distribution for the energy levels
throughout the source and the respective populations of the upper and lower energy levels 
given by the excitation temperature T$_{ex}$.
Since NH$_2$D has hyperfine components resolved in frequency,
it is possible to obtain an estimate of both the excitation temperature and the total opacity of 
the rotational line. These parameters were retrieved using the hyperfine
method of CLASS\footnote{http://www.iram.fr/IRAMFR/GILDAS/}. The values obtained were T$_{ex}$ = 6K and $\tau$ = 5.24
which agree within $\sim$20\% with the averaged values T$_{ex}$ = 6.5K and $\tau$ = 6.4 
derived in the present work. Using, e.g., Eq. (1) of \citet{lis2002},  both sets of values lead to column density
estimates that agree within 15\%.
On the other hand, the hyperfine components of $^{15}$NH$_2$D are not resolved and using this methodology,
it is not possible to estimate both T$_{ex}$ and $\tau$ from the observations. It was thus assumed that
the excitation temperature of $^{15}$NH$_2$D is equal to that of NH$_2$D, which then enables 
to estimate the opacity of the line from the observed spectra. However, within the current
methodology, we obtain an excitation temperature for $^{15}$NH$_2$D which is 30\% lower than that of NH$_2$D,
which is a consequence of larger line trapping effects for the more opaque isotopologue.
This difference in the value of the excitation temperature for $^{15}$NH$_2$D is important when
estimating the line opacity. As an example, using T$_{ex}$ = 5K rather than the value of 6K used in 
\citet{gerin2009}, leads to an increase of the line opacity by a factor of 1.5. This subsequently
leads to reestimate the $^{15}$NH$_2$D column density which is increased by a factor of 2. 
By examining Eq. (1) of \citet{lis2002}, we can see that the strong dependence of the column density 
on the excitation temperature comes from the term $\textrm{exp} (E_u/ k_b \, T_{ex})$, and is due to the fact 
that the excitation temperature of the $1_{11}--1_{01}$ transition
is low by comparison with the energy of the upper level ($E_u \sim 20.7$ K). This strong dependence of the 
$^{15}$NH$_2$D column density on the excitation temperature explains why the $^{14}$N/$^{15}$N column density
ratio was estimated to be $470^{+170}_{-100}$ in \citet{gerin2009}, while the current estimate is $230^{+105}_{-55}$. 

The o--NH$_2$D map in Fig. \ref{fig:oNH2D_map} shows the peak intensity close to 
the B1--bN core. The emission is elongated towards the position of B1--bS, but the line 
intensities are lower towards this position. It was found by \citet{huang2013}, while discussing the 
deuterium fractionation of N$_2$H$^+$, that the deuterium enrichment is a factor of two higher towards 
the B1--bN position compared to B1--bS. The current NH$_2$D observations suggest a similar  
trend for this species. The fit of the SED performed by \citet{hirano2013} suggests that B1--bS has a 
temperature higher by a few Kelvins compared to B1--bN, with a temperature close to 20K for the former
source. The lower level of deuterium enrichment towards B1--bS could thus be related to the fact that 
the temperature of the dust is close to the temperature at which CO desorbs. A higher amount of CO
in the gas phase would drastically reduce the amount deuterated isotopologues of H$_3^+$ and would thus 
limit the deuteration fraction. 

\subsection{CN}

\subsubsection{Spectroscopy and collisional rate coefficients}

Since CN is an open shell molecule, its rotational energy levels are split into fine structure levels due to the interaction with the 
electron spin. These levels are further split because of the interaction with the nuclear spin of the nitrogen atom. 
In the case of the $^{13}$CN and C$^{15}$N isotopologues, the 1/2 spins of the $^{13}$C and $^{15}$N nuclei 
lead to magnetic coupling, which have associated coupling constants sufficiently large, so that the degeneracy 
break can be spectroscopically resolved.

For CN, we have used the CN / He collisional rate coefficients of \citet{lique2011}. For either  
$^{13}$CN or C$^{15}$N, the energy pattern is modified compared to CN, because of the large magnetic coupling
constants associated with the $^{13}$C or $^{15}$N nuclei. Since all these molecules have 
specific energy structures, it is not possible to directly use the CN / He  collisional rate coefficients to obtain the 
$^{13}$CN or C$^{15}$N rates. Hence, in the present study, we derived the rate coefficients for the rarest 
isotopologues from the diffusion matrix elements obtained 
for the CN / He system but accounting for their specificities, while performing 
the recoupling of the matrix elements. The description is given
in the Appendix and these rates will be made available through the BASECOL 
database \citep{dubernet2012}.

In the case of CN, rate coefficients for the CN / p--H$_2$ system have recently been calculated by \citet{kalugina2012}.
They show that individual state--to--state rate coefficients with He or H$_2$ can differ by up to a factor 3.
However, these differences mainly affect the rates of lowest magnitude, while the highest ones
scale accordingly to the factor expected from the differences in the reduced masses of the two collisional systems.
For consistency between the various isotopologues, we used the CN / He collisional rate 
coefficients of \citet{lique2011} for the main isotopologue.

\subsubsection{Results}

\begin{figure*}
\begin{center}
  \subfigure[]{\includegraphics[angle=0,scale=0.35]{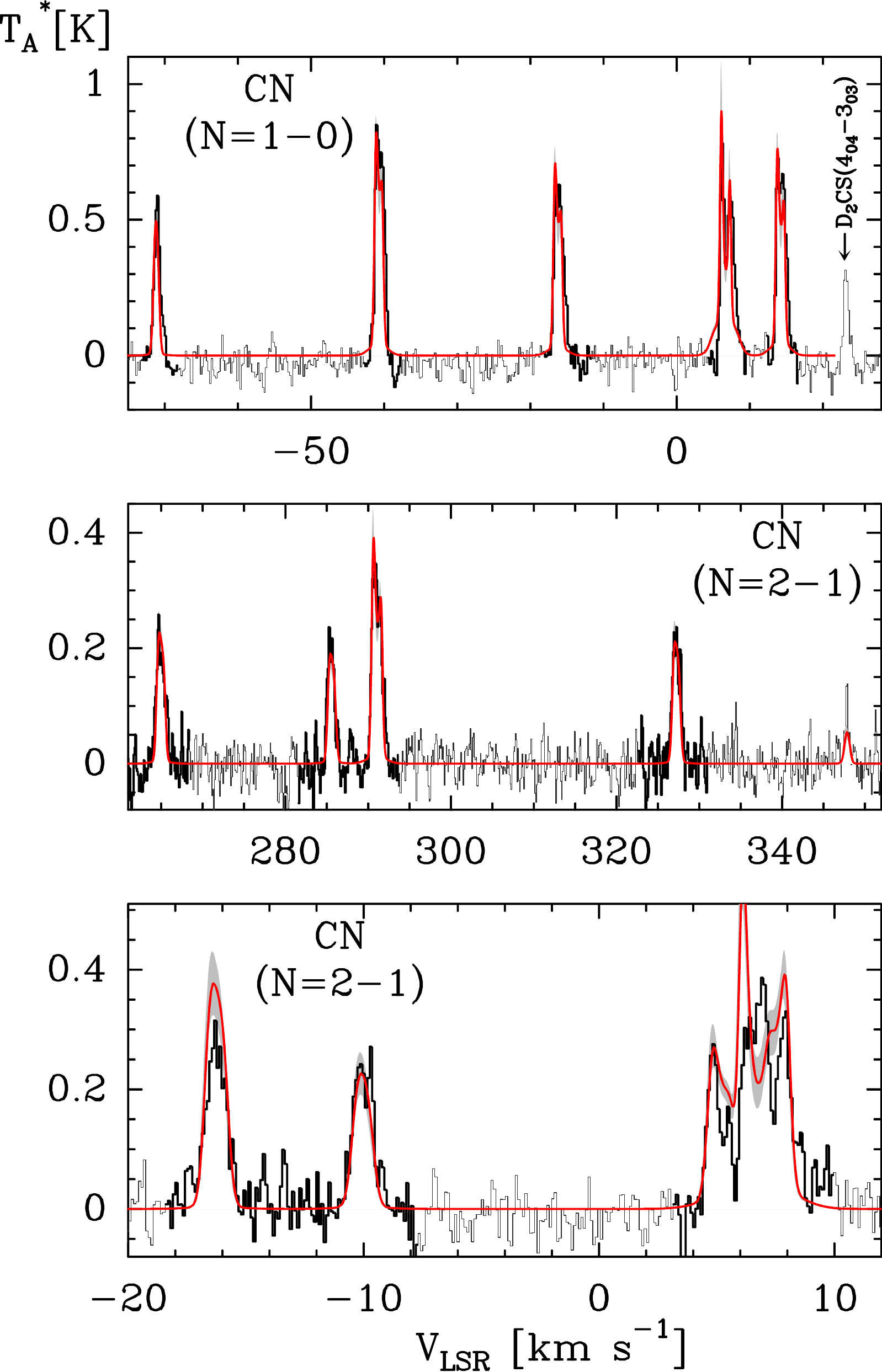}} 
  \subfigure[]{\includegraphics[angle=0,scale=0.35]{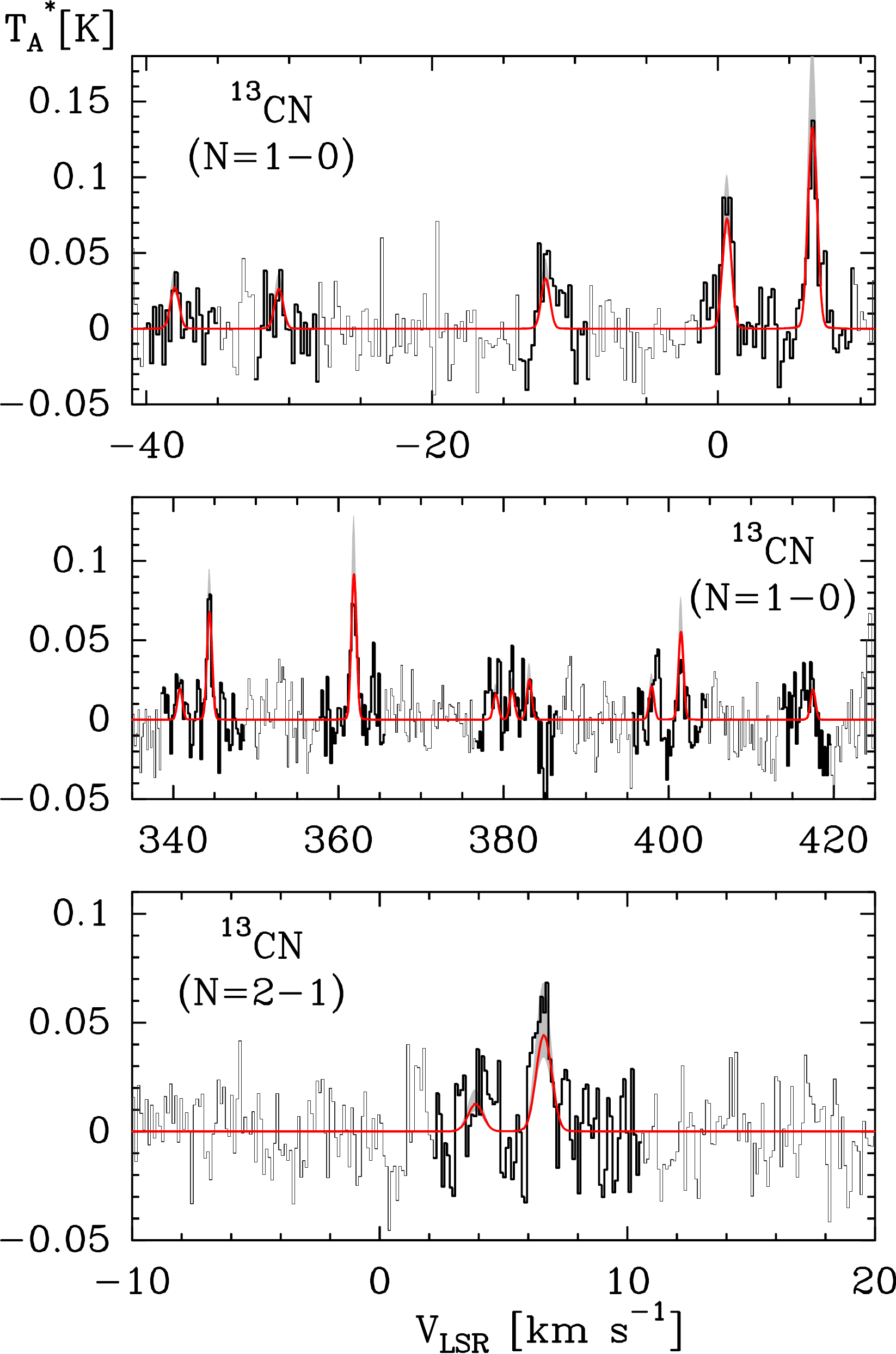}} 
  \subfigure[]{\includegraphics[angle=0,scale=0.35]{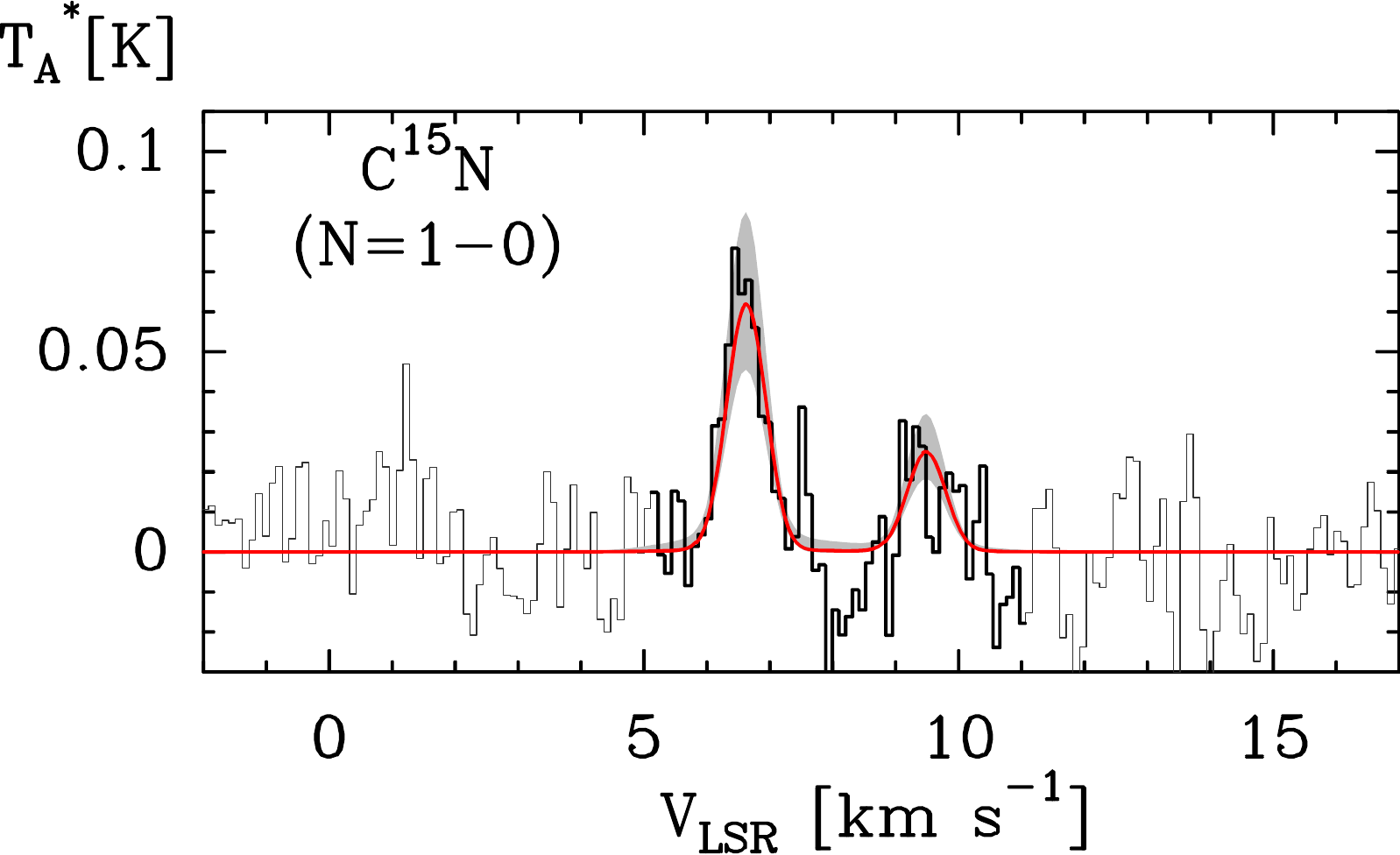}} 
\end{center}
\caption{
Observed (histograms) and modeled (red lines) spectra for 
\textbf{(a)} CN ($N$=1--0) (top panel), and CN ($N$=2--1) (middle and bottom panel)
\textbf{(b)} $^{13}$CN ($N$=1--0) (top and middle panel), and $^{13}$CN ($N$=2--1) (bottom panel)
\textbf{(c)} C$^{15}$N ($N$=1--0)
.}
\label{fig:CN}
\end{figure*}

The various isotopologue abundance profiles derived from the modeling are shown in Fig. \ref{profil_abondance-CN}.
In Fig. \ref{fig:CN} we compare the model with the observations of the $N$=1--0 and $N$=2--1 
spectra of CN and $^{13}$CN and of the $N$=1--0 transition of C$^{15}$N. Note that in the $N$=1--0 CN spectrum,
a transition due to the doubly deuterated thioformaldehyde isotopologue is seen at a 
velocity offset $v\sim$22 km s$^{-1}$ \citep{marcelino2005}. 
The column densities for each isotopologue, 
as well as the column density ratios, are given in Table \ref{table-lineparameters}.
In this study, we report the first observations made for the various CN isotopologues toward B1b.

\section{Discussion}\label{discussion}

\begin{figure}
\begin{center}
\includegraphics[angle=270,scale=0.35]{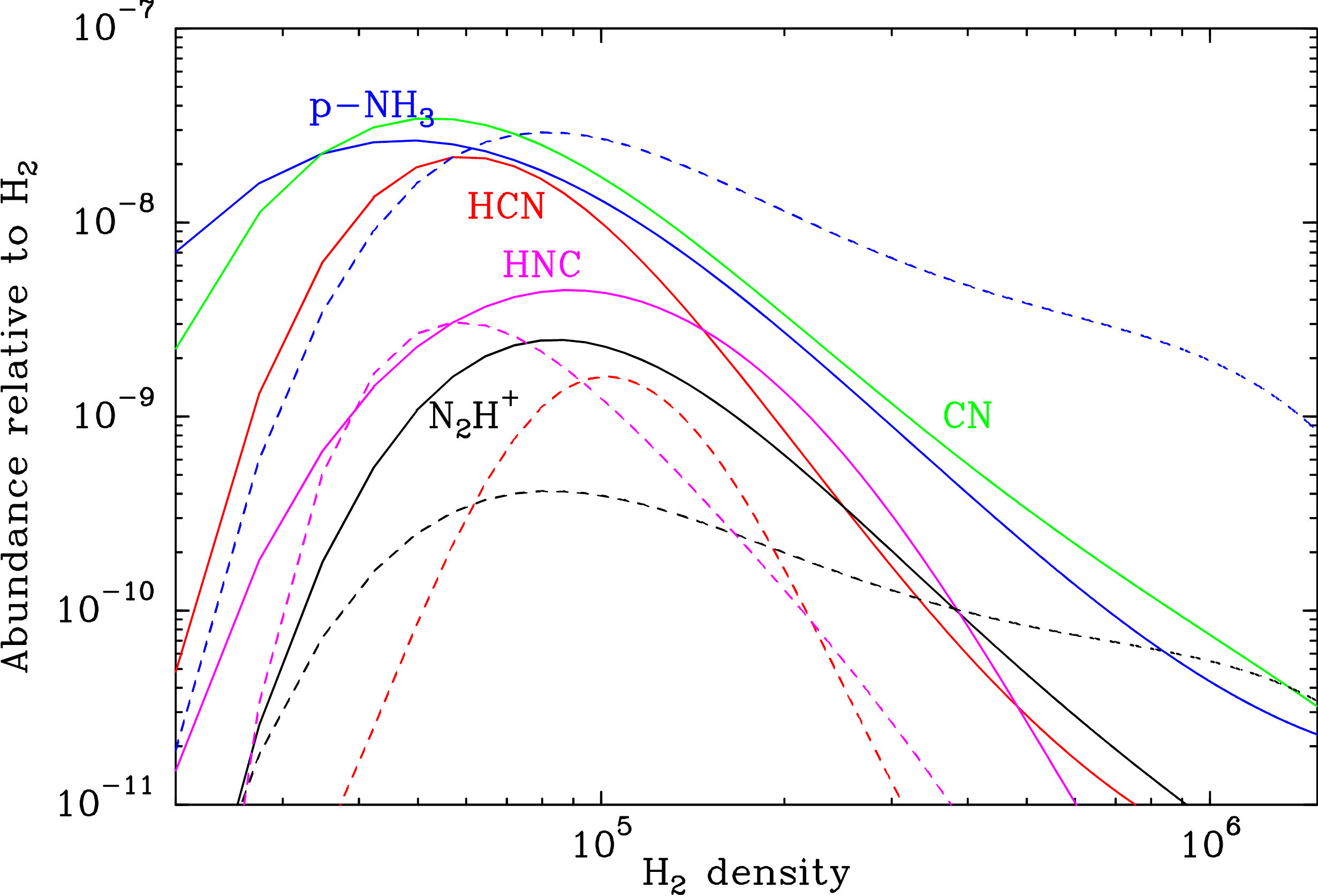}
\includegraphics[angle=270,scale=0.35]{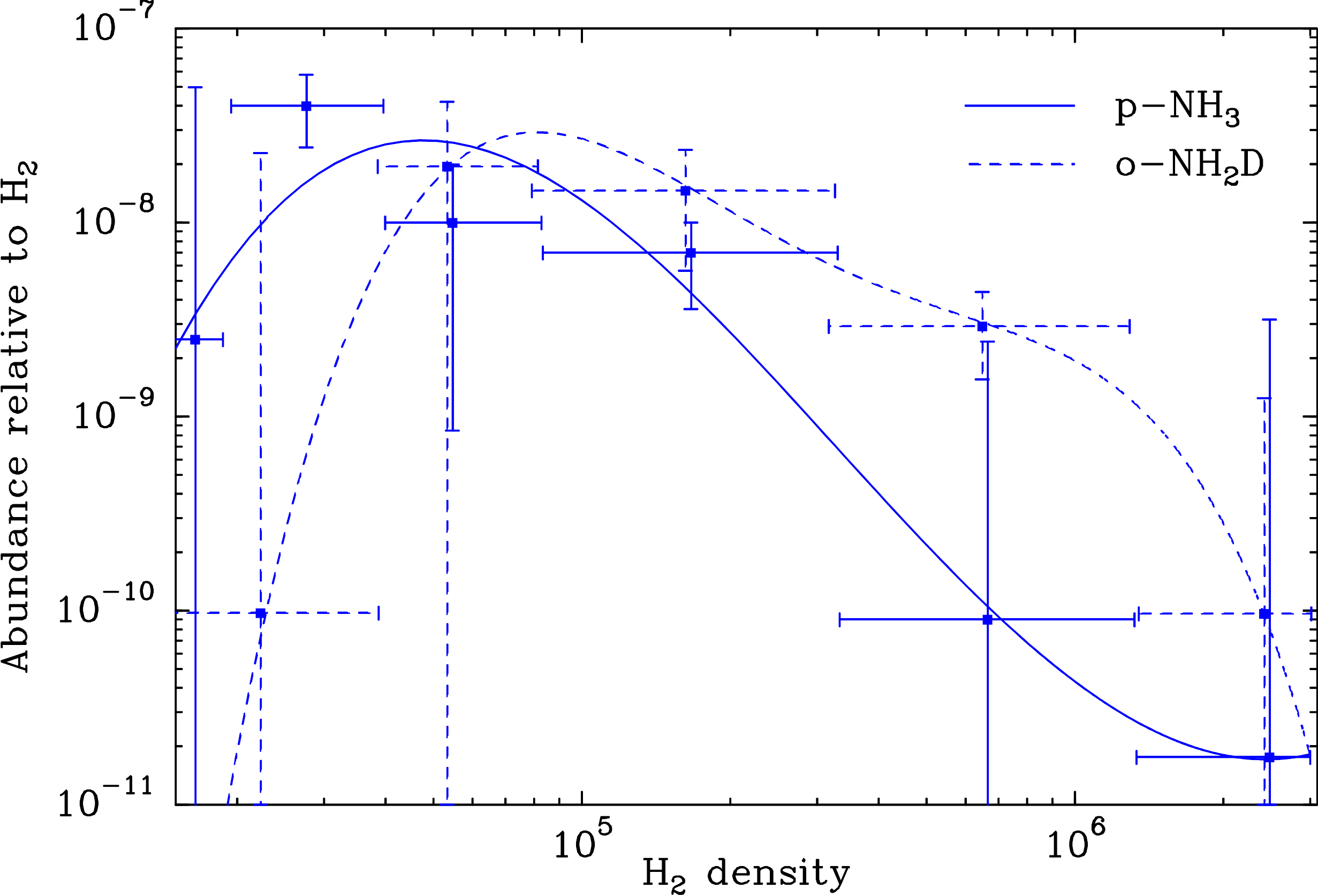}
\includegraphics[angle=270,scale=0.35]{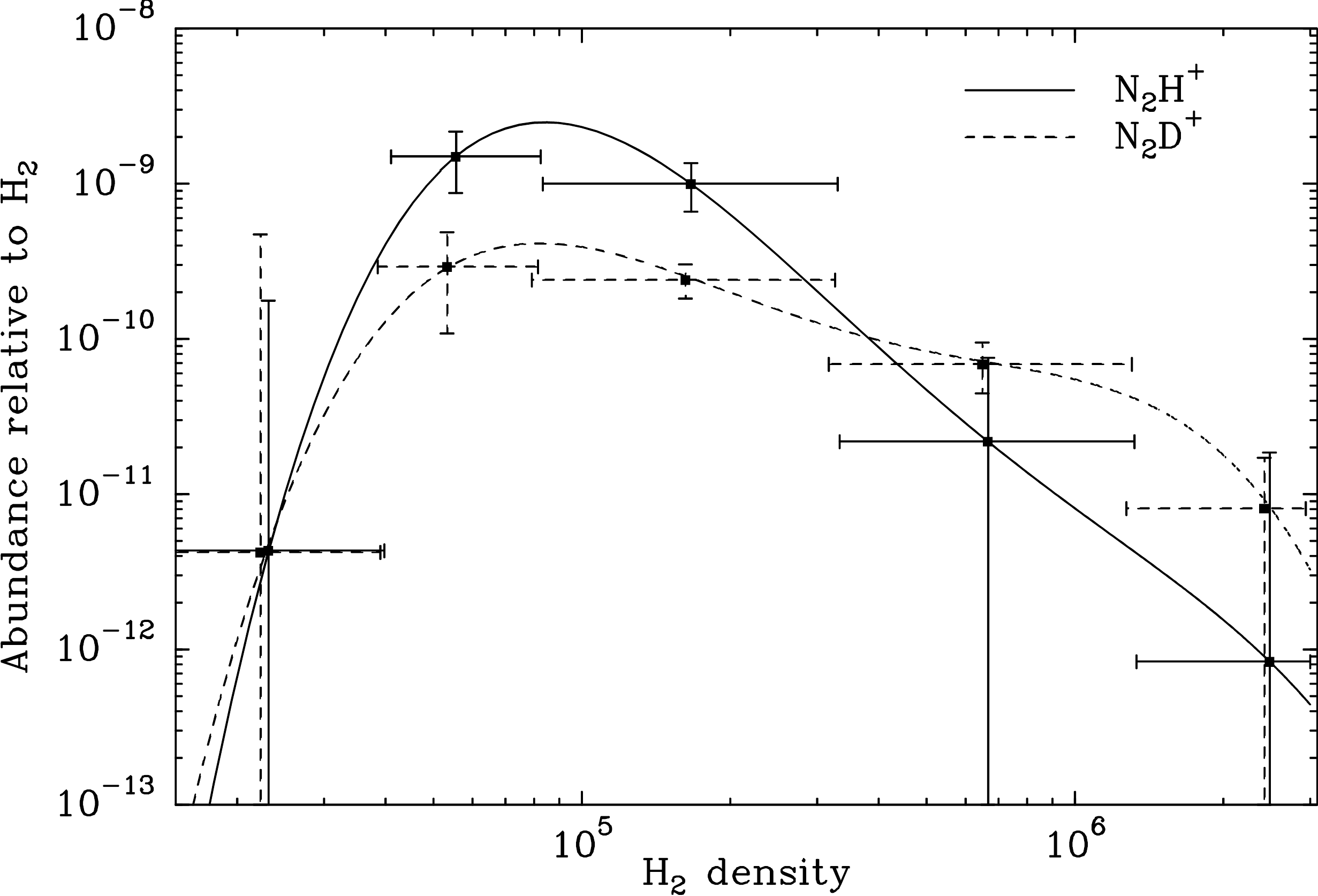}
\caption{Abundances of the various molecules observed in this study as a function of the H$_2$ density (upper panel).
The full lines show the abundances of the main species and the dashed lines the abundances of the deuterated isotopologues.
The abundances are obtained by fitting the step functions used in the RT modelling. The central and bottom
panels give the initial points used in the fit as well as the error bars for NH$_3$ and N$_2$H$^+$.
 \label{fig:abondances}}
\end{center}
\end{figure}

In the upper panel of Fig. \ref{fig:abondances}, we report the abundance profiles of the main isotopologues as a function
of the H$_2$ density. These abundances are obtained from a polynomial fit to the discrete values reported
in Fig. \ref{profil_abondance-N2H+}-\ref{profil_abondance-NH3},
where all the regions of the model, except the region with $r >  120''$ are considered. In Figures
\ref{profil_abondance-N2H+} to \ref{profil_abondance-NH3}, it can be seen that the molecular abundances 
have large errorbars when the abundance is much
below the maximum abundance found for a given molecule. This means that in Fig. \ref{fig:abondances}, we expect 
the abundances to be accurate within a factor of a few around $10^5$ cm$^{-3}$, while around $10^4$ cm$^{-3}$
and $10^6$ cm$^{-3}$, the error will be typically of an order of magnitude. 
This can be seen in the middle and bottom panel of Fig. \ref{fig:abondances} 
where the abundances used in the fit, as well as their estimated error bars, are reported for the 
hydrogenated and deuterated isotopologues of NH$_3$ and N$_2$H$^+$. 
The abundances reported in this figure
can be directly compared to the prediction of the gas--phase chemical model in Fig. 2 of \citet{gerin2009}. 
While the molecular abundances inferred from the observations agree with the predictions of this chemical 
model at low densities,
the current modeling shows that all the molecules are expected to suffer depletion at densities above 
$10^{5}$ cm$^{-3}$ and are then considerably lower than predicted by the gas--phase chemistry.

\subsection{Molecular depletion and consequences on the N$_2$H$^+$ and NH$_3$ excitation}

\citet{johnstone2010} re--analyzed the N$_2$H$^+$  and 850 $\mu$m continuum observations 
of \citet{kirk2007} by combining them with the p--NH$_3$ observations of \citet{rosolowsky2008},
for a number of prestellar and protostellar objects in the Perseus molecular cloud. They found that 
the column density ratio of these two species is relatively constant regardless of the evolutionnary 
stage of the object, with N(p--NH$_3$)/N(N$_2$H$^+$) = 22 $\pm$ 10. Our results for the B1b region, i.e. 
N(p--NH$_3$)/N(N$_2$H$^+$) = 18, are thus in good agreement with the earlier results. Additionally, 
\citet{johnstone2010} found that
the p--NH$_3$ abundance is rather constant in all the prestellar cores, i.e. $\chi$(p--NH$_3$) $\sim$ 10$^{-8}$, 
while there is a larger scatter for the protostellar objects, where it ranges from 
10$^{-8}$ at low H$_2$ column densities and decreases by an order of magnitude at larger 
column densities.

\citet{johnstone2010} also derived a tight correlation between the excitation temperatures of the N$_2$H$^+$ ($J$=1--0) 
and p--NH$_3$ (1,1) transitions, with a tendency for the p--NH$_3$ excitation temperature to be higher
by 1K. Their sample is consistent with constant excitation temperatures, independently of the central density
of the objects. Again, our current results for the B1b region, reported in Table \ref{table-lineparameters}, 
i.e. T$_{ex}$(N$_2$H$^+$) $\sim$ 6.8 K and T$_{ex}$(p--NH3) $\sim$ 8.1 K,
confirm the tendency outlined by \citet{johnstone2010}. As pointed
out by these authors, the similarity of the excitation temperatures for these two species is puzzling when
considering the critical densities of the respective transitions. More precisely, the excitation temperatures 
are below the estimated kinetic temperatures of the studied 
regions which indicates subthermal excitation. Indeed, \citet{johnstone2010} found that the dust 
emission at 850 $\mu$m can be reproduced 
assuming T$_d$ $\sim$ 11K. For densities higher than $10^4$ cm$^{-3}$, the kinetic temperature is 
believed to be well coupled to the dust temperature \citep{goldsmith2001}.
However, considering the differences in the 
critical densities, the p--NH$_3$
excitation temperature should be higher than that of N$_2$H$^+$, if the molecules trace the same volume
of gas. As pointed out by these authors, similar excitation temperatures would require densities large enough
for the lines to be thermalized. Hence, they stress that the excitation temperatures for both species may have been
underestimated. Our current results shed some light on the puzzling behaviour reported by \citet{johnstone2010}. 
Indeed, considering the spatial distribution of $\chi$(p--NH$_3$) and $\chi$(N$_2$H$^+$) shown in 
Fig. \ref{profil_abondance-N2H+} and \ref{profil_abondance-NH3}, we 
see that N$_2$H$^+$ has an abundance distribution more centrally peaked than p--NH$_3$. Since the 
innermost region of the cloud is warmer and denser, this enhances the mean excitation temperature of N$_2$H$^+$
with respect to that of p--NH$_3$, thus reducing the impact of the differing critical densities. Moreover, our models
show that both molecules are removed from the gas phase in the innermost part of the cloud. If we consider that
both molecules suffer depletion above a given density, it entails that the properties derived from
both molecules will characterize the gas which is below this density threshold. 
 On the other hand, the dust will still trace the total 
volume of gas. Thus, if depletion affects both p--NH$_3$ and N$_2$H$^+$, it can explain why \citet{johnstone2010}
conclude that the p--NH$_3$ and N$_2$H$^+$ excitation temperatures are independent of the central 
H$_2$ density.

\citet{tobin2011} presented interferometric observations of NH$_3$ and N$_2$H$^+$ in a sample of class 0 and class I protostars.
In particular, they observed that the position of the maximum intensity in these two tracers was offset from the position
of the continuum peak, hence from the position of the protostar. Apart from depletion, an alternative way to explain the 
disappearance of these molecules from the gas phase can be chemical destruction. 
Above 20K, CO will be released from the ice
mantles and will react with N$_2$H$^+$ to form HCO$^+$. Hence, this reaction will efficiently destroy N$_2$H$^+$.
Such a process has been invoked by \citet{tobin2013} in order to explain the central hole observed for both the N$_2$H$^+$ and
N$_2$D$^+$ abundances in the protostellar object L1157. Additionally, in this object, N$_2$H$^+$ is found to be more centrally
peaked than N$_2$D$^+$, which might be a consequence of variations in the H$_2$ ortho--to--para ratio.
In the case
of NH$_3$, an efficient destruction can come from the reaction with HCO$^+$ 
\citep{woon2009}. Thus, chemical destruction could explain
the shape of the abundance profiles we obtain for NH$_3$ and N$_2$H$^+$ in the B1b region. 
Two arguments can be invoked against such an possibility. First, the 
observations of  H$^{13}$CO$^+$ performed by \citet{hirano1999} showed that at the positions of B1-bS and B1-bN, 
the abundance of this molecule was reduced by a factor of the order $\sim$ 3-5 with respect to the H$^{13}$CO$^+$ abundance 
of the surrounding envelope. Second, the dust temperature obtained from the SED fitting is found to be below 20K.
Hence, the hypothesis of depletion seems to be favored in order to explain the central decrease in the N$_2$H$^+$ and NH$_3$
abundances.

The fact that NH$_3$ is strongly depleted onto dust grains in the innermost region of B1b is strengthened by the 
analysis of the NH$_3$ ice features. At the position of the \textit{Spitzer} source which is 
offset by $\sim$15$\arcsec$ from the central position of our model, 
\citet{bottinelli2010} inferred a total column density of NH$_3$ locked in the ice of $\sim$$7 \, 10^{17}$  cm$^{-2}$, while 
we derive a column density N(p--NH$_3$) $\sim$$5.5 \, 10^{14}$ cm$^{-2}$ in the gas phase, at the central position 
of our observations.
Converting this column density to the total amount of NH$_3$ requires
knowledge of the respective abundances of ortho-- and para--NH$_3$. It is expected that if NH$_3$ is formed in the gas phase, the ratio
will be of the order of unity. On the other hand, a formation that would occur on 
dust grains would raise the ortho--to--para ratio above unity \citep{umemoto1999}.
If we assume a similar amount of ortho and para--NH$_3$, this leads to a total gas phase column density of 
N(NH$_3$) $\sim 1 \, 10^{15}$ cm$^{-2}$. Since the observations of \citet{bottinelli2010} are offset with respect to our core center,
we can infer that the gas phase NH$_3$ abundance is a factor $>$$700$ lower than inferred from the ice feature observations.

In principle, the comparison of the respective abundances in the gas and solid phases should shed light on the evolutionary stage 
of the source. Indeed, it is predicted
theoretically that the relative abundance of NH$_3$ in both phases will vary with time during the 
collapse \citep[e.g.][]{rodgers2003,lee2004,aikawa2008,aikawa2012}. In particular,
after the formation of a protostar, the NH$_3$ molecules locked in the ices will be released into the gas phase, 
because of the warm--up, and the ratio of the NH$_3$ gas--to--dust abundances will increase with time.
However, in the present case, this point can only be discussed qualitatively, because of the simplicity of the geometry
we assumed to describe the B1b region. We consider a 1D--spherical model while two objects, B1--bS and B1-bN are
present and separated by a projected distance of only $\sim$20$\arcsec$. Moreover, if we consider 
e.g. Fig. 2 of \citet{aikawa2012}, we see that $\sim$500 years after the formation of a protostar, the increase of the 
NH$_3$ abundance in the gas 
phase occurs at the 10 AU scale (i.e. $\sim$0.04$\arcsec$ at 235 pc), and at the 100 AU scale at $t = 10^4$ years. 
This makes the FHSC stage ($t$ = -560 yr in \citet{aikawa2012} where $t = 0$ corresponds to the formation of the 
second hydrostatic core, i.e. the protostar) hardly distinguishable from the stage that harbours a newly born protostar, 
at least in single dish observations. 
Thus, the abundance profile we derived for the gas--phase p--NH$_3$ abundance would be consistent with the predictions
of \citet{aikawa2012} from $t = -560$ yr to $t = 10^4$ yr and it is thus impossible to infer from these observations the 
evolutionary stage of the two sources in the B1b region, i.e. to disentangle the FHSC stage from the 
class 0 stage.

\subsection{Deuteration}

The abundance profiles reported in Fig. \ref{fig:abondances} show that at the innermost radii, the behaviour of the 
p--NH$_3$ and 
o--NH$_2$D abundances changes, with a ratio $\chi$(o--NH$_2$D) / $\chi$(p--NH$_3$) below unity for densities lower than
$\sim 6 \, 10^4$ cm$^{-3}$, and above unity for higher densities. We found that this enhancement is higher than an 
order of magnitude at the core center. Hence, this behaviour will still be true if we consider the total column densities of 
NH$_3$ and NH$_2$D and if we consider standard ratios to relate the abundances of the ortho and para species.
Fig. 3 of \citet{aikawa2012} shows that
a ratio $\chi$(NH$_2$D)/$\chi$(NH$_3$) $> 1$ is expected at small radii, but only at an evolutionary stage close to the FHSC stage. At times
greater than 500 yrs after the formation of the first hydrostatic core, the NH$_3$ abundance always exceeds the NH$_2$D abundance throughout the envelope. This is due to the fact that when the gas temperature becomes high enough, 
the backward reaction that leads to the deuterated isotopologues of H$_3^+$ becomes efficient enough to inhibit the deuteration
of this ion.
Finally, the NH$_2$D map in Fig. \ref{fig:oNH2D_map} shows that the NH$_2$D peak 
is close to the position of the B1--bN core. This would suggest that this core is closer to the FHSC stage than B1--bS.

\subsection{Nitrogen fractionation}

To date, there have been few studies estimating the abundance ratio
of $^{14}$N and $^{15}$N isotopologues in dark clouds. 
In the source L1544, the $^{14}$N/$^{15}$N ratio was estimated
to be $1000\pm200$ by \citet{bizzocchi2010,bizzocchi2013} for the N$_2$H$^+$ isotopologues. Toward the same
source, \citet{hily-blant2013} estimated $140 < $ $^{14}$N/$^{15}$N $< 360$ for the HCN molecule, using a double--isotope 
analysis of H$^{13}$CN and HC$^{15}$N. In the same study, a similar analysis was also performed 
towards L183 leading to $140 <$ $^{14}$N/$^{15}$N $< 250$. \citet{ikeda2002} observed HCN and CO isotopologues in a sample
of three dark clouds in the Taurus molecular cloud. They derived values of $151\pm16$ and $>813$ for L1521E and L1498 respectively
using the $^{12}$C/$^{13}$C ratio derived from the analysis of the CO isotopologues. In the case of L1498, this ratio was 
determined to be $^{12}$C/$^{13}$C$ >117$ 
from the analysis of $^{13}$C$^{18}$O and  $^{12}$C$^{18}$O lines, a value significantly higher than the standard value in
the local ISM. Assuming a ratio of 60 rather than 117 would lead to HCN / HC$^{15}$N = $472 \pm 55$.
Moreover, both \citet{ikeda2002} and \citet{hily-blant2013} found that the H$^{13}$CN/HC$^{15}$N abundance 
ratio varies as a function position. As pointed out by \citet{ikeda2002}, 
this result can be interpreted either as due to an incomplete mixing
of the parental gas, or variations in the chemistry of nitrogen fractionation within the clouds. 

The NH$_3$ and NH$_2$D data considered in the present study were already analyzed by \citet{lis2010} and
\citet{gerin2009}, using a LTE approach. In the case of NH$_3$, the isotopologue abundance ratio
was found to be $334 \pm 50$, in good agreement with the present determination of $300^{+55}_{-40}$. However, in the case of NH$_2$D, 
we obtain a value of $230^{+105}_{-55}$ compared to the earlier estimate of 
$470^{+170}_{-100}$. As explained in Sect. \ref{nh2d}, 
this comes from the fact that the assumption of equal excitation temperatures for NH$_2$D
and $^{15}$NH$_2$D made in \citet{gerin2009} is not totally valid.
Moreover, we found that variations 
as small as $\sim$20\% in the excitation temperatures lead a factor 2 variations in the column density ratio.
Apart from B1b, \citet{lis2010} and \citet{gerin2009} reported column density ratio 
estimates in other dark clouds. A value of $344\pm173$ was obtained for NH$_3$ in NGC1333 and for NH$_2$D, values
in the range 350 $<$ $^{14}$N/$^{15}$N $<$ 850 were obtained in a sample of four clouds. In the latter case, as for B1b, the 
column density ratio may have been overestimated by a factor $\sim$2.

\citet{tennekes2006} reported observations of various HCN and HNC isotopologues towards Cha--MMS1. They analyzed
the emission from these molecules both with a non--LTE non--local radiative transfer modelling and with the LTE approach. In their 
non--LTE analysis, they fixed the isotopologue abundances assuming $^{12}$C/$^{13}$C = 20 and $^{14}$N/$^{15}$N = 280,
 on the base of their LTE analysis. 
At that time, it was not yet recognized that the HCN and HNC collisional rate coefficients could differ by a significant factor. 
Indeed, in their LTE analysis, they assumed a single excitation temperature for 
all the HCN and HNC isotopologues. However, due to differing HCN and HNC rate coefficients and because of line trapping effects
that affect the main isotopologues, this assumption is not valid as can be seen in Table \ref{table-lineparameters}.

An estimate of the $^{14}$N/$^{15}$N galactic gradient was reported by \citet{dahmen1995} from HCN observations 
and \citet{adande2012} from CN and HNC observations. The first study used a double isotope 
method with a $^{12}$C/$^{13}$C ratio based
on H$_2$CO observations. However, H$_2$CO might undergo carbon fractionation effects and the results from \citet{dahmen1995} were
then reanalyzed by \citet{adande2012} assuming a different $^{12}$C/$^{13}$C ratio as derived from the CN observations of \citet{milam2005}.
From both HCN, HNC and CN, they obtain a value for the local ISM of $^{14}$N/$^{15}$N = $290\pm40$. 
The sample of clouds used in these
studies have typical kinetic temperatures in the range 25K $<$ T$_K$ $<$ 90K \citep{adande2012}. At such high kinetic 
temperatures, fractionation effects for the nitrogen are unlikely to occur and the value quoted should thus be representative 
of the elemental atomic abundance ratio. This estimate is in good agreement with the estimate of 
$^{14}$N/$^{15}$N  = 237$_{-21}^{+27}$ obtained
toward diffuse clouds of the local ISM observed in absorption toward compact extragalactic sources \citep{lucas1998}.

\begin{figure}
\begin{center}
  \includegraphics[angle=270,scale=0.35]{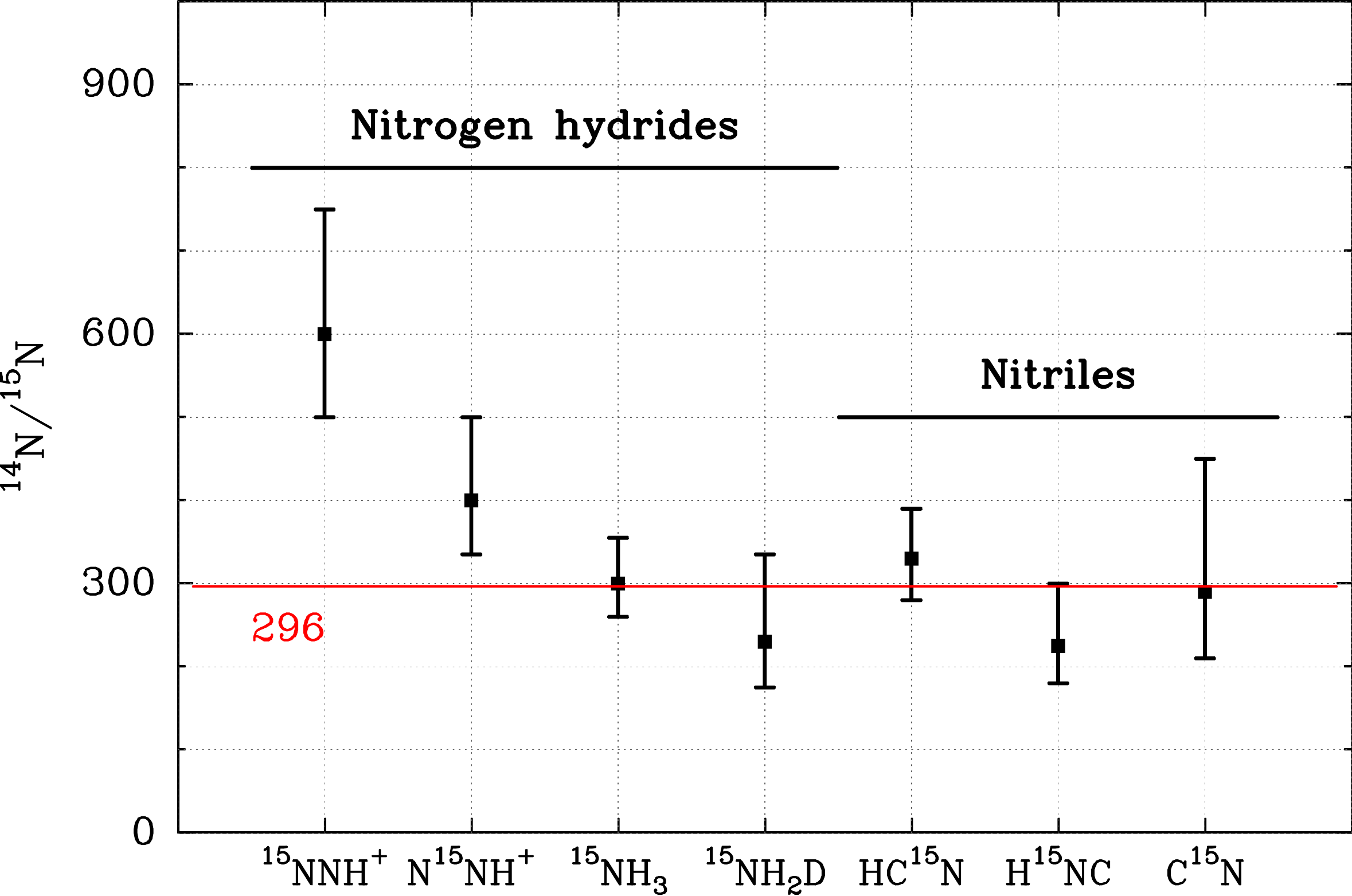}
\end{center}
\caption{$^{14}$N/$^{15}$N isotopologue abundance ratio for the molecules of the nitrogen hydride and nitrile families.}
\label{ratios-nitrogen}
\end{figure}

The values obtained in the current study for the column density ratios are summarized in Fig. \ref{ratios-nitrogen}. In this
figure, we also report the mean ratio ($\sim$296) computed from all the values except that 
of $^{15}$NNH$^+$, for which we only obtained a lower limit. In the figure, for this molecule, we indicate this lower limit as well
as approximate error bars which are obtained by scaling that of N$^{15}$NH$^+$. The values reported for the nitriles
come from the double isotope method rather than from a direct estimate (see Sect. \ref{hcn}). 
By examining Fig. \ref{ratios-nitrogen}, it can be seen that given the errorbars, the column density ratios are consistent  
with a constant value. The only exception is N$_2$H$^+$, and particularly $^{15}$NNH$^+$, for which a higher
 $^{14}$N/$^{15}$N ratio is obtained. The current result support the conclusion that no fractionation
affects the nitrogen chemistry in the B1b region.  
The mean value we obtain for B1b is in good agreement with the estimate of $290\pm40$ \citep{adande2012} obtained toward
warmer sources in the local ISM. This latter value could be representative of the gas associated with the Perseus 
molecular cloud. But, it was also recognized \citep{ikeda2002,adande2012} that the elemental 
$^{14}$N/$^{15}$N atomic ratio can vary from source to source. The absence of large differences between the nitriles 
and the nitrogen hydrides obtained in the current study suggest that no fractionation effects occur in B1b 
and a ratio $^{14}$N/$^{15}$N $\sim$ 300 should thus be representative of this region.

However, two things must be kept in mind while considering this conclusion. The first point is 
that the N$_2$H$^+$ column density ratio show a tendency towards higher values, in particular for 
$^{15}$NNH$^+$, for which we derived a lower limit N(N$_2$H$^+$) / N($^{15}$NNH$^+$)  $>$ 600.
The second point concerns the nitriles. For this family, we derived the main isotopologue column densities
from the $^{13}$C isotopologues assuming a ratio $^{12}$C/$^{13}$C = 60. This value is representative
of the local ISM, but as for the nitrogen case, can be affected by source--to--source variations.

The astrochemical modelling by \citet{wirstrom2012} suggests that the nitrogen hydride and nitrile families should show
different degrees of fractionation. Such a finding is supported by the results of \citet{ikeda2002} and \citet{hily-blant2013} where 
low $^{14}$N/$^{15}$N ratios were derived for HCN. Such differential fractionation effects 
are appealing since they would enable to reconcile some of the observations made in the solar system. However, in the current
study, no differential fractionation is found between the nitrogen hydride and nitrile families. Regarding the absence of 
fractionation in B1b, a possible explanation may be that this region contains two cores in 
a late evolutionary stage. The gas temperature in the core center is estimated to be $\sim$17K, somewhat higher 
than the lowest temperatures $\sim$7K that can be reached during the prestellar phase. 
Similarly to our definition of the mean excitation temperature given in Sect. \ref{modelling:molecules}, we can define a mean gas temperature,
averaged over all the molecules considered in this work, that describes the mean temperature of the gas
from which the emission originates. We find that $\overline{T}_K = 13.7^{+1.9}_{-2.6}$ K.
At such high temperatures, 
the nitrogen fractionation might already be ineffective. 
Indeed, as shown by \citet{rodgers2008}, decreasing the gas temperature from 10K to 7K
can lead to an increase by a factor $\sim$3 in the $^{15}$N--enrichment for the NH$_3$ and N$_2$ molecules. This effect
is partly due to the low exothermicity of the exchange reactions that involve the $^{15}$N$^+$ ions, which are
typically in the range $\Delta E = 20-30$K. As a consequence, small variations of the temperature around the canonical
value of 10K, commonly used to describe prestellar cores, could result in substantial variations in $^{15}$N--fractionation
effects, because of the variations of the backward reaction rates. 
However, to date, a quantitative estimate of the amount of $^{15}$N--fractionation for temperatures higher than 10K 
has never been addressed by astrochemical models. 
The strong dependence of $^{15}$N--fractionation effects with temperature would imply that the $^{14}$N/$^{15}$N ratio
should vary within the sources with temperature gradients. However, this issue cannot be addressed 
in the present study, since to test this hypothesis it would be necessary to constrain the models on the basis of 
emission maps of the $^{15}$N--substituted species. 
Finally, the current model shows that at the core center,
all the molecules are affected by depletion onto dust grains. Thus, it cannot be discarded that fractionation effects 
could have occurred earlier in the life of the B1b cloud. However, the $^{15}$N enrichment would nowadays be unobservable, 
since the molecules formed in the previous colder evolutionary stages would have been subsequently 
incorporated into the ice mantles that surround the dust grains.

As discussed by \citet{wirstrom2012}, in order to link the prestellar phase
chemistry with the observations of the solar system, the astrochemical models have to be able to produce high 
degrees of fractionation for both $^{15}$N and D substituted isotopologues. Another prerequisite is that they also 
have to account for the fact that the meteoritic hot--spots for these two
isotopes are not necessarily coexistents.
In the present case, while no $^{15}$N--fractionation is seen for both the nitriles and nitrogen hydrides in B1b, 
we conclude that the nitrogen hydrides are highly fractioned in deuterium. The D--fractionation for the nitriles is also found to 
be high, even if lower than for the nitrogen hydride family. Hence, the gas depleted at this evolutionary stage 
will create hot--spots in D and will not be associated with high $^{15}$N enrichments.

\section{Conclusion} \label{conclusion}

We report a model based on continuum emission and molecular line observations for the B1b region,
located in the Perseus molecular cloud.
First, we modelled the 350 $\mu$m and 1.2 mm continuum radial profiles to obtain an estimate of the temperature 
and H$_2$ density distributions in the source. These density and temperature 
profiles were subsequently used as an input for the non--local radiative transfer modeling of the molecular lines.
The molecular data include D--, $^{13}$C-- and $^{15}$N--substituted isotopologues of molecules that 
belong to the nitrogen hydride family, i.e. N$_2$H$^+$ and NH$_3$, as well as molecules from the nitrile family, 
i.e. HCN, HNC and CN. For each molecule, the radial abundance profile was divided in several 
zones, in which the molecular abundance was considered as a free parameter. The best estimate for the 
abundance profile was then obtained from a Levenberg--Marquardt algorithm, which minimizes the $\chi^2$
between models and observations. Additionally, since our $^{13}$C-- and $^{15}$N-- observations are
only towards a single position, we assumed that the various isotopologues have abundance 
profiles that only differ by a scaling factor.

A common feature obtained for all the molecules is that in the innermost part of B1b, all the molecules
have abundances, which are at least one order of magnitude lower than the peak abundance found in the envelope. Since
all the molecules seem affected, this feature
is interpreted as depletion onto dust grains rather than due to chemical destruction, a process that might be invoked 
for N$_2$H$^+$ since the innermost region of B1b has a temperature which is close to the CO thermal desorption temperature. 
We derived high degrees of deuteration with the D/H ratio increasing inwards.
On the other hand, we derived similar $^{14}$N/$^{15}$N abundance ratios for all the molecules
and independently of the chemical family, with an average ratio of $\sim$300. This value is consistent with a recent 
estimate of the $^{14}$N/$^{15}$N atomic elemental abundance ratio obtained for warmer sources in local ISM.
 We thus conclude
that the ratio $^{14}$N/$^{15}$N$\sim$300 inferred from the molecules is representative of the
atomic elemental abundances of the parental molecular cloud. We propose that the absence of 
$^{15}$N--fractionation in B1b is a consequence of the relatively high gas temperature of this region, as compared to
the low temperatures that can be reached during the prestellar phase. The current state of the chemistry in B1b would
not produce any $^{15}$N--fractionation. On the other hand, it cannot be excluded by the current observations 
that conditions favorable to produce a $^{15}$N--enrichment were met during earlier evolutionary stages 
of the cloud. Since the molecules are found to be affected by depletion onto dust grains, 
the products of such a chemistry could have been incorporated into the ice mantles that surround the dust grains.

Given that $^{15}$N--fractionation effects can have strong implications for our 
understanding of the diversity observed in the various bodies of the solar system, it would be important to extend such a study 
focusing on earlier evolutionary stages with lower gas temperature, more favorable for producing nitrogen 
fractionation. Additionally, it was found in the current study that a direct estimate of the fractionation in the 
nitriles, i.e. HCN and HNC, is difficult because of the large opacities of the lines. 
This difficulty is not found for CN since this molecule is an open--shell molecule, which leads to a redistribution of the total
opacity of the rotational line over a larger number of radiative transitions.
Observations of less 
abundant species from this family, like HC$_3$N, would thus be particularly helpful. Moreover, 
we had to use the double isotope method to estimate the HCN and HNC column densities, 
from the H$^{13}$CN and HN$^{13}$C isotopologues. We assumed the standard $^{12}$C/$^{13}$C elemental ratio,
which introduces uncertainties due to possible source--to--source variations. It would thus be helpful to complement
the set of molecular observations with other carbon bearing molecules, like e.g. C$^{18}$O.

Finally, we have shown that the isotopic ratios are sensitive to the assumptions
made in deriving the column densities. Even relatively small
differences in excitation temperatures of different
isotopologues may lead to large variations in the derived isotopic ratios.
The use of a sophisticated radiative transfer method is therefore recommended.

\begin{acknowledgements}
This work is based on observations carried out with the IRAM 30m Telescope. IRAM is supported 
by INSU/CNRS (France), MPG (Germany) and IGN (Spain).
This work is based upon observations carried out at the Caltech Submillimeter Observatory, which is operated 
by the California Institute of Technology under cooperative agreement with the National Science 
Foundation (AST-0838261).
This paper was partially supported
within the programme CONSOLIDER INGENIO 2010, under grant \textit{Molecular
Astrophysics: The Herschel and ALMA Era.- ASTROMOL} (Ref.: CSD2009-
00038). We also thank the Spanish MICINN for funding support through
grants AYA2006-14876 and AYA2009-07304. 
\end{acknowledgements}

\appendix
\section{Inelastic rate coefficients for the CN isotopologues}

We used the two dimensional CN--He PES of \cite{lique2010} to determine hyperfine resolved excitation and de-excitation cross sections and rate coefficients of C$^{15}$N and $^{13}$CN molecules by He. CN--He rate coefficients obtained from this PES were found to be in excellent agreement with the experimental rotational (with unresolved fine and hyperfine structure) rate coefficients of \cite{fei1994} . This agreement demonstrate the accuracy of the PES and suggests that the rate coefficients obtained from this potential will be accurate enough to allow for improved astrophysical modeling of the ISM.

The Alexander's description \citep{alexander1982} of the inelastic scattering between an atom and a  diatomic molecule in a $^2\Sigma^+$ electronic state and a fully-quantum close-coupling calculations was used in order to obtain the $S^{J}(jl;j'l')$ diffusion matrix between fine structure levels of CN. $J$ and $l$ denote the total angular momentum ($\vec{J} = \vec{j} + \vec{l}$) and the orbital angular momentum quantum numbers, respectively. The nuclear spin-free $S^{J}(jl;j'l')$ matrix were computed recently in \cite{lique2011} and we used these matrices in the following. 

For C$^{15}$N, The coupling between the nuclear spin ($I=1/2$) of the $^{15}$N atom and the molecular rotation results in a weak splitting \citep{alexander1985} of each rotational level $j$. Each hyperfine level is designated by a quantum number $F$ ($F=I+j$) varying between $|I-j|$ and $I+j$. The integral cross sections corresponding to transitions between hyperfine levels of the C$^{15}$N molecule can be obtained from scattering S-matrix between fine structure levels using the re-coupling method of \cite{alexander1985}. Inelastic cross sections associated with a transition from the initial hyperfine level ($NjF$) to a final hyperfine level ($N'j'F'$) were thus obtained as follow :

\begin{eqnarray}
\sigma_{NjF \to N'j'F'}  & = & \frac{\pi}{k^{2}_{NjF}} (2F'+1) \sum_{K} \nonumber \\
& & \times \left\{ 
\begin{array}{ccc}
j & j' & K \\
F' & F & I 
\end{array}
\right\}^2
P^K(j \to j')
\end{eqnarray}
The $P^{K}(j \to j')$ are the tensor opacities defined by  :
\begin{equation}
P^{K}(j \to j')=\frac{1}{2K+1}\sum_{ll'}|T^{K}(jl;j'l')|^{2}
\end{equation}
The reduced T-matrix elements (where $T = 1 - S$) are defined by \cite{alexander1983}: 
\begin{eqnarray}
T^{K}(jl;j'l') & = & (-1)^{-j-l'}(2K+1)\sum_{J}(-1)^{J}(2J+1) \nonumber\\
& & \times  \left\{\begin{array}{ccc}
l' & j' & J \\ 
j & l & K 
\end{array}\right\}
T^{J}(jl;j'l')
\end{eqnarray}

For $^{13}$CN, the situation is more complex  since both the $^{13}$C and N atoms have non zero nuclear spin. The coupling between the nuclear spin ($I_1=1/2$) of the $^{13}$C atom and the molecular rotation is first taken into account. Each hyperfine level is designated by a quantum number $F_1$ ($F-1=I_1+j$) varying between $|I_1-j|$ and $I_1+j$. The coupling between the nuclear spin ($I_2=1$) of the N atom and the first hyperfine state is then taken into account. Each hyperfine level is designated by a quantum number $F$ ($F=I_2+F_1$) varying between $|I_2-F_1|$ and $I_2+F_1$.   The integral cross sections corresponding to transitions between hyperfine levels of the $^{13}$CN molecule can be obtained from scattering S-matrix between fine structure levels also using the re-coupling method of \cite{alexander1985}. Inelastic cross sections associated with a transition from the  initial hyperfine level ($NjF_1F$) to a final hyperfine level ($N'j'F_1'F'$) were thus obtained as follow \citep{daniel2004}:

\begin{eqnarray}
\sigma_{NjF_1F \to N'j'F_1'F'}  =  \frac{\pi}{k^{2}_{NjF}} (2F_1+1)(2F_1'+1)(2F'+1) \nonumber \\
\times \sum_{K}  \left\{ 
\begin{array}{ccc}
F_1 & F_1' & K \\
F' & F & I_2 
\end{array}
\right\}^2
 \left\{ 
\begin{array}{ccc}
j & j' & K \\
F_1' & F_1 & I_1 
\end{array}
\right\}^2
P^K(j \to j')
\end{eqnarray}

From the rotationally inelastic cross sections $\sigma_{\alpha \to \beta} (E_{c})$, one can obtain the corresponding thermal rate coefficients at temperature $T$ by an average over the collision energy ($E_c$) \citep{smith1980}:
\begin{eqnarray}
\label{thermal_average}
k_{\alpha \to \beta}(T) & = & \left(\frac{8}{\pi\mu k_B^3 T^3}\right)^{\frac{1}{2}}  \nonumber\\
&  & \times  \int_{0}^{\infty} \sigma_{\alpha \to \beta}\, E_{c}\, e^{-\frac{E_c}{k_BT}}\, dE_{c}
\end{eqnarray}
where $k_B$ is Boltzmann's constant and $\mu$ is the reduced mass of the CN--He complex. $\alpha$ and $\beta$ designate the initial and final states of the molecule, respectively. 

The complete set of (de)excitation rate coefficients for rotational transitions considered in this studies is available on-line from the BASECOL website\footnote{http://www.obspm.fr/basecol/}.

\end{document}